\documentclass[12pt]{article}
\usepackage{amsmath,amsthm}
\usepackage{eucal}
\usepackage{eufrak}
\DeclareMathAlphabet{\bb}{U}{msb}{m}{n}
\gdef\C{\bb C}
\gdef\dZ{\bb Z}

\gdef\R{\bb R}
\gdef\K{\bb K}
\gdef\BH{\bb H}
\gdef\F{\bb F}

\DeclareMathOperator{\End}{End}
\DeclareMathOperator{\spin}{{\bf Spin}}
\DeclareMathOperator{\pin}{{\bf Pin}}

\DeclareMathOperator{\Id}{Id}

\DeclareMathOperator{\Aut}{Aut}
\DeclareMathOperator{\sAut}{{\sf Aut}}
\DeclareMathOperator{\sExt}{{\sf Ext}}
\DeclareMathOperator{\Ker}{Ker}

\DeclareMathOperator{\Ext}{Ext}

\newcommand{\s}{\!}

\newcommand{\cA}{\mathcal{A}}

\newcommand{\cE}{\mathcal{E}}
\newcommand{\cN}{\mathcal{N}}

\newcommand{\M}{{\bf\sf M}}
\newcommand{\bsA}{{\bf\sf A}}
\newcommand{\sA}{{\sf A}}

\newcommand{\sI}{{\sf I}}
\newcommand{\sW}{{\sf W}}
\newcommand{\sE}{{\sf E}}
\newcommand{\sC}{{\sf C}}
\newcommand{\sF}{{\sf F}}
\newcommand{\sT}{{\sf T}}
\newcommand{\sS}{{\sf S}}

\newcommand{\sK}{{\sf K}}

\newcommand{\bi}{{\bf i}}
\newcommand{\bj}{{\bf j}}
\newcommand{\bk}{{\bf k}}
\newcommand{\bx}{{\bf x}}

\newcommand{\Lip}{\boldsymbol{\Gamma}}

\newcommand{\cl}{C\kern -0.2em \ell}

\newcommand{\p}{\prime}
\newcommand{\e}{\mbox{\bf e}}

\newtheorem{theorem}{Theorem}

\newtheorem{prop}{Proposition}

\begin{document}
\title{Group Theoretical Interpretation of the $CPT$-theorem}
\author{V.~V. Varlamov\\
{\small\it Department of Mathematics, Siberia State University of Industry,}\\
{\small\it Kirova 42, Novokuznetsk 654007, Russia}}
\date{}
\maketitle
\begin{abstract}
An algebraic description of basic discrete symmetries (space inversion $P$,
time reversal $T$, charge conjugation $C$ and their combinations $PT$, $CP$,
$CT$, $CPT$) is studied. Discrete subgroups $\{1,\,P,\,T,\,PT\}$ of
orthogonal groups of multidimensional spaces over the fields of real and
complex numbers are considered in terms of fundamental automorphisms of
Clifford algebras. The fundamental automorphisms form a finite group of
order 4. The charge conjugation is represented by a pseudoautomorphism
of the Clifford algebra. Such a description allows one to extend the
automorphism group. It is shown that an extended automorphism group
($CPT$-group) forms a finite group of order 8. The group structure and
isomorphisms between the extended automorphism groups and finite groups
are studied in detail. It is proved that there exist 64 different
realizations of $CPT$-group. An extension of universal coverings
(Clifford-Lipschitz groups) of the orthogonal groups is given in terms
of $CPT$-structures which include well-known Shirokov-D\c{a}browski
$PT$-structures as a particular case.
\end{abstract}
\section{Introduction}
Importance of discrete transformations is well--known, many textbooks
on quantum theory began with description of the discrete symmetries, and
famous L\"{u}ders--Pauli $CPT$--Theorem is a keystone of quantum field
theory. 
A fundamental notion of antimatter immediately relates with the
charge conjugation $C$. The requirement of invariance concerning the
each of the discrete transformations gives rise to certain consequences
which can be verified in experience. So, from the invariance concerning
the $CPT$-transformation it follows an identity of the masses and full life
times of particles and antiparticles, from the invariance concerning the
time reversal $T$ we have certain relations between the forward and
reverse reaction cross-sections. In turn, from the invariance concerning
the charge conjugation $C$ it follows an absence of the reactions
forbidden by the conservation law of charge parity, and from the
invariance concerning the space inversion $P$ and time reversal $T$
it follows an absence of the electric dipol moment of particles.
As follows from experience, for the processes defined by strong and 
electromagnetic interactions there exists an invariance with respect to
the all discrete transformations. In contrast to strong and electromagnetic
interactions, for the weak interaction, as shown by experiment, 
there is no invariance concerning space inversion $P$, but there is invariance
with respect to $CP$--transformation. Moreover, there are experimental
evidences confirming $CP$--violation (a decay of the neutral $K$--mesons).
It is clear that the analysis of the discrete symmetries allows us
to reveal the most profound structural characteristics of the matter.

However, usual practice of definition of the discrete symmetries
from the analysis of relativistic wave equations does not give a full and
consistent theory of the discrete transformations. In the standard approach,
except a well studied case of the spin $j=1/2$ (Dirac equation), a situation
with the discrete symmetries remains unclear for the fields of higher spin
$j>1/2$. It is obvious that a main reason of this is an absence of a fully
adequate formalism for description of higher--spin fields (all widely
accepted higher--spin formalisms such as Rarita--Schwinger approach \cite{RS41},
Bargmann--Wigner \cite{BW48} and Gel'fand--Yaglom \cite{GY48} multispinor
theories, and also Joos--Weinberg $2(2j+1)$--component formalism 
\cite{Joo62,Wein} have many intrinsic contradictions and difficulties).
The first attempt of going out from this situation was initiated by
Gel'fand, Minlos and Shapiro in 1958 \cite{GMS}. In the
Gel'fand--Minlos--Shapiro approach, the discrete symmetries are represented
by outer involutory automorphisms of the Lorentz group (there are also other
realizations of the discrete symmetries via the outer automorphisms, see
\cite{Mic64,Kuo71,Sil92}).
At present, the
Gel'fand--Minlos--Shapiro ideas have been found further development in the
works of Buchbinder, Gitman and Shelepin \cite{BGS00,GS01}, where the
discrete symmetries are represented by both outer and inner automorphisms of
the Poincar\'{e} group.

Discrete symmetries $P$ and $T$ transform (reflect) space and time
(two the most fundamental notions in physics), but in the Minkowski
4--dimensional space--time continuum \cite{Min} space and time are not
separate and independent. By this reason a transformation of one (space or
time) induces a transformation of another. Therefore, discrete symmetries
should be expressed by such trnsformations of the continuum, that transform
all its structure totally with a full preservation of discrete nature. The
only possible candidates on the role of such transformations are
automorphisms. In such a way, the idea of representation of the discrete
symmetries via the automorphisms of the Lorentz group (`rotation' group of the
4--dimensional continuum) is appearred in the Gel'fand--Minlos--Shapiro
approach, or via the automorphisms of the Poincar\'{e} group (motion
group of the 4--dimensional continuum) in the Buchbinder--Gitman--Shelepin
approach.

In 1909, Minkowski showed \cite{Min} that a causal structure of the world
is described by a 4--dimensional pseudo--Euclidean geometry. In accordance
with \cite{Min} the quadratic form $x^2+y^2+z^2-c^2t^2$ remains invariant
under the action of linear transformations of the four variables $x,y,z$ and 
$t$,
which form a general Lorentz group $G$. As is known, the general Lorentz group
$G$ consists of a proper orthochronous Lorentz group $G_0$ and three reflections
(discrete transformations) $P,\,T,\,PT$, where $P$ and $T$ are space and
time reversal, and $PT$ is a so--called full reflection. The discrete
transformations $P,\,T$ and $PT$ added to an identical transformation
form a finite group. Thus, the general Lorentz group may be represented by
a semidirect product $G_0\odot\{1,P,T,PT\}$. 

In 1958, Shirokov pointed out \cite{Shi58,Shi60} that an universal covering of the
inhomogeneous Lorentz group has eight inequivalent realizations. Later on,
in the eighties this idea was applied to a general orthogonal group
$O(p,q)$ by D\c{a}browski \cite{Dab88}.
As is known, the orthogonal
group $O(p,q)$ of the real space $\R^{p,q}$ is represented by the semidirect
product of a connected component $O_0(p,q)$ and a discrete subgroup
$\{1,P,T,PT\}$.
Further, a double covering of the orthogonal group $O(p,q)$ is a
Clifford--Lipschitz group $\pin(p,q)$ which is completely constructed within
a Clifford algebra $\cl_{p,q}$. In accordance with squares of elements of the
discrete subgroup ($a=P^2,\,b=T^2,\,c=(PT)^2$) there exist eight double
coverings (D\c{a}browski groups \cite{Dab88}) of the orthogonal group
defining by the signatures $(a,b,c)$, where $a,b,c\in\{-,+\}$. Such in brief is
a standard description scheme of the discrete transformations.
However, in this scheme there is one essential flaw. Namely, the
Clifford--Lipschitz group is an intrinsic notion of the algebra $\cl_{p,q}$
(a set of the all invertible elements of $\cl_{p,q}$), whereas the discrete
subgroup is introduced into the standard scheme in an external way, and the
choice of the signature $(a,b,c)$ of the discrete subgroup is not
determined by the signature of the space $\R^{p,q}$. Moreover, it is suggest
by default that for any signature $(p,q)$ of the vector space there exist
the all eight kinds of the discrete subgroups. It is obvious that a
consistent description of the double coverings of $O(p,q)$ in terms of
the Clifford--Lipschitz groups $\pin(p,q)\subset\cl_{p,q}$ can be obtained
only in the case when the discrete subgroup $\{1,P,T,PT\}$ is also defined
within the algebra $\cl_{p,q}$. Such a description has been given in the
works \cite{Var99,Var00,Var03}, where the discrete symmetries are
represented by fundamental automorphisms of the Clifford algebras.
So, the space inversion $P$, time reversal $T$ and their
combination $PT$ correspond to an automorphism $\star$
(involution), an antiautomorphism $\widetilde{\phantom{cc}}$ (reversion) and
an antiautomorphism $\widetilde{\star}$ (conjugation), respectively.
Group theoretical structure of the discrete transformations is a central
point in this work. The fundamental automorphisms of the Clifford algebras
are compared to elements of the finite group formed by the discrete
transformations. In its turn, a set of the fundamental automorphisms,
added by an identical automorphism, forms a finite group 
$\Aut(\cl)$, for which in virtue of the Wedderburn--Artin Theorem
there exists a matrix representation. In such a way, an isomorphism
$\{1,P,T,PT\}\simeq\Aut(\cl)$ plays a central role and allows us
to use methods of the
Clifford algebra theory at the study of a group theoretical structure
of the discrete transformations. First of all, it allows us to classify the
discrete groups into Abelian 
$\dZ_2\otimes\dZ_2$, $\dZ_4$ and non--Abelian
$D_4$, $Q_4$ finite groups, and also we establish a dependence between the
finite groups and signature of the spaces in case of real numbers.
It is shown that
the division ring structure of $\cl_{p,q}$ imposes hard restrictions on
existence and choice of the discrete subgroup, and the signature
$(a,b,c)$ depends upon the signature of the underlying space $\R^{p,q}$.
Moreover, this description allows us to incorporate the
Gel'fand--Minlos--Shapiro automorphism theory into 
Shirokov--D\c{a}browski scheme and further to unite them on the basis
of the Clifford algebra theory.

Other important discrete symmetry is the charge conjugation $C$. In contrast
with the transformations $P$, $T$, $PT$, the operation $C$ is not
space--time discrete symmetry. This transformation is first appearred
on the representation spaces of the Lorentz group and its nature is
strongly different from other discrete symmetries. By this reason in the
section 3 the charge conjugation $C$ is represented by a
pseudoautomorphism $\cA\rightarrow\overline{\cA}$ which is not fundamental
automorphism of the Clifford algebra. All spinor representations of the
pseudoautomorphism $\cA\rightarrow\overline{\cA}$ are given in
Theorem \ref{tpseudo}. An introduction of the transformation
$\cA\rightarrow\overline{\cA}$ allows us to extend the automorphism group
$\Aut(\cl)$ of the Clifford algebra. It is shown that automorphisms
$\cA\rightarrow\cA^\star$, $\cA\rightarrow\widetilde{\cA}$,
$\cA\rightarrow\widetilde{\cA^\star}$, $\cA\rightarrow\overline{\cA}$,
$\cA\rightarrow\overline{\cA^\star}$, 
$\cA\rightarrow\overline{\widetilde{\cA}}$ and
$\cA\rightarrow\overline{\widetilde{\cA^\star}}$ form a finite group of
order 8 (an extended automorphism group $\Ext(\cl)=\{\Id,\star,
\widetilde{\phantom{cc}},\widetilde{\star},\overline{\phantom{cc}},
\overline{\star},\overline{\widetilde{\phantom{cc}}},
\overline{\widetilde{\star}}\}$). The group $\Ext(\cl)$ is isomorphic
to a $CPT$-group $\{1,P,T,PT,C,CP,CT,CPT\}$. There exist isomorphisms
between $\Ext(\cl)$ and finite groups of order 8. A full number of different
realizations of $\Ext(\cl)$ is equal to 64. This result allows us to define
extended universal coverings ($CPT$-structures) of the orthogonal groups.
It is shown that the eight Shirokov-D\c{a}browski $PT$-structures present
a particular case of general $CPT$-structures.

\section{Clifford--Lipschitz groups}\index{group!Clifford-Lipschitz}
\label{Sec:1.4}
Theory of spinors and spinor representations (universal coverings of the
groups) is closely related with the foundations of quantum mechanics.
As is known, any quantized system $S$ corresponds to a complex separabel Hilbert
space\index{space!Hilbert}
$\mathcal{H}$. At this point, a physical state is represented by a
vector $\mid x\rangle\in\mathcal{H}$ and $\langle x\mid x\rangle=1$.
In turn, any physical observable $a$ (for example, energy, electric charge
and so on) corresponds to a self-conjugated operator $A$ in the space
$\mathcal{H}$. The spectrum of $A$ coincides with a set of all possible
values of $a$. Quantum mechanics does not give a definite value of the
magnitude $a$ in the state $\mid x\rangle$. We have here only a
mathematical expectation of $a$:
\[
\langle x\mid A\mid x\rangle=\text{Sp} AP_x,
\]
where $P_x$ is a Hermitean projection operator ($P^+_x=P_x$) on the state
$\mid x\rangle$. The unit
eigenvectors of $P_x$ are distinguished by the scalar phase factor. All
these vectors lead to identical physical predictions and, therefore,
they describe one and the same state. In turn, the operators $P_x$
in itself are observables. Indeed, the magnitude
\begin{equation}\label{QM}
\text{Sp}P_xP_y=|\langle x\mid y\rangle|^2
\end{equation}
is a location probability of the physical system $S$ in the state
$\mid x\rangle$ (or $\mid y\rangle$).

For the each quantized system $S$ there exists 
an invariance group\index{group!invariance} $G$.
It means that $G$ acts on the system $S$ and there exists an isomorphism
between $S$ and $g(S)$, $g\in G$. Let $P_{gx_i}$ be the state obtained
from $P_{x_i}$ via the transformation $g\in G$. Then, the group $G$ is
the invariance group if all probabilities in (\ref{QM}) are invariant:
\[
\forall\mid x\rangle\in\mathcal{H},\quad\forall g\in G,\quad
\text{Sp}P_{gx_1}P_{gx_2}=\text{Sp}P_{x_1}P_{x_2},
\]
or
\[
|\langle gx_1\mid gx_2\rangle|^2=|\langle x_1\mid x_2\rangle|^2.
\]
It means that $G$ acts on $\mathcal{H}$ isometrically. As it shown in
\cite{Weyl,Wigner,Barg64}, the mapping 
$\mid x\rangle\rightarrow\mid gx\rangle$ is either unitary operator
$U(g)$ or antiunitary operator $V(g)$ on the Hilbert space $\mathcal{H}$.
Let $\mathcal{V}(\mathcal{H})$ be a group of both unitary and antiunitary
operators on $\mathcal{H}$, and let $\mathcal{U}(\mathcal{H})$ be a subgroup
of unitary operators. The group $\mathcal{U}(\mathcal{H})$ is a subgroup of
index 2 of the group $\mathcal{V}(\mathcal{H})$, and by this reason
$\mathcal{U}(\mathcal{H})$ is an invariant subgroup of 
$\mathcal{V}(\mathcal{H})$. The transformations $\mathcal{U}(g)$ for
$g\in G$ generate a subgroup $\cE(G)$ of the group 
$\mathcal{V}(\mathcal{H})$, which is an extension of $G$ by means of the
group $U$ (multiplication of the vectors of $\mathcal{H}$ by the
phase factor $e^{i\alpha}$, where $\alpha$ is a real number; at this point
the state remains unaltered), that is,
\[
G\overset{f}{\longrightarrow}\Aut U,
\]
where a kernel\index{kernel}
of the mapping $\Ker f$ is an invariant subgroup
$G_+\subset G$ of index 2, which acts as the group of unitary
transformations, and a nontrivial element of the mapping image $\text{Im}f$
is the complex conjugation. Therefore, one can say that $G_+$ acts as a
linear unitary projective representation\index{representation!projective}. 
For example, if $G_+$ is a
rotation group\index{group!rotation}
$SO(3)$, then its projective representations are coincide
with linear irreducible unitary representations of the group $SU(2)$
(the group $SU(2)$ is an universal covering of $SO(3)$). In turn, universal
coverings of the orthogonal groups are completely defined within
Clifford--Lipschitz groups.

The Lipschitz group $\Lip_{p,q}$, also called the Clifford group, introduced
by Lipschitz in 1886 \cite{Lips}, may be defined as the subgroup of
invertible elements $s$ of the algebra $\cl_{p,q}$:
\[
\Lip_{p,q}=\left\{s\in\cl^+_{p,q}\cup\cl^-_{p,q}\;|\;\forall \bx\in\R^{p,q},\;
s\bx s^{-1}\in\R^{p,q}\right\}.
\]
The set $\Lip^+_{p,q}=\Lip_{p,q}\cap\cl^+_{p,q}$ is called {\it special
Lipschitz group}\index{group!special Lipschitz} \cite{Che55}.

Let $N:\;\cl_{p,q}\rightarrow\cl_{p,q},\;N(\bx)=\bx\widetilde{\bx}$.
If $\bx\in\R^{p,q}$, then $N(\bx)=\bx(-\bx)=-\bx^2=-Q(\bx)$. Further, the
group $\Lip_{p,q}$ has a subgroup
\begin{equation}\label{Pin}
\pin(p,q)=\left\{s\in\Lip_{p,q}\;|\;N(s)=\pm 1\right\}.
\end{equation}
Analogously, {\it a spinor group}\index{group!spinor}
$\spin(p,q)$ is defined by the set
\begin{equation}\label{Spin}
\spin(p,q)=\left\{s\in\Lip^+_{p,q}\;|\;N(s)=\pm 1\right\}.
\end{equation}
It is obvious that $\spin(p,q)=\pin(p,q)\cap\cl^+_{p,q}$.
The group $\spin(p,q)$ contains a subgroup
\begin{equation}\label{Spin+}
\spin_+(p,q)=\left\{s\in\spin(p,q)\;|\;N(s)=1\right\}.
\end{equation}
It is easy to see that the groups $O(p,q),\,SO(p,q)$ and $SO_+(p,q)$ are
isomorphic, respectively, 
to the following quotient groups\index{group!quotient}
\[
O(p,q)\simeq\pin(p,q)/\dZ_2,\quad
SO(p,q)\simeq\spin(p,q)/\dZ_2,\quad
SO_+(p,q)\simeq\spin_+(p,q)/\dZ_2,
\]
\begin{sloppypar}\noindent
where the kernel\index{kernel}
$\dZ_2=\{1,-1\}$. Thus, the groups $\pin(p,q)$, $\spin(p,q)$
and $\spin_+(p,q)$ are the double coverings of the groups $O(p,q),\,SO(p,q)$
and $SO_+(p,q)$, respectively.\end{sloppypar}

Further, since $\cl^+_{p,q}\simeq\cl^+_{q,p}$, then
\[
\spin(p,q)\simeq\spin(q,p).
\]
In contrast with this, the groups $\pin(p,q)$ and $\pin(q,p)$ are 
non--isomorphic. Denote $\spin(n)=\spin(n,0)\simeq\spin(0,n)$.
\begin{theorem}[{\rm\cite{Cor84}}]\label{t3}
The spinor groups
\[
\spin(2),\;\;\spin(3),\;\;\spin(4),\;\;\spin(5),\;\;\spin(6)
\]
are isomorphic to the unitary groups
\[
U(1),\;\;Sp(1)\sim SU(2),\;\;SU(2)\times SU(2),\;\;Sp(2),\;\;SU(4).
\]
\end{theorem} 
%In accordance with Theorem \ref{t1} and decompositions (\ref{e5'})
Over the field $\F=\R$ in the case of $p-q\equiv 1,5\pmod{8}$
the algebra $\cl_{p,q}$ is isomorphic to a direct
sum of two mutually annihilating 
simple ideals\index{ideal!mutually annihilating} $\frac{1}{2}(1\pm\omega)
\cl_{p,q}$: $\cl_{p,q}\simeq\frac{1}{2}(1+\omega)\cl_{p,q}\oplus\frac{1}{2}
(1-\omega)\cl_{p,q}$, where $\omega=\e_{12\ldots p+q},\,p-q\equiv 1,5
\pmod{8}$. At this point, the each ideal is isomorpic to $\cl_{p,q-1}$ or
$\cl_{q,p-1}$. Therefore, for the Clifford--Lipschitz groups we have the
following isomorphisms
\begin{eqnarray}
\pin(p,q)&\simeq&\pin(p,q-1)\bigcup\pin(p,q-1)\nonumber\\
&\simeq&\pin(q,p-1)\bigcup\pin(q,p-1).\label{Pinodd1}
\end{eqnarray}
Or, since $\cl_{p,q-1}\simeq\cl^+_{p,q}\subset\cl_{p,q}$, then 
according to (\ref{Spin+})
\[
\pin(p,q)\simeq\spin(p,q)\bigcup\spin(p,q)
\]
if $p-q\equiv 1,5\pmod{8}$.

Further, when $p-q\equiv 3,7\pmod{8}$, the algebra
$\cl_{p,q}$ is isomorphic to a complex algebra $\C_{p+q-1}$. Therefore,
for the $\pin$ groups we obtain
\begin{eqnarray}
\pin(p,q)&\simeq&\pin(p,q-1)\bigcup\e_{12\ldots p+q}\pin(p,q-1)\nonumber\\
&\simeq&\pin(q,p-1)\bigcup\e_{12\ldots p+q}\pin(q,p-1)\label{e13}
\end{eqnarray}
if $p-q\equiv 1,5\pmod{8}$ and correspondingly
\begin{equation}\label{e14}
\pin(p,q)\simeq\spin(p,q)\cup\e_{12\ldots p+q}\spin(p,q).
\end{equation}
In case of $p-q\equiv 3,7\pmod{8}$ we have isomorphisms which are analoguos
to (\ref{e13})-(\ref{e14}), since $\omega\cl_{p,q}\sim\cl_{p,q}$.
Generalizing these results, we obtain the following
\begin{theorem}\label{t4}
Let $\pin(p,q)$ and $\spin(p,q)$ be the Clifford-Lipschitz groups of the
invertible elements of the algebras $\cl_{p,q}$ with odd dimensionality,
$p-q\equiv 1,3,5,7\pmod{8}$. Then
\begin{eqnarray}
\pin(p,q)&\simeq&\pin(p,q-1)\bigcup\omega\pin(p,q-1)\nonumber\\
&\simeq&\pin(q,p-1)\bigcup\omega\pin(q,p-1)\nonumber
\end{eqnarray}
and
\[
\pin(p,q)\simeq\spin(p,q)\bigcup\omega\spin(p,q),
\]
where $\omega=\e_{12\ldots p+q}$ is a volume element of $\cl_{p,q}$.
\end{theorem}
In case of low dimensionalities from Theorem \ref{t3} and Theorem
\ref{t4} it immediately follows
\begin{theorem}\label{t5}
For $p+q\leq 5$ and $p-q\equiv 3,5\pmod{8}$,
\begin{eqnarray}
\pin(3,0)&\simeq&SU(2)\cup iSU(2),\nonumber\\
\pin(0,3)&\simeq&SU(2)\cup eSU(2),\nonumber\\
\pin(5,0)&\simeq&Sp(2)\cup eSp(2),\nonumber\\
\pin(0,5)&\simeq&Sp(2)\cup iSp(2).\nonumber
\end{eqnarray}
\end{theorem}
\begin{proof}\begin{sloppypar}\noindent
Indeed, in accordance with Theorem \ref{t4} $\pin(3,0)\simeq\spin(3)
\cup\e_{123}\spin(3)$. Further, from Theorem \ref{t3} we have
$\spin(3)\simeq SU(2)$, and a square of the element $\omega=\e_{123}$ is
equal to $-1$, therefore, $\omega\sim i$. Thus, $\pin(3,0)\simeq SU(2)\cup
iSU(2)$. For the group $\pin(0,3)$ a square of $\omega$ is equal to $+1$,
therefore, $\pin(0,3)\simeq SU(2)\cup eSU(2)$, $e$ is a double unit.
As expected, $\pin(3,0)\not\simeq\pin(0,3)$. The isomorphisms for the
groups $\pin(5,0)$ and $\pin(0,5)$ are analogously proved.\end{sloppypar}
\end{proof}

In turn, over the field $\F=\C$ there exists a complex Clifford--Lipschitz
group\index{group!complex Clifford-Lipschitz}
\[
\Lip_n=\left\{s\in\C^+_n\cup\C^-_n\;|\;\forall\bx\in\C^n,\;s\bx s^{-1}\in
\C^n\right\}.
\]
The group $\Lip_n$ has a subgroup
\begin{equation}\label{CL}
\pin(n,\C)=\left\{s\in\Lip_n\;|\; N(s)=\pm 1\right\}.
\end{equation}
$\pin(n,\C)$ is an universal covering of 
the complex orthogonal group\index{group!complex orthogonal}
$O(n,\C)$. When $n\equiv 1\pmod{2}$ we have
\begin{equation}\label{PinoddC}
\pin(n,\C)\simeq\pin(n-1,\C)\bigcup\e_{12\cdots n}\pin(n-1,\C).
\end{equation}

On the other hand, there exists a more detailed version of the $\pin$--group
(\ref{Pin}) proposed by D\c{a}browski in 1988 \cite{Dab88}. In general,
there are eight double coverings of the orthogonal group 
$O(p,q)$ \cite{Dab88,BD89}:
\[
\rho^{a,b,c}:\;\;\pin^{a,b,c}(p,q)\longrightarrow O(p,q),
\]
where $a,b,c\in\{+,-\}$. As is known, the group $O(p,q)$ consists of four
connected components: identity connected component $O_0(p,q)$, and three
components corresponding to space inversion $P$, time reversal
$T$, and the combination of these two $PT$, i.e., $O(p,q)=(O_0(p,q))\cup
P(Q_0(p,q))\cup T(O_0(p,q))\cup PT(O_0(p,q))$. Further, since the
four--element group (reflection group) $\{1,\,P,\,T,\,PT\}$ is isomorphic to
the finite group $\dZ_2\otimes\dZ_2$ 
(Gauss--Klein viergruppe\index{group!Gauss-Klein} \cite{Sal81a,Sal84}), then
$O(p,q)$ may be represented by 
a semidirect product\index{product!semidirect} $O(p,q)\simeq O_0(p,q)
\odot(\dZ_2\otimes\dZ_2)$. The signs of $a,b,c$ correspond to the signs of the
squares of the elements in $\pin^{a,b,c}(p,q)$ which cover space inversion
$P$, time reversal $T$ and a combination of these two
$PT$ ($a=-P^2,\,b=T^2,\,c=-(PT)^2$ in D\c{a}browski's notation \cite{Dab88} and
$a=P^2,\,b=T^2,\,c=(PT)^2$ in Chamblin's notation \cite{Ch94} which we will
use below).
An explicit form of the group $\pin^{a,b,c}(p,q)$ is given by the following
semidirect product
\begin{equation}\label{Pinabc}
\pin^{a,b,c}(p,q)\simeq\frac{(\spin_+(p,q)\odot C^{a,b,c})}{\dZ_2},
\end{equation}
where $C^{a,b,c}$ are the four double coverings of
$\dZ_2\otimes\dZ_2$. 
All the eight double coverings of the orthogonal group
$O(p,q)$ are given in the following table:
\begin{center}
{\renewcommand{\arraystretch}{1.4}
%{\renewcommand{\arraystretch}{0.85}
\begin{tabular}{|c|l|l|}\hline
$a$ $b$ $c$ & $C^{a,b,c}$ & Remark \\ \hline
$+$ $+$ $+$ & $\dZ_2\otimes\dZ_2\otimes\dZ_2$ & $PT=TP$\\
$+$ $-$ $-$ & $\dZ_2\otimes\dZ_4$ & $PT=TP$\\
$-$ $+$ $-$ & $\dZ_2\otimes\dZ_4$ & $PT=TP$\\
$-$ $-$ $+$ & $\dZ_2\otimes\dZ_4$ & $PT=TP$\\ \hline
$-$ $-$ $-$ & $Q_4$ & $PT=-TP$\\
$-$ $+$ $+$ & $D_4$ & $PT=-TP$\\
$+$ $-$ $+$ & $D_4$ & $PT=-TP$\\
$+$ $+$ $-$ & $D_4$ & $PT=-TP$\\ \hline
\end{tabular}
}
\end{center}
Here $\dZ_4$, $Q_4$, and $D_4$ are complex\index{group!complex}, 
quaternion\index{group!quaternionic}, and
dihedral groups\index{group!dihedral}, respectively.
According to \cite{Dab88} the group $\pin^{a,b,c}(p,q)$ satisfying the
condition
$PT=-TP$ is called {\it Cliffordian}\index{group!Cliffordina}, 
and respectively {\it non--Cliffordian}\index{group!non-Cliffordian} 
when $PT=TP$. 
\section{Discrete symmetries and Clifford algebras}
In Clifford algebra $\cl$ there exist four fundamental automorphisms.\\[0.2cm]
1) {\bf Identity}: An automorphism $\cA\rightarrow\cA$ and 
$\e_{i}\rightarrow\e_{i}$.\\
This automorphism is an identical automorphism of the algebra $\cl$. 
$\cA$ is an arbitrary element of $\cl$.\\[0.2cm]
2) {\bf Involution}: An automorphism $\cA\rightarrow\cA^\star$ and 
$\e_{i}\rightarrow-\e_{i}$.\\
In more details, for an arbitrary element $\cA\in\cl$ there exists a
decomposition
$
\cA=\cA^{\p}+\cA^{\p\p},
$
where $\cA^{\p}$ is an element consisting of homogeneous odd elements, and
$\cA^{\p\p}$ is an element consisting of homogeneous even elements,
respectively. Then the automorphism
$\cA\rightarrow\cA^{\star}$ is such that the element
$\cA^{\p\p}$ is not changed, and the element $\cA^{\p}$ changes sign:
$
\cA^{\star}=-\cA^{\p}+\cA^{\p\p}.
$
If $\cA$ is a homogeneous element, then
\begin{equation}\label{auto16}
\cA^{\star}=(-1)^{k}\cA,
\end{equation}
where $k$ is a degree of the element. It is easy to see that the
automorphism $\cA\rightarrow\cA^{\star}$ may be expressed via the volume
element $\omega=\e_{12\ldots p+q}$:
\begin{equation}\label{auto17}
\cA^{\star}=\omega\cA\omega^{-1},
\end{equation}
where
$\omega^{-1}=(-1)^{\frac{(p+q)(p+q-1)}{2}}\omega$. When $k$ is odd, the basis
elements 
$\e_{i_{1}i_{2}\ldots i_{k}}$ the sign changes, and when $k$ is even, the sign
is not changed.\\[0.2cm]
3) {\bf Reversion}: An antiautomorphism $\cA\rightarrow\widetilde{\cA}$ and
$\e_i\rightarrow\e_i$.\\
The antiautomorphism $\cA\rightarrow\widetilde{\cA}$ is a reversion of the
element $\cA$, that is the substitution of each basis element
$\e_{i_{1}i_{2}\ldots i_{k}}\in\cA$ by the element
$\e_{i_{k}i_{k-1}\ldots i_{1}}$:
\[
\e_{i_{k}i_{k-1}\ldots i_{1}}=(-1)^{\frac{k(k-1)}{2}}
\e_{i_{1}i_{2}\ldots i_{k}}.
\]
Therefore, for any $\cA\in\cl_{p,q}$ we have
\begin{equation}\label{auto19}
\widetilde{\cA}=(-1)^{\frac{k(k-1)}{2}}\cA.
\end{equation}
4) {\bf Conjugation}: An antiautomorpism $\cA\rightarrow\widetilde{\cA^\star}$
and $\e_i\rightarrow-\e_i$.\\
This antiautomorphism is a composition of the antiautomorphism
$\cA\rightarrow\widetilde{\cA}$ with the automorphism
$\cA\rightarrow\cA^{\star}$. In the case of a homogeneous element from
the formulae (\ref{auto16}) and (\ref{auto19}), it follows
\begin{equation}\label{20}
\widetilde{\cA^{\star}}=(-1)^{\frac{k(k+1)}{2}}\cA.
\end{equation}

One of the most fundamental theorems in the theory of associative algebras
is
\begin{theorem}[{\rm Wedderburn--Artin}]
Any finite--dimensional associative simple algebra $\mathfrak{A}$ over the
field $\F$ is isomorphic to a full matrix algebra $\M_n(\K)$, where a
natural number $n$ defined unambiguously, and a division ring $\K$ defined
with an accuracy of isomorphism.
\end{theorem}
In accordance with this theorem all properties of the initial algebra
$\mathfrak{A}$ are isomorphically transferred to the matrix algebra 
$\M_n(\K)$. Later on we will widely use this theorem. In its turn, for the
Clifford algebra $\cl_{p,q}$ over the field $\F=\R$ we have an isomorphism
$\cl_{p,q}\simeq\End_{\K}(I_{p,q})\simeq\M_{2^m}(\K)$, where $m=\frac{p+q}{2}$,
$I_{p,q}=\cl_{p,q}f$ is a minimal left ideal of $\cl_{p,q}$, and
$\K=f\cl_{p,q}f$ is a division ring of $\cl_{p,q}$. The primitive idempotent
of the algebra $\cl_{p,q}$ has a form
\[
f=\frac{1}{2}(1\pm\e_{\alpha_1})\frac{1}{2}(1\pm\e_{\alpha_2})\cdots\frac{1}{2}
(1\pm\e_{\alpha_k}),
\]
where $\e_{\alpha_1},\e_{\alpha_2},\ldots,\e_{\alpha_k}$ are commuting
elements with square 1 of the canonical basis of $\cl_{p,q}$ generating
a group of order $2^k$. The values of $k$ are defined by a formula
$k=q-r_{q-p}$, where $r_i$ are the Radon--Hurwitz numbers \cite{Rad22,Hur23},
values of which form a cycle of the period 8: $r_{i+8}=r_i+4$. The values of
all $r_i$ are
\begin{center}
\begin{tabular}{lcccccccc}
$i$  & 0 & 1 & 2 & 3 & 4 & 5 & 6 & 7\\ \hline
$r_i$& 0 & 1 & 2 & 2 & 3 & 3 & 3 & 3
\end{tabular}.
\end{center}
The all Clifford algebras $\cl_{p,q}$ over the field $\F=\R$ are divided
into eight different types with a following division ring structure:\\[0.3cm]
{\bf I}. Central simple algebras.
\begin{description}
\item[1)] Two types $p-q\equiv 0,2\pmod{8}$ with a division ring 
$\K\simeq\R$.
\item[2)] Two types $p-q\equiv 3,7\pmod{8}$ with a division ring
$\K\simeq\C$.
\item[3)] Two types $p-q\equiv 4,6\pmod{8}$ with a division ring
$\K\simeq\BH$.
\end{description}
{\bf II}. Semi--simple algebras.
\begin{description}
\item[4)] The type $p-q\equiv 1\pmod{8}$ with a double division ring
$\K\simeq\R\oplus\R$.
\item[5)] The type $p-q\equiv 5\pmod{8}$ with a double quaternionic 
division ring $\K\simeq\BH\oplus\BH$.
\end{description}
Over the field $\F=\C$ there is an isomorphism $\C_n\simeq\M_{2^{n/2}}(\C)$
and there are two different types of complex Clifford algebras $\C_n$:
$n\equiv 0\pmod{2}$ and $n\equiv 1\pmod{2}$.

In virtue of the Wedderburn--Artin theorem the all fundamental automorphisms
of $\cl$ are transferred to the matrix algebra. Matrix representations of the
fundamental automorphisms of $\C_n$ were first obtained by Rashevskii in 1955
\cite{Rash}: 1) Involution: $\sA^\star=\sW\sA\sW^{-1}$, where $\sW$ is a
matrix of the automorphism $\star$ (matrix representation of the volume
element $\omega$); 2) Reversion: $\widetilde{\sA}=\sE\sA^{\sT}\sE^{-1}$, where
$\sE$ is a matrix of the antiautomorphism $\widetilde{\phantom{cc}}$
satisfying the conditions $\cE_i\sE-\sE\cE^{\sT}_i=0$ and 
$\sE^{\sT}=(-1)^{\frac{m(m-1)}{2}}\sE$, here $\cE_i=\gamma(\e_i)$ are matrix
representations of the units of the algebra $\cl$; 3) Conjugation:
$\widetilde{\sA^\star}=\sC\sA^{\sT}\sC^{-1}$, where $\sC=\sE\sW^{\sT}$ 
is a matrix of
the antiautomorphism $\widetilde{\star}$ satisfying the conditions
$\sC\cE^{\sT}+\cE_i\sC=0$ and
$\sC^{\sT}=(-1)^{\frac{m(m+1)}{2}}\sC$.

In the recent paper \cite{Var99} it has been shown that space
reversal $P$, time reversal $T$ and combination $PT$ are correspond 
respectively to the fundamental automorphisms 
$\cA\rightarrow\cA^\star$, $\cA\rightarrow\widetilde{\cA}$ and
$\cA\rightarrow\widetilde{\cA^{\star}}$. 
\begin{prop}\label{prop1}
Let $\cl_{p,q}$ ($p+q=2m$) be a Clifford algebra over the field $\F=\R$ and
let $\pin(p,q)$ be a double covering of the orthogonal group $O(p,q)=O_0(p,q)
\odot\{1,P,T,PT\}\simeq O_0(p,q)\odot(\dZ_2\otimes\dZ_2)$ of transformations
of the space $\R^{p,q}$, where $\{1,P,T,PT\}\simeq\dZ_2\otimes\dZ_2$ is a
group of discrete transformations of $\R^{p,q}$, $\dZ_2\otimes\dZ_2$ is the
Gauss--Klein group. Then there is an isomorphism between the group
$\{1,P,T,PT\}$ and an automorphism group\index{group!automorphism}
$\{\Id,\star,\widetilde{\phantom{cc}},
\widetilde{\star}\}$ of the algebra $\cl_{p,q}$. In this case, space
inversion $P$, time reversal $T$ and combination $PT$ are correspond 
to the fundamental automorphisms $\cA\rightarrow\cA^\star,\,
\cA\rightarrow\widetilde{\cA}$ and $\cA\rightarrow\widetilde{\cA^\star}$.
\end{prop} 
An equivalence of the multiplication tables of the groups
$\{1,P,T,PT\}$ and 
$\Aut(\cl)=\{\Id,\star,\widetilde{\phantom{cc}},\widetilde{\star}\}$
proves this isomorphism
(in virtue of the commutativity $\widetilde{(\cA^\star)}=
(\widetilde{\cA})^\star$ and the 
involution conditions $(\star)^2=(\widetilde{\phantom{cc}
})^2=\Id$):
\[
{\renewcommand{\arraystretch}{1.4}
%{\renewcommand{\arraystretch}{0.85}
\begin{tabular}{|c||c|c|c|c|}\hline
        & $\Id$ & $\star$ & $\widetilde{\phantom{cc}}$ & $\widetilde{\star}$\\ \hline\hline
$\Id$   & $\Id$ & $\star$ & $\widetilde{\phantom{cc}}$ & $\widetilde{\star}$\\ \hline
$\star$ & $\star$ & $\Id$ & $\widetilde{\star}$ & $\widetilde{\phantom{cc}}$\\ \hline
$\widetilde{\phantom{cc}}$ & $\widetilde{\phantom{cc}}$ &$\widetilde{\star}$
& $\Id$ & $\star$ \\ \hline
$\widetilde{\star}$ & $\widetilde{\star}$ & $\widetilde{\phantom{cc}}$ &
$\star$ & $\Id$\\ \hline
\end{tabular}
\;\sim\;
\begin{tabular}{|c||c|c|c|c|}\hline
    & $1$ & $P$ & $T$ & $PT$\\ \hline\hline
$1$ & $1$ & $P$ & $T$ & $PT$\\ \hline
$P$ & $P$ & $1$ & $PT$& $T$\\ \hline
$T$ & $T$ & $PT$& $1$ & $P$\\ \hline
$PT$& $PT$& $T$ & $P$ & $1$\\ \hline
\end{tabular}
}
\]
Further, in the case $P^2=T^2=(PT)^2=\pm 1$ and $PT=-TP$ there is an
isomorphism between the group $\{1,P,T,PT\}$ and an automorphism group
$\sAut(\cl)=\{\sI,\sW,\sE,\sC\}$. So, for the Dirac algebra $\C_4$ in the
canonical $\gamma$--basis there exists a standard (Wigner) representation
$P=\gamma_0$ and $T=\gamma_1\gamma_3$ \cite{BLP89}, therefore,
$\{1,P,T,PT\}=\{1,\gamma_0,\gamma_1\gamma_3,\gamma_0\gamma_1\gamma_3\}$.
On the other hand, in the $\gamma$--basis an automorphism group
of $\C_4$ has a form $\sAut(\C_4)=\{\sI,\sW,\sE,\sC\}=
\{\sI,\gamma_0\gamma_1\gamma_2\gamma_3,\gamma_1\gamma_3,\gamma_0\gamma_2\}$.
It has been shown \cite{Var99} that $\{1,P,T,PT\}=
\{1,\gamma_0,\gamma_1\gamma_3,\gamma_0\gamma_1\gamma_3\}\simeq\sAut(\C_4)
\simeq\dZ_4$, where $\dZ_4$ is a complex group with the signature
$(+,-,-)$. Generalizations of these results onto the algebras $\C_n$
are contained in the following two theorems:
\begin{theorem}[{\rm\cite{Var99}}]\label{taut}
Let $\sAut=\{\sI,\,\sW,\,\sE,\,\sC\}$ be 
the automorphism group\index{group!automorphism} of the algebra
$\C_{p+q}$ $(p+q=2m)$, where 
$\sW=\cE_1\cE_2\cdots\cE_m\cE_{m+1}\cE_{m+2}\cdots\cE_{p+q}$,
and $\sE=\cE_1\cE_2\cdots\cE_m$, $\sC=\cE_{m+1}\cE_{m+2}\cdots\cE_{p+q}$ if
$m\equiv 1\pmod{2}$, and $\sE=\cE_{m+1}\cE_{m+2}\cdots\cE_{p+q}$, 
$\sC=\cE_1\cE_2\cdots
\cE_m$ if $m\equiv 0\pmod{2}$. Let $\sAut_-$ and $\sAut_+$ be the automorphism 
groups, in which the all elements correspondingly commute
$(m\equiv 0\pmod{2})$ and anticommute $(m\equiv 1\pmod{2})$.
Then over the field $\F=\C$ there are only two non--isomorphic groups:
$\sAut_-\simeq\dZ_2\otimes\dZ_2$ for the signature $(+,\,+,\,+)$ if
$n\equiv 0,1\pmod{4}$ and
$\sAut_+\simeq Q_4/\dZ_2$ for the signature
 $(-,\,-,\,-)$ if
$n\equiv 2,3\pmod{4}$.
\end{theorem}
\begin{theorem}[{\rm\cite{Var99}}]\label{t10}
Let $\pin^{a,b,c}(p,q)$ be a double covering of 
the complex orthogonal group\index{group!complex orthogonal}
$O(n,\C)$ of the space $\C^n$ associated with the complex algebra
$\C_n$.
Squares of the
symbols $a,b,c\in
\{-,+\}$ are correspond to squares of the elements of the finite group 
$\bsA=\{\sI,\sW,\sE,\sC\}:\;a=\sW^2,\,b=\sE^2,\,c=\sC^2$, where $\sW,\sE$
 and $\sC$
are correspondingly the matrices of the fundamental automorphisms $\cA\rightarrow
\cA^\star,\,\cA\rightarrow\widetilde{\cA}$ and $\cA\rightarrow
\widetilde{\cA^\star}$ of $\C_{n}$. Then over the field
$\F=\C$ for the algebra $\C_n$ there exist 
two non--isomorphic double coverings of the group
$O(n,\C)$:\\
1) Non--Cliffordian groups\index{group!non-Cliffordian}
\[
\pin^{+,+,+}(n,\C)\simeq\frac{(\spin_+(n,\C)\odot\dZ_2\otimes\dZ_2\otimes\dZ_2)}
{\dZ_2},
\]
if $n\equiv 0\pmod{4}$ and
\[
\pin^{+,+,+}(n,\C)\simeq\pin^{+,+,+}(n-1,\C)\bigcup\e_{12\ldots n}
\pin^{+,+,+}(n-1,\C),
\]
if $n\equiv 1\pmod{4}$.\\
2) Cliffordian groups\index{group!Cliffordian}
\[
\pin^{-,-,-}(n,\C)\simeq\frac{(\spin_+(n,\C)\odot Q_4)}{\dZ_2},
\]
if $n\equiv 2\pmod{4}$ and
\[
\pin^{-,-,-}(n,\C)\simeq\pin^{-,-,-}(n-1,\C)\bigcup\e_{12\ldots n}
\pin^{-,-,-}(n-1,\C),
\]
if $n\equiv 3\pmod{4}$.
\end{theorem}
A consideration of the discrete symmetries over the field of real numbers
is a much more complicated problem. First of all, in contrast to the field
of complex numbers over the field $\F=\R$ there exist eight different types
of the Clifford algebras and five division rings, which, in virtue of the
Wedderburn--Artin Theorem\index{theorem!Wedderbarn-Artin}, 
impose hard restrictions on existence and choice
of the matrix representations for the fundamental automorphisms.
\begin{theorem}\label{tautr}\begin{sloppypar}\noindent
Let $\cl_{p,q}$ be a Clifford algebra over a field $\F=\R$ and let
$\sAut(\cl_{p,q})=\{\sI,\sW,\sE,\sC\}$ be a group of fundamental
automorphisms\index{automorphism!fundamental}
of the algebra $\cl_{p,q}$. Then for eight types of the 
algebras $\cl_{p,q}$ there exist, depending upon a division ring structure
of $\cl_{p,q}$, following isomorphisms between finite groups and groups
$\sAut(\cl_{p,q})$ with different signatures
$(a,b,c)$, where $a,b,c\in\{-,+\}$:\\[0.2cm]
1) $\K\simeq\R$, types $p-q\equiv 0,2\pmod{8}$.\\
If $\sE=\cE_{p+1}\cE_{p+2}\cdots\cE_{p+q}$ and $\sC=\cE_1\cE_2\cdots\cE_p$,
then Abelian groups $\sAut_-(\cl_{p,q})\simeq\dZ_2\otimes\dZ_2$
with the signature $(+,+,+)$ and $\sAut_-(\cl_{p,q})\simeq\dZ_4$ with the
signature
$(+,-,-)$ exist at $p,q\equiv 0\pmod{4}$ and $p,q\equiv 2\pmod{4}$, 
respectively,
for the type $p-q\equiv 0\pmod{8}$, and also Abelian groups
$\sAut_-(\cl_{p,q})\simeq\dZ_4$ with the signature $(-,-,+)$ and
$\sAut_-(\cl_{p,q})
\simeq\dZ_4$ with the signature $(-,+,-)$ exist 
at $p\equiv 0\pmod{4},\,
q\equiv 2\pmod{4}$ and $p\equiv 2\pmod{4},\,q\equiv 0\pmod{4}$ for the type
$p-q\equiv 2\pmod{8}$, respectively.\\
If $\sE=\cE_1\cE_2\cdots\cE_p$ and $\sC=\cE_{p+1}\cE_{p+2}\cdots\cE_{p+q}$,
then non--Abelian groups $\sAut_+(\cl_{p,q})\simeq D_4/\dZ_2$ with the
signature $(+,-,+)$ and $\sAut_+(\cl_{p,q})\simeq D_4/\dZ_2$ with the
signature
$(+,+,-)$ exist at $p,q\equiv 3\pmod{4}$ and $p,q\equiv 1\pmod{4}$, 
respectively,
for the type $p-q\equiv 0\pmod{8}$, and also non--Abelian groups
$\sAut_+(\cl_{p,q})\simeq Q_4/\dZ_2$ with $(-,-,-)$ and 
$\sAut_+(\cl_{p,q})\simeq
D_4/\dZ_2$ with $(-,+,+)$ exist at $p\equiv 3\pmod{4},\,q\equiv 1
\pmod{4}$ and $p\equiv 1\pmod{4},\,q\equiv 3\pmod{4}$ for the type
$p-q\equiv 2\pmod{8}$, respectively.\\[0.2cm]
2) $\K\simeq\BH$, types $p-q\equiv 4,6\pmod{8}$.\\
If $\sE=\cE_{j_1}\cE_{j_2}\cdots\cE_{j_k}$ is a product of $k$
skewsymmetric matrices (among which $l$ matrices have a square $+\sI$
and $t$ matrices have a square $-\sI$)
and $\sC=\cE_{i_1}\cE_{i_2}\cdots\cE_{i_{p+q-k}}$ is a product of $p+q-k$
symmetric matrices (among which $h$ matrices have a square $+\sI$ and
$g$ have a square $-\sI$),
then at $k\equiv 0\pmod{2}$ for the type $p-q\equiv 4\pmod{8}$ there exist
Abelian groups $\sAut_-(\cl_{p,q})\simeq\dZ_2\otimes\dZ_2$ with $(+,+,+)$
and $\sAut_-(\cl_{p,q})\simeq\dZ_4$ with $(+,-,-)$ if
$l-t,\,h-g\equiv 0,1,4,5\pmod{8}$ and
$l-t,\,h-g\equiv 2,3,6,7\pmod{8}$, respectively. And also at
$k\equiv 0\pmod{2}$ for the type $p-q\equiv 6\pmod{8}$ there exist
$\sAut_-(\cl_{p,q})\simeq\dZ_4$ with $(-,+,-)$ and 
$\sAut_-(\cl_{p,q})\simeq\dZ_4$
with $(-,-,+)$ if $l-t\equiv 0,1,4,5\pmod{8},\,
h-g\equiv 2,3,6,7\pmod{8}$ and $l-t\equiv 2,3,6,7\pmod{8},\,
h-g\equiv 0,1,4,5\pmod{8}$,respectively.\\
Inversely, if $\sE=\cE_{i_1}\cE_{i_2}\cdots\cE_{i_{p+q-k}}$ is a product of
$p+q-k$ symmetric matrices and 
$\sC=\cE_{j_1}\cE_{j_2}\cdots\cE_{j_k}$ is a product of $k$ skewsymmetric
matrices, then at $k\equiv 1\pmod{2}$
for the type $p-q\equiv 4\pmod{8}$ there exist non--Abelian groups
$\sAut_+(\cl_{p,q})
\simeq D_4/\dZ_2$ with $(+,-,+)$ and $\sAut_+(\cl_{p,q})\simeq D_4/\dZ_2$ with
$(+,+,-)$ if $h-g\equiv 2,3,6,7\pmod{8},\,l-t\equiv
0,1,4,5\pmod{8}$ and $h-g\equiv 0,1,4,5\pmod{8},\,l-t\equiv 2,3,6,7\pmod{8}$,
respectively.
And also at $k\equiv 1\pmod{2}$ for the type $p-q\equiv 6\pmod{8}$ there exist
$\sAut_+(\cl_{p,q})\simeq Q_4/\dZ_2$ with $(-,-,-)$ and $\sAut_+(\cl_{p,q})
\simeq D_4/\dZ_2$ with $(-,+,+)$ if $h-g,
\,l-t\equiv 2,3,6,7\pmod{8}$ and $h-g,\,l-t\equiv 0,1,4,5
\pmod{8}$, respectively.\\[0.2cm]
3) $\K\simeq\R\oplus\R,\,\K\simeq\BH\oplus\BH$, types $p-q\equiv 1,5\pmod{8}$.\\
For the algebras $\cl_{0,q}$ of the types $p-q\equiv 1,5\pmod{8}$ there exist
Abelian automorphism groups with the signatures
$(-,-,+)$, $(-,+,-)$ and non--Abelian automorphism groups with the signatures
$(-,-,-)$, $(-,+,+)$. Correspondingly, for the algebras $\cl_{p,0}$ of the
types $p-q\equiv 1,5\pmod{8}$ there exist Abelian groups with
$(+,+,+)$, $(+,-,-)$ and non--Abelian groups with $(+,-,+)$,
$(+,+,-)$. In general case for $\cl_{p,q}$, the types $p-q\equiv 1,5\pmod{8}$
admit all eight automorphism groups.\\[0.2cm]
4) $\K=\C$, types $p-q\equiv 3,7\pmod{8}$.\\
The types $p-q\equiv 3,7\pmod{8}$ admit the Abelian group $\sAut_-(\cl_{p,q})
\simeq\dZ_2\otimes\dZ_2$ with the signature $(+,+,+)$ if $p\equiv 0\pmod{2}$ and
$q\equiv 1\pmod{2}$, and also non--Abelian group 
$\sAut_+(\cl_{p,q})\simeq
Q_4/\dZ_2$ with the signature $(-,-,-)$ if $p\equiv 1\pmod{2}$ and
$q\equiv 0\pmod{2}$. \end{sloppypar}
\end{theorem} 
\begin{theorem}\label{tgroupr}\begin{sloppypar}\noindent
Let $\pin^{a,b,c}(p,q)$ be a double covering of the orthogonal group
$O(p,q)$ of the real space $\R^{p,q}$ associated with the algebra
$\cl_{p,q}$.
The squares of symbols $a,b,c\in
\{-,+\}$ correspond to the squares of the elements of a finite group
$\sAut(\cl_{p,q})=\{\sI,\sW,\sE,\sC\}:\;a=\sW^2,\,b=\sE^2,\,c=\sC^2$, 
where $\sW,\sE$ and $\sC$
are the matrices of the 
fundamental automorphisms\index{automorphism!fundamental} $\cA\rightarrow
\cA^\star,\,\cA\rightarrow\widetilde{\cA}$ and $\cA\rightarrow
\widetilde{\cA^\star}$ of the algebra $\cl_{p,q}$, respectively.
Then over the field $\F=\R$ 
in dependence on a division ring structure of the algebra $\cl_{p,q}$,
there exist eight double coverings of the orthogonal group $O(p,q)$:\\[0.2cm]
1) A non--Cliffordian group\index{group!non-Cliffordian}\end{sloppypar}
\[
\pin^{+,+,+}(p,q)\simeq\frac{(\spin_0(p,q)\odot\dZ_2\otimes\dZ_2\otimes\dZ_2)}
{\dZ_2},
\]
exists if $\K\simeq\R$ and the numbers $p$ and $q$ form the type 
$p-q\equiv 0\pmod{8}$ and $p,q\equiv 0\pmod{4}$, and also if
$p-q\equiv 4\pmod{8}$ and $\K\simeq\BH$. The algebras $\cl_{p,q}$ with the
rings $\K\simeq\R\oplus\R,\,\K\simeq\BH\oplus\BH$ ($p-q\equiv 1,5\pmod{8}$)
admit the group $\pin^{+,+,+}(p,q)$ if in the direct sums there are
addendums of the type
$p-q\equiv 0\pmod{8}$ or $p-q\equiv 4\pmod{8}$. The types $p-q\equiv 3,7
\pmod{8}$, $\K\simeq\C$ admit a non--Cliffordian group $\pin^{+,+,+}(p+q-1,
\C)$ if $p\equiv 0\pmod{2}$ and $q\equiv 1\pmod{2}$. Further, 
non--Cliffordian groups
\[
\pin^{a,b,c}(p,q)\simeq\frac{(\spin_0(p,q)\odot(\dZ_2\otimes\dZ_4)}{\dZ_2},
\]
with $(a,b,c)=(+,-,-)$ exist if $p-q\equiv 0\pmod{8}$, 
$p,q\equiv 2\pmod{4}$ and $\K\simeq\R$, and also if
$p-q\equiv 4\pmod{8}$ and $\K\simeq\BH$. Non--Cliffordian
groups with the signatures
$(a,b,c)=(-,+,-)$ and $(a,b,c)=(-,-,+)$ exist over the ring
$\K\simeq\R$ ($p-q\equiv 2\pmod{8}$) if $p\equiv
2\pmod{4},\,q\equiv 0\pmod{4}$ and $p\equiv 0\pmod{4},\,q\equiv 2\pmod{4}$,
respectively,
and also these groups exist over the ring $\K\simeq\BH$ if
$p-q\equiv 6\pmod{8}$. 
The algebras $\cl_{p,q}$ with the rings
$\K\simeq\R\oplus\R,\,\K\simeq\BH\oplus\BH$ ($p-q\equiv 1,5\pmod{8}$)
admit the group $\pin^{+,-,-}(p,q)$ if in the direct sums there are addendums
of the type $p-q\equiv 0\pmod{8}$ or $p-q\equiv 4\pmod{8}$, and also admit the
groups $\pin^{-,+,-}(p,q)$ and $\pin^{-,-,+}(p,q)$ if in the direct sums
there are addendums of the type $p-q\equiv 2\pmod{8}$ 
or $p-q\equiv 6\pmod{8}$.\\[0.2cm]
2) A Cliffordian group\index{group!Cliffordian}
\[
\pin^{-,-,-}(p,q)\simeq\frac{(\spin_0(p,q)\odot Q_4)}{\dZ_2},
\]
exists if $\K\simeq\R$ ($p-q\equiv 2\pmod{8}$) and $p\equiv 3\pmod{4},\,
q\equiv 1\pmod{4}$, and also if $p-q\equiv 6\pmod{8}$ and $\K\simeq\BH$.
The algebras $\cl_{p,q}$ with the rings 
$\K\simeq\R\oplus\R,\,\K\simeq\BH\oplus\BH$ ($p-q\equiv 1,5\pmod{8}$)
admit the group $\pin^{-,-,-}(p,q)$ if in the direct sums there are 
addendums of the type
$p-q\equiv 2\pmod{8}$ or $p-q\equiv 6\pmod{8}$. The types $p-q\equiv 3,7
\pmod{8}$, $\K\simeq\C$ admit a Cliffordian group $\pin^{-,-,-}(p+q-1,\C)$,
if $p\equiv 1\pmod{2}$ and $q\equiv 0\pmod{2}$. Further, Cliffordian groups
\[
\pin^{a,b,c}(p,q)\simeq\frac{(\spin_0(p,q)\odot D_4)}{\dZ_2},
\]
with $(a,b,c)=(-,+,+)$ exist if $\K\simeq\R$ ($p-q\equiv 2\pmod{8}$)
and $p\equiv 1\pmod{4},\,q\equiv 3\pmod{4}$,
and also if $p-q\equiv 6\pmod{8}$ and $\K\simeq\BH$. Cliffordian groups with
the signatures
$(a,b,c)=(+,-,+)$ and $(a,b,c)=(+,+,-)$ exist over the ring
$\K\simeq\R$ ($p-q\equiv 0\pmod{8}$) if
$p,q\equiv 3\pmod{4}$ and $p,q\equiv 1\pmod{4}$, respectively,
and also these groups
exist over the ring $\K\simeq\BH$ if $p-q\equiv 4\pmod{8}$.
The algebras $\cl_{p,q}$ with the rings 
$\K\simeq\R\oplus\R,\,\K\simeq\BH\oplus\BH$ ($p-q\equiv 1,5\pmod{8}$)
admit the group $\pin^{-,+,+}(p,q)$ if in the direct sums there are addendums
of the type $p-q\equiv 2\pmod{8}$ or $p-q\equiv 6\pmod{8}$, and also admit the
groups $\pin^{+,-,+}(p,q)$ and $\pin^{+,+,-}(p,q)$ if in the direct sums there
are addendums of the type $p-q\equiv 0\pmod{8}$ or $p-q\equiv 4\pmod{8}$.
\end{theorem}
%\chapter{$CPT$--theorem}
\section{Pseudoautomorphism $\cA\longrightarrow\overline{\cA}$
and charge conjugation}
As is known, the algebra $\C_n$ is associated with a complex vector
space $\C^n$. Let $n=p+q$, then an extraction operation of the real subspace
$\R^{p,q}$ in $\C^n$  forms the foundation of definition of the discrete
transformation known in physics as
{\it a charge conjugation} $C$. Indeed, let
$\{\e_1,\ldots,\e_n\}$ be an orthobasis in the space $\C^n$, $\e^2_i=1$.
Let us remain the first $p$ vectors of this basis unchanged, and other $q$
vectors multiply by the factor $i$. Then the basis
\begin{equation}\label{6.23}
\left\{\e_1,\ldots,\e_p,i\e_{p+1},\ldots,i\e_{p+q}\right\}
\end{equation}
allows us to extract the subspace $\R^{p,q}$ in $\C^n$. Namely,
for the vectors $\R^{p,q}$ we take the vectors of
$\C^n$ which decompose on the basis
(\ref{6.23}) with real coefficients. In this way, we obtain a real vector
space $\R^{p,q}$ endowed (in general case) with a non--degenerate
quadratic form\index{form!quadratic}
\[
Q(x)=x^2_1+x^2_2+\ldots+x^2_p-x^2_{p+1}-x^2_{p+2}-\ldots-x^2_{p+q},
\]
where $x_1,\ldots,x_{p+q}$ are coordinates of the vector $\bx$ 
in the basis (\ref{6.23}).
It is easy to see that the extraction of
$\R^{p,q}$ in $\C^n$ induces an extraction of
{\it a real subalgebra}\index{subalgebra!real}
$\cl_{p,q}$ in $\C_n$. Therefore, any element
$\cA\in\C_n$ can be unambiguously represented in the form
\[
\cA=\cA_1+i\cA_2,
\]
where $\cA_1,\,\cA_2\in\cl_{p,q}$. The one-to-one mapping
\begin{equation}\label{6.24}
\cA\longrightarrow\overline{\cA}=\cA_1-i\cA_2
\end{equation}
transforms the algebra $\C_n$ into itself with preservation of addition
and multiplication operations for the elements $\cA$; the operation of
multiplication of the element $\cA$ by the number transforms to an operation
of multiplication by the complex conjugate number.
Any mapping of $\C_n$ satisfying these conditions is called
{\it a pseudoautomorphism}.\index{pseudoautomorphism}
Thus, the extraction of the subspace
$\R^{p,q}$ in the space $\C^n$ induces in the algebra $\C_n$ 
a pseudoautomorphism $\cA\rightarrow\overline{\cA}$ \cite{Rash,Ras58}.

Let us consider a spinor representation of the pseudoautomorphism
$\cA\rightarrow\overline{\cA}$ of the algebra $\C_n$ when $n\equiv 0\s\pmod{2}$.
In the spinor representation the every element $\cA\in\C_n$ should be
represented by some matrix $\sA$, and the pseudoautomorphism (\ref{6.24})
takes a form of the pseudoautomorphism of the full 
matrix algebra\index{algebra!matrix}
$\M_{2^{n/2}}$:
\[
\sA\longrightarrow\overline{\sA}.
\]\begin{sloppypar}\noindent
On the other hand, a transformation replacing the matrix $\sA$ by the
complex conjugate matrix, $\sA\rightarrow\dot{\sA}$, is also some
pseudoautomorphism of the algebra $\M_{2^{n/2}}$. The composition of the two
pseudoautomorpisms $\dot{\sA}\rightarrow\sA$ and
$\sA\rightarrow\overline{\sA}$, $\dot{\sA}\rightarrow\sA\rightarrow
\overline{\sA}$, is an internal automorphism\index{automorphism!internal}
$\dot{\sA}\rightarrow\overline{\sA}$ of the full matrix algebra $\M_{2^{n/2}}$:
\end{sloppypar}
\begin{equation}\label{6.25}
\overline{\sA}=\Pi\dot{\sA}\Pi^{-1},
\end{equation}
where $\Pi$ is a matrix of the pseudoautomorphism 
$\cA\rightarrow\overline{\cA}$ in the spinor representation.
The sufficient condition for definition of the pseudoautomorphism
$\cA\rightarrow\overline{\cA}$ is a choice of the matrix
$\Pi$ in such a way that the transformation 
$\sA\rightarrow\Pi\dot{\sA}\Pi^{-1}$ transfers into itself the matrices
$\cE_1,\ldots,\cE_p,i\cE_{p+1},\ldots,i\cE_{p+q}$
(the matrices of the spinbasis of $\cl_{p,q}$), that is,
\begin{equation}\label{6.26}
\cE_i\longrightarrow\cE_i=\Pi\dot{\cE}_i\Pi^{-1}\quad
(i=1,\ldots,p+q).
\end{equation}
\begin{theorem}\label{tpseudo}
Let $\C_n$ be a complex Clifford algebra for $n\equiv 0\s\pmod{2}$
and let $\cl_{p,q}\subset\C_n$ be its subalgebra with a real division ring
$\K\simeq\R$ when $p-q\equiv 0,2\s\pmod{8}$ and quaternionic division ring
$\K\simeq\BH$ when $p-q\equiv 4,6\s\pmod{8}$, $n=p+q$. Then in dependence
on the division ring structure of the real subalgebra $\cl_{p,q}$ the matrix
$\Pi$ of the pseudoautomorphism $\cA\rightarrow\overline{\cA}$ 
has the following form:\\[0.2cm]
1) $\K\simeq\R$, $p-q\equiv 0,2\s\pmod{8}$.\\[0.1cm]
The matrix $\Pi$ for any spinor representation over the ring $\K\simeq\R$
is proportional to the unit matrix.\\[0.2cm]
2) $\K\simeq\BH$, $p-q\equiv 4,6\s\pmod{8}$.\\[0.1cm]
$\Pi=\cE_{\alpha_1}\cE_{\alpha_2}\cdots\cE_{\alpha_a}$ when 
$a\equiv 0\s\pmod{2}$ and
$\Pi=\cE_{\beta_1}\cE_{\beta_2}\cdots\cE_{\beta_b}$ when $b\equiv 1\s\pmod{2}$,
where $a$ complex matrices $\cE_{\alpha_t}$ 
and $b$ real matrices $\cE_{\beta_s}$ form a basis of the spinor
representation of the algebra $\cl_{p,q}$ over the ring $\K\simeq\BH$,
$a+b=p+q,\,0<t\leq a,\,0<s\leq b$. At this point,
\begin{eqnarray}
\Pi\dot{\Pi}&=&\phantom{-}\sI\quad\text{if $a,b\equiv 0,1\s\pmod{4}$},
\nonumber\\
\Pi\dot{\Pi}&=&-\sI\quad\text{if $a,b\equiv 2,3\s\pmod{4}$},\nonumber
\end{eqnarray}
where $\sI$ is the unit matrix.
\end{theorem}
\begin{proof}\begin{sloppypar}\noindent
The algebra $\C_n$ ($n\equiv 0\s\pmod{2}$, $n=p+q$) in virtue of
$\C_n=\C\otimes\cl_{p,q}$ and definition of the division ring
$\K\simeq f\cl_{p,q}f$ 
($f$ is a primitive idempotent of the algebra $\cl_{p,q}$)
has four different real subalgebras: $p-q\equiv 0,2\s\pmod{8}$
for the real division ring $\K\simeq\R$ and $p-q\equiv 4,6\s\pmod{8}$ for
the quaternionic division ring $\K\simeq\BH$.\\[0.2cm]
1) $\K\simeq\R$.\\[0.1cm]
Since for the types $p-q\equiv 0,2\s\pmod{8}$ there is an isomorphism
$\cl_{p,q}\simeq\M_{2^{\frac{p+q}{2}}}(\R)$ (Wedderburn--Artin Theorem), then
all the matrices $\cE_i$ of the spinbasis of $\cl_{p,q}$ are real and
$\dot{\cE}_i=\cE_i$. Therefore, in this case the condition (\ref{6.26})
can be written as follows\end{sloppypar}
\[
\cE_i\longrightarrow\cE_i=\Pi\cE_i\Pi^{-1},
\]
whence $\cE_i\Pi=\Pi\cE_i$. Thus, for the algebras $\cl_{p,q}$ of the types
$p-q\equiv 0,2\s\pmod{8}$ the matrix $\Pi$ of the pseudoautomorphism
$\cA\rightarrow\overline{\cA}$ commutes with all the matrices $\cE_i$.
It is easy to see that
$\Pi\sim\sI$.\\[0.2cm]
2) $\K\simeq\BH$.\\[0.1cm]
In turn, for the quaternionic types $p-q\equiv 4,6\s\pmod{8}$ there is an
isomorphism $\cl_{p,q}\simeq\M_{2^{\frac{p+q}{2}}}(\BH)$. Therefore, among
the matrices of the spinbasis of the algebra $\cl_{p,q}$ there are matrices
$\cE_\alpha$ satisfying the condition $\dot{\cE}_\alpha=-\cE_\alpha$. 
Let $a$ be a quantity of the complex matrices, then the spinbasis of $\cl_{p,q}$
is divided into two subsets. The first subset
$\{\dot{\cE}_{\alpha_t}=-\cE_{\alpha_t}\}$ contains complex matrices,
$0<t\leq a$, and the second subset
$\{\dot{\cE}_{\beta_s}=\cE_{\beta_s}\}$ contains real matrices,
$0<s\leq p+q-a$. In accordance with a spinbasis structure of the algebra
$\cl_{p,q}\simeq\M_{2^{\frac{p+q}{2}}}(\BH)$, the condition (\ref{6.26})
can be written as follows
\[
\cE_{\alpha_t}\longrightarrow-\cE_{\alpha_t}=\Pi\cE_{\alpha_t}\Pi^{-1},\quad
\cE_{\beta_s}\longrightarrow\cE_{\beta_s}=\Pi\cE_{\beta_s}\Pi^{-1}.
\]
Whence
\begin{equation}\label{6.27}
\cE_{\alpha_t}\Pi=-\Pi\cE_{\alpha_t},\quad
\cE_{\beta_s}\Pi=\Pi\cE_{\beta_s}.
\end{equation}
Thus, for the quaternionic types $p-q\equiv 4,6\s\pmod{8}$ the matrix
$\Pi$ of the pseudoautomorphism $\cA\rightarrow\overline{\cA}$ anticommutes
with a complex part of the spinbasis of $\cl_{p,q}$ and commutes with
a real part of the same spinbasis. From (\ref{6.27}) it follows that a
structure of the matrix $\Pi$ is analogous to the structure of
the matrices $\sE$ and $\sC$ of the antiautomorphisms
$\cA\rightarrow\widetilde{\cA}$ and
$\cA\rightarrow\widetilde{\cA^\star}$, correspondingly 
(see Theorem \ref{tautr}), that is, the matrix
$\Pi$ of the pseudoautomorphism $\cA\rightarrow\overline{\cA}$ of the algebra
$\C_n$ is a product of only complex matrices, or only real matrices
of the spinbasis of the subalgebra $\cl_{p,q}$.

So, let $0<a<p+q$ and let $\Pi=\cE_{\alpha_1}\cE_{\alpha_2}\cdots
\cE_{\alpha_a}$ be a matrix of $\cA\rightarrow\overline{\cA}$,
then permutation conditions of the matrix $\Pi$ 
with the matrices $\cE_{\beta_s}$
of the real part ($0<s\leq p+q-a$) and with the matrices
$\cE_{\alpha_t}$ of the complex part ($0<t\leq a$) have the form
\begin{equation}\label{6.28}
\Pi\cE_{\beta_s}=(-1)^a\cE_{\beta_s}\Pi,
\end{equation}
\begin{eqnarray}
\Pi\cE_{\alpha_t}&=&(-1)^{a-t}\sigma(\alpha_t)\cE_{\alpha_1}\cE_{\alpha_2}
\cdots\cE_{\alpha_{t-1}}\cE_{\alpha_{t+1}}\cdots\cE_{\alpha_a},\nonumber\\
\cE_{\alpha_t}\Pi&=&(-1)^{t-1}\sigma(\alpha_t)\cE_{\alpha_1}\cE_{\alpha_2}
\cdots\cE_{\alpha_{t-1}}\cE_{\alpha_{t+1}}\cdots\cE_{\alpha_a},\label{6.29}
\end{eqnarray}
that is, when $a\equiv 0\s\pmod{2}$ the matrix $\Pi$ commutes with the real
part and anticommutes with the complex part of the spinbasis of $\cl_{p,q}$.
Correspondingly, when $a\equiv 1\s\pmod{2}$ the matrix $\Pi$ anticommutes
with the real part and commutes with the complex part. Further, let
$\Pi=\cE_{\beta_1}\cE_{\beta_2}\cdots\cE_{\beta_{p+q-a}}$ be a product of the
real matrices, then
\begin{eqnarray}
\Pi\cE_{\beta_s}&=&(-1)^{p+q-a-s}\sigma(\beta_s)\cE_{\beta_1}\cE_{\beta_2}
\cdots\cE_{\beta_{s-1}}\cE_{\beta_{s+1}}\cdots\cE_{\beta_{p+q-a}},\nonumber\\
\cE_{\beta_s}\Pi&=&(-1)^{s-1}\sigma(\beta_s)\cE_{\beta_1}\cE_{\beta_2}
\cdots\cE_{\beta_{s-1}}\cE_{\beta_{s+1}}\cdots\cE_{\beta_{p+q-a}},\label{6.30}
\end{eqnarray}
\begin{equation}\label{6.31}
\Pi\cE_{\alpha_t}=(-1)^{p+q-a}\cE_{\alpha_t}\Pi,
\end{equation}
that is, when $p+q-a\equiv 0\s\pmod{2}$ the matrix $\Pi$ anticommutes with
the real part and commutes with the complex part 
of the spinbasis of $\cl_{p,q}$. Correspondingly, when
$p+q-a\equiv 1\s\pmod{2}$ the matrix $\Pi$ commutes with the real part and
anticommutes with the complex part.

The comparison of the conditions (\ref{6.28})--(\ref{6.29}) 
with the condition (\ref{6.27}) shows that the matrix
$\Pi=\cE_{\alpha_1}\cE_{\alpha_2}\cdots\cE_{\alpha_a}$ exists only at
$a\equiv 0\s\pmod{2}$, that is, $\Pi$ is a product of the complex matrices
$\cE_{\alpha_t}$ of the even number. In turn, a comparison of
(\ref{6.30})--(\ref{6.31}) with
(\ref{6.27}) shows that the matrix $\Pi=\cE_{\beta_1}\cE_{\beta_2}\cdots
\cE_{\beta_{p+q-a}}$ exists only at $p+q-a\equiv 1\s\pmod{2}$, that is,
$\Pi$ is a product of the real matrices $\cE_{\beta_s}$ of the odd number.

Let us calculate now the product $\Pi\dot{\Pi}$. 
Let $\Pi=\cE_{\beta_1}\cE_{\beta_2}\cdots
\cE_{\beta_{p+q-a}}$ be a product of the $p+q-a$ real matrices.
Since $\dot{\cE}_{\beta_s}=\cE_{\beta_s}$, then
$\dot{\Pi}=\Pi$ and $\Pi\dot{\Pi}=\Pi^2$. Therefore,
\begin{equation}\label{6.32}
\Pi\dot{\Pi}=(\cE_{\beta_1}\cE_{\beta_2}\cdots\cE_{\beta_{p+q-a}})^2=
(-1)^{\frac{(p+q-a)(p+q-a-1)}{2}}\cdot\sI.
\end{equation}
Further, let $\Pi=\cE_{\alpha_1}\cE_{\alpha_2}\cdots\cE_{\alpha_a}$ be a
product of the $a$ complex matrices. Then
$\dot{\cE}_{\alpha_t}=-\cE_{\alpha_t}$ and $\dot{\Pi}=(-1)^a\Pi=\Pi$, since
$a\equiv 0\s\pmod{2}$. Therefore,
\begin{equation}\label{6.33}
\Pi\dot{\Pi}=(\cE_{\alpha_1}\cE_{\alpha_2}\cdots\cE_{\alpha_a})^2=
(-1)^{\frac{a(a-1)}{2}}\cdot\sI.
\end{equation}
Let $p+q-a=b$ be a quantity of the real matrices $\cE_{\beta_s}$ of the
spinbasis of $\cl_{p,q}$, then $p+q=a+b$. Since $p+q$ is always even number
for the quaternionic types $p-q\equiv 4,6\s\pmod{8}$, then $a$ and $b$ 
are simultaneously even or odd numbers. Thus, from (\ref{6.32}) and (\ref{6.33})
it follows
\[
\Pi\dot{\Pi}=\begin{cases}
\phantom{-}\sI,& \text{if $a,b\equiv 0,1\s\!\!\pmod{4}$},\\
-\sI,& \text{if $a,b\equiv 2,3\s\!\!\pmod{4}$},
\end{cases}
\]
which required to be proved.
\end{proof}
%\begin{rem}
In the present form of quantum field theory complex fields correspond
to charged particles. Thus, the extraction of the subalgebra $\cl_{p,q}$ with
the real ring $\K\simeq\R$ in $\C_n$, $p-q\equiv 0,2\s\pmod{8}$,
corresponds to physical fields describing {\it truly neutral particles}
such as photon and neutral mesons ($\pi^0,\,\eta^0,\,\rho^0,\,
\omega^0,\,\varphi^0,\,K^0$). In turn, the subalgebras $\cl_{p,q}$ with the
ring $\K\simeq\BH$, $p-q\equiv 4,6\s\pmod{8}$, correspond to charged or
neutral fields.
%\end{rem}

\section{Extended automorphism groups}
An introduction of the pseudoautomorphism $\cA\rightarrow\overline{\cA}$
allows us to extend the automorphism set of the complex Clifford algebra
$\C_n$. Namely, we add to the four fundamental automorphisms
$\cA\rightarrow\cA$, $\cA\rightarrow\cA^\star$, 
$\cA\rightarrow\widetilde{\cA}$, $\cA\rightarrow\widetilde{\cA^\star}$
the pseudoautomorphism $\cA\rightarrow\overline{\cA}$ and
following three combinations:\\
1) A pseudoautomorphism $\cA\rightarrow\overline{\cA^\star}$. This 
transformation is a composition of the pseudoautomorphism
$\cA\rightarrow\overline{\cA}$ with 
the automorphism $\cA\rightarrow\cA^\star$.\\
2) A pseudoantiautomorphism\index{pseudoantiautomorphism}
$\cA\rightarrow\overline{\widetilde{\cA}}$.
This transformation is a composition of $\cA\rightarrow\overline{\cA}$ with
the antiautomorphism $\cA\rightarrow\widetilde{\cA}$.\\
3) A pseudoantiautomorphism $\cA\rightarrow\overline{\widetilde{\cA^\star}}$
(a composition of $\cA\rightarrow\overline{\cA}$ with the antiautomorphism
$\cA\rightarrow\widetilde{\cA^\star}$).

Thus, we obtain an automorphism set of $\C_n$ consisting of the
eight transformations. Let us show that the set
$\{\Id,\,\star,\,\widetilde{\phantom{cc}},\,\widetilde{\star},\,
\overline{\phantom{cc}},\,\overline{\star},\,
\overline{\widetilde{\phantom{cc}}},\,\overline{\widetilde{\star}}\}$
forms a finite group of order 8 and 
let us give a physical interpretation of this
group.
\begin{prop}
Let $\C_n$ be a Clifford algebra over the field $\F=\C$ and let
$\Ext(\C_n)=
\{\Id,\,\star,\,\widetilde{\phantom{cc}},\,\widetilde{\star},\,
\overline{\phantom{cc}},\,\overline{\star},\,
\overline{\widetilde{\phantom{cc}}},\,\overline{\widetilde{\star}}\}$
be an extended automorphism group\index{group!automorphism!extended}
of the algebra $\C_n$. Then there is
an isomorphism between $\Ext(\C_n)$ and the full $CPT$--group
of the discrete transformations,
$\Ext(\C_n)\simeq\{1,\,P,\,T,\,PT,\,C,\,CP,\,CT,\,CPT\}\simeq
\dZ_2\otimes\dZ_2\otimes\dZ_2$. In this case, space inversion $P$, time
reversal $T$, full reflection $PT$, charge conjugation $C$, transformations
$CP$, $CT$ and the full $CPT$--transformation correspond to the automorphism
$\cA\rightarrow\cA^\star$, antiautomorphisms $\cA\rightarrow\widetilde{\cA}$,
$\cA\rightarrow\widetilde{\cA^\star}$, pseudoautomorphisms
$\cA\rightarrow\overline{\cA}$, $\cA\rightarrow\overline{\cA^\star}$,
pseudoantiautomorphisms $\cA\rightarrow\overline{\widetilde{\cA}}$ and
$\cA\rightarrow\overline{\widetilde{\cA^\star}}$, respectively.
\end{prop}
\begin{proof}\begin{sloppypar}\noindent
The group $\{1,\,P,\,T,\,PT,\,C,\,CP,\,CT,\,CPT\}$ at the conditions
$P^2=T^2=(PT)^2=C^2=(CP)^2=(CT)^2=(CPT)^2=1$ and commutativity of all the
elements forms an Abelian group of order 8, which is isomorphic to a cyclic
group $\dZ_2\otimes\dZ_2\otimes\dZ_2$. 
The Cayley tableau\index{tableau!Cayley}
of this group has a form\end{sloppypar}
\begin{center}{\renewcommand{\arraystretch}{1.4}
\begin{tabular}{|c||c|c|c|c|c|c|c|c|}\hline
     & $1$  & $P$  & $T$  & $PT$ & $C$  & $CP$ & $CT$ & $CPT$ \\ \hline\hline
$1$  & $1$  & $P$  & $T$  & $PT$ & $C$  & $CP$ & $CT$ & $CPT$ \\ \hline
$P$  & $P$  & $1$  & $PT$ & $T$  & $CP$ & $C$  & $CPT$& $CT$\\ \hline
$T$  & $T$  & $PT$ & $1$  & $P$  & $CT$ & $CPT$& $C$  & $CP$\\ \hline
$PT$ & $PT$ & $T$  & $P$  & $1$  & $CPT$& $CT$ & $CP$ & $C$\\ \hline
$C$  & $C$  & $CP$ & $CT$ & $CPT$& $1$  & $P$  & $T$  & $PT$\\ \hline
$CP$ & $CP$ & $C$  & $CPT$& $CT$ & $P$  & $1$  & $PT$ & $T$\\ \hline
$CT$ & $CT$ & $CPT$& $C$  & $CP$ & $T$  & $PT$ & $1$  & $P$\\ \hline
$CPT$& $CPT$& $CT$ & $CP$ & $C$  & $PT$ & $T$  & $P$  & $1$\\ \hline
\end{tabular}
}
\end{center}
In turn, for the extended automorphism group
$\{\Id,\,\star,\,\widetilde{\phantom{cc}},\,\widetilde{\star},\,
\overline{\phantom{cc}},\,\overline{\star},\,
\overline{\widetilde{\phantom{cc}}},\,\overline{\widetilde{\star}}\}$
in virtue of commutativity $\widetilde{\left(\cA^\star\right)}=
\left(\widetilde{\cA}\right)^\star$, 
$\overline{\left(\cA^\star\right)}=\left(\overline{\cA}\right)^\star$,
$\overline{\left(\widetilde{\cA}\right)}=
\widetilde{\left(\overline{\cA}\right)}$,
$\overline{\left(\widetilde{\cA^\star}\right)}=
\widetilde{\left(\overline{\cA}\right)^\star}$ and an involution property
$\star\star=\widetilde{\phantom{cc}}\widetilde{\phantom{cc}}=
\overline{\phantom{cc}}\;\overline{\phantom{cc}}=\Id$ we have a following
Cayley tableau\index{tableau!Cayley}
\begin{center}{\renewcommand{\arraystretch}{1.4}
\begin{tabular}{|c||c|c|c|c|c|c|c|c|}\hline
  & $\Id$ & $\star$ & $\widetilde{\phantom{cc}}$ & $\widetilde{\star}$ &
$\overline{\phantom{cc}}$ & $\overline{\star}$ & 
$\overline{\widetilde{\phantom{cc}}}$ &
$\overline{\widetilde{\star}}$ \\ \hline\hline
$\Id$ & $\Id$ & $\star$ & $\widetilde{\phantom{cc}}$ & $\widetilde{\star}$ &
$\overline{\phantom{cc}}$ & $\overline{\star}$ & 
$\overline{\widetilde{\phantom{cc}}}$ &
$\overline{\widetilde{\star}}$ \\ \hline
$\star$ & $\star$ & $\Id$ & $\widetilde{\star}$ & $\widetilde{\phantom{cc}}$ &
$\overline{\star}$ & $\overline{\phantom{cc}}$ &
$\overline{\widetilde{\star}}$ & $\overline{\widetilde{\phantom{cc}}}$\\ \hline
$\widetilde{\phantom{cc}}$ & 
$\widetilde{\phantom{cc}}$ & $\overline{\star}$ & $\Id$ &
$\star$ & $\overline{\widetilde{\phantom{cc}}}$ & $\overline{\widetilde{\star}}$ &
$\overline{\phantom{cc}}$ & $\overline{\star}$\\ \hline
$\widetilde{\star}$ & $\widetilde{\star}$ & $\widetilde{\phantom{cc}}$ &
$\star$ & $\Id$ & $\overline{\widetilde{\star}}$ & 
$\overline{\widetilde{\phantom{cc}}}$ &
$\overline{\star}$ & $\overline{\phantom{cc}}$\\ \hline
$\overline{\phantom{cc}}$ & $\overline{\phantom{cc}}$ & $\overline{\star}$ &
$\overline{\widetilde{\phantom{cc}}}$ & $\overline{\widetilde{\star}}$ & $\Id$ &
$\star$ & $\widetilde{\phantom{cc}}$ & $\widetilde{\star}$\\ \hline
$\overline{\star}$ & $\overline{\star}$ & $\overline{\phantom{cc}}$ &
$\overline{\widetilde{\star}}$ & 
$\overline{\widetilde{\phantom{cc}}}$ & $\star$ &
$\Id$ & $\widetilde{\star}$ & $\widetilde{\phantom{cc}}$\\ \hline
$\overline{\widetilde{\phantom{cc}}}$ & 
$\overline{\widetilde{\phantom{cc}}}$ &
$\overline{\widetilde{\star}}$ & 
$\overline{\phantom{cc}}$ & $\overline{\star}$ &
$\widetilde{\phantom{cc}}$ & $\widetilde{\star}$ & $\Id$ & $\star$\\ \hline
$\overline{\widetilde{\star}}$ & $\overline{\widetilde{\star}}$ &
$\overline{\widetilde{\phantom{cc}}}$ & $\overline{\star}$ & 
$\overline{\phantom{cc}}$ &
$\widetilde{\star}$ & $\widetilde{\phantom{cc}}$ & $\star$ & $\Id$\\ \hline
\end{tabular}
}
\end{center}
The identity of multiplication tables proves the group isomorphism
\[
\{1,\,P,\,T,\,PT,\,C,\,CP,\,CT,\,CPT\}\simeq
\{\Id,\,\star,\,\widetilde{\phantom{cc}},\,\widetilde{\star},\,
\overline{\phantom{cc}},\,\overline{\star},\,
\overline{\widetilde{\phantom{cc}}},\,\overline{\widetilde{\star}}\}\simeq
\dZ_2\otimes\dZ_2\otimes\dZ_2.
\]
\end{proof}

Further, in the case of $P^2=T^2=\ldots=(CPT)^2=\pm 1$ and anticommutativity
of the elements we have an isomorphism between the $CPT$--group and a group
$\sExt(\C_n)$. The elements of $\sExt(\C_n)$ are spinor
representations of the automorphisms of the algebra $\C_n$.
The Wedderburn--Artin Theorem allows us to define any spinor representaions for
the automorphisms of $\C_n$.
\subsection{Pseudoautomorphism $\cA\rightarrow\overline{\cA^\star}$}
Let us find a spinor representation of 
the pseudoautomorphism\index{pseudoautomorphism}
$\cA\rightarrow\overline{\cA^\star}$. The transformation
$\cA\rightarrow\overline{\cA^\star}$ is a composition of the
pseudoautomorphism $\cA\rightarrow\overline{\cA}$ and the automorphism
$\cA\rightarrow\cA^\star$. Under action of $\cA\rightarrow\cA^\star$ we have
$\e_i\rightarrow-\e_i$, where $\e_i$ are the units of $\cl_{p,q}$.
In turn, under action of $\cA\rightarrow\overline{\cA}$ the units $\e_i$
remain unaltered, $\e_i\rightarrow\e_i$. Therefore, under action of the
pseudoautomorphism $\cA\rightarrow\overline{\cA^\star}$ we obtain
$\e_i\rightarrow-\e_i$.

As it shown previously, the transformations $\cA\rightarrow\cA^\star$
and $\cA\rightarrow\overline{\cA}$ in the spinor representation are defined
by the expressions $\sA^\star=\sW\sA\sW^{-1}$ and 
$\overline{\sA}=\Pi\dot{\sA}\Pi^{-1}$. The order of the composition of these
transformations is not important ($\overline{\cA^\star}=
\left(\overline{\cA}\right)^\star=\overline{\left(\cA^\star\right)}$).
Indeed, if $\sW$ is a real matrix, then
\[
\overline{\sA^\star}=\sW\Pi\dot{\sA}\Pi^{-1}\sW^{-1}=
\Pi\left(\sW\sA\sW^{-1}\right)^{\cdot}\Pi^{-1},
\]
or
\begin{equation}\label{CPT1}
\overline{\sA^\star}=(\sW\Pi)\dot{\sA}(\sW\Pi)^{-1}=
(\Pi\sW)\dot{\sA}(\Pi\sW)^{-1}.
\end{equation}
Otherwise, we have $\overline{\left(\sA^\star\right)}=
\Pi\dot{\sW}\dot{\sA}\dot{\sW}^{-1}\Pi^{-1}$. Let us assume that $\sW$ is a
complex matrix, then $\dot{\sW}=-\sW$ and, therefore,
$\overline{\left(\sA^\star\right)}=\Pi(-\sW)\dot{\sA}(-\sW^{-1})\Pi^{-1}=
(\Pi\sW)\dot{\sA}(\Pi\sW)^{-1}$. Thus, the relation (\ref{CPT1}) is always
fulfilled.

Let $\sK=\Pi\sW$ be a matrix of the pseudoautomorphism
$\cA\rightarrow\overline{\cA^\star}$. Then (\ref{CPT1}) can be written as
follows
\begin{equation}\label{CPT2}
\overline{\sA^\star}=\sK\dot{\sA}\sK^{-1}.
\end{equation}
Since under action of the pseudoautomorphism 
$\cA\rightarrow\overline{\cA^\star}$ we have $\e_i\rightarrow-\e_i$,
in the spinor representation we must demand
$\cE_i\rightarrow-\cE_i$ also, or
\begin{equation}\label{CPT3}
\cE_i\longrightarrow -\cE_i=\sK\dot{\cE}_i\sK^{-1}.
\end{equation}
In the case of real subalgebras $\cl_{p,q}$ with the ring $\K\simeq\R$ we have
$\dot{\cE}_i=\cE_i$ and the relation (\ref{CPT3}) takes a form
\[
\cE_i\longrightarrow -\cE_i=\sK\cE_i\sK^{-1},
\]
whence
\[
\cE_i\sK=-\sK\cE_i,
\]
that is, the matrix $\sK$ is always anticommutes with the matrices of the
spinbasis. However, for the ring $\K\simeq\R$ the matrix $\Pi$ of
$\cA\rightarrow\overline{\cA}$ is proportional to the unit matrix,
$\Pi\sim\sI$ (Theorem \ref{tpseudo}). Therefore, in this case we have
$\sK\sim\sW$.

In the case of real subalgebras $\cl_{p,q}$ with the quaternionic ring
$\K\simeq\BH$ the spinbasis is divided into two parts:
a complex part $\left\{\dot{\cE}_{\alpha_t}=-\cE_{\alpha_t}\right\}$,
$(0<t\leq a)$, where $a$ is a number of the complex matrices of the spinbasis,
and a real part $\left\{\dot{\cE}_{\beta_s}=\cE_{\beta_s}\right\}$,
where $p+q-a$ is a number of the real matrices, $(0<s\leq p+q-a)$.
Then, in accordance with the spinbasis structure of the algebra
$\cl_{p,q}\simeq\M_{2^{\frac{p+q}{2}}}(\BH)$, the relation (\ref{CPT3})
can be written as follows
\[
\cE_{\alpha_t}\longrightarrow\cE_{\alpha_t}=\sK\cE_{\alpha_t}\sK^{-1},\quad
\cE_{\beta_s}\longrightarrow -\cE_{\beta_s}=\sK\cE_{\beta_s}\sK^{-1}.
\]
Whence
\begin{equation}\label{CPT4}
\cE_{\alpha_t}\sK=\sK\cE_{\alpha_t},\quad
\cE_{\beta_s}\sK=-\sK\cE_{\beta_s}.
\end{equation}
Thus, for the quaternionic types $p-q\equiv 4,6\pmod{8}$ the matrix $\sK$ of
the pseudoautomorphism $\cA\rightarrow\overline{\cA^\star}$ commutes with
the complex part and anticommutes with the real part of the spinbasis of
$\cl_{p,q}$. Hence it follows that a structure of the matrix $\sK$ is
analogous to the structure of the matrix $\Pi$ of the pseudoautomorphism
$\cA\rightarrow\overline{\cA}$ (see Theorem \ref{tpseudo}), that is,
the matrix $\sK$ of $\cA\rightarrow\overline{\cA^\star}$ is a product of
only complex or only real matrices.

So, let $0< a\leq p+q$ and let 
$\sK=\cE_{\alpha_1}\cE_{\alpha_2}\cdots\cE_{\alpha_a}$ be the matrix of the
pseudoautomorphism $\cA\rightarrow\overline{\cA^\star}$, then permutation
conditions of the matrix $\sK$ with the matrices $\cE_{\beta_s}$ of the
real part ($0< s\leq p+q-a$) and the matrices $\cE_{\alpha_t}$ of the
complex part ($0< t \leq a$) have the form
\begin{gather}
\sK\cE_{\beta_s}=(-1)^a\cE_{\beta_s}\sK,\label{CPT5}\\
\sK\cE_{\alpha_t}=(-1)^{a-t}\sigma(\alpha_t)
\cE_{\alpha_1}\cE_{\alpha_2}\cdots
\cE_{\alpha_{t-1}}\cE_{\alpha_{t+1}}\cdots\cE_{\alpha_a},\nonumber\\
\cE_{\alpha_t}\sK=(-1)^{t-1}\sigma(\alpha_t)
\cE_{\alpha_1}\cE_{\alpha_2}\cdots
\cE_{\alpha_{t-1}}\cE_{\alpha_{t+1}}\cdots\cE_{\alpha_a},\label{CPT6}
\end{gather}
that is, at $a\equiv 0\pmod{2}$ $\sK$ commutes with the real part and
anticommutes with the complex part of the spinbasis. Correspondingly,
at $a\equiv 1\pmod{2}$ $\sK$ anticommutes with the real and commutes with
the complex part. Further, let
$\sK=\cE_{\beta_1}\cE_{\beta_2}\cdots\cE_{\beta_{p+q-a}}$ be a product of the
real matrices of the spinbasis, then
\begin{eqnarray}
\sK\cE_{\beta_s}&=&(-1)^{p+q-a-s}\sigma(\beta_s)
\cE_{\beta_1}\cE_{\beta_2}\cdots\cE_{\beta_{s-1}}\cE_{\beta_{s+1}}\cdots
\cE_{\beta_{p+q-a}},\nonumber\\
\cE_{\beta_s}\sK&=&(-1)^{s-1}\sigma(\beta_s)\cE_{\beta_1}\cE_{\beta_2}\cdots
\cE_{\beta_{s-1}}\cE_{\beta_{s+1}}\cdots\cE_{\beta_{p+q-a}},\label{CPT7}
\end{eqnarray}
\begin{equation}\label{CPT8}
\sK\cE_{\alpha_t}=(-1)^{p+q-a}\cE_{\alpha_t}\sK,
\end{equation}
that is, at $p+q-a\equiv 0\pmod{2}$ the matrix $\sK$ anticommutes with the
real part and commutes with the complex part of the spinbasis.
Correspondingly, at $p+q-a\equiv 1\pmod{2}$ $\sK$ commutes with the real
and anticommutes with the complex part.
\begin{sloppypar}
A comparison of the conditions (\ref{CPT5})--(\ref{CPT6}) with (\ref{CPT4})
shows that the matrix 
$\sK=\cE_{\alpha_1}\cE_{\alpha_2}\cdots\cE_{\alpha_a}$ exists only if
$a\equiv 1\pmod{2}$. In turn, a comparison of the conditions
(\ref{CPT7})--(\ref{CPT8}) with (\ref{CPT4}) shows that the matrix
$\sK=\cE_{\beta_1}\cE_{\beta_2}\cdots\cE_{\beta_{p+q-a}}$ exists only if
$p+q-a\equiv 0\pmod{2}$.\end{sloppypar}

Let us find now squares of the matrix $\sK$. In accordance with obtained
conditions there exist two possibilities:\\
1) $\sK=\cE_{\alpha_1}\cE_{\alpha_2}\cdots\cE_{\alpha_a}$,
$a\equiv 1\pmod{2}$.
\[
\sK^2=\begin{cases}
+\sI, & \text{if $a_+-a_-\equiv 1,5\pmod{8}$},\\
-\sI, & \text{if $a_+-a_-\equiv 3,7\pmod{8}$},
\end{cases}
\]\begin{sloppypar}\noindent
where $a_+$ and $a_-$ are numbers of matrices with `$+$'- and `$-$'-squares in
the product $\cE_{\alpha_1}\cE_{\alpha_2}\cdots\cE_{\alpha_a}$.\\
2) $\sK=\cE_{\beta_1}\cE_{\beta_2}\cdots\cE_{\beta_{p+q-a}}$,
$p+q-a\equiv 0\pmod{2}$.\end{sloppypar}
\[
\sK^2=\begin{cases}
+\sI, & \text{if $b_+-b_-\equiv 0,4\pmod{8}$},\\
-\sI, & \text{if $b_+-b_-\equiv 2,6\pmod{8}$},
\end{cases}
\]\begin{sloppypar}\noindent
where $b_+$ and $b_-$ are numbers of matrices with `$+$'- and `$-$'-squares in
the product $\cE_{\beta_1}\cE_{\beta_2}\cdots\cE_{\beta_{p+q-a}}$,
respectively.\end{sloppypar}
\subsection{Pseudoantiautomorphism 
$\cA\rightarrow\overline{\widetilde{\cA}}$}
The pseudoantiautomorphism\index{pseudoantiautomorphism}
$\cA\rightarrow\overline{\widetilde{\cA}}$ is the
composition of the pseudoautomorphism $\cA\rightarrow\overline{\cA}$ with
the antiautomorphism $\cA\rightarrow\widetilde{\cA}$. Under action of
$\cA\rightarrow\widetilde{\cA}$ the units $\e_i$ remain unaltered,
$\e_i\rightarrow\e_i$. Analogously, under action of 
$\cA\rightarrow\overline{\cA}$ we have $\e_i\rightarrow\e_i$. Therefore,
under action of the pseudoantiautomorphism 
$\cA\rightarrow\overline{\widetilde{\cA}}$ the units $\e_i$ remain
unaltered also, $\e_i\rightarrow\e_i$.

The spinor representations of the transformations 
$\cA\rightarrow\widetilde{\cA}$ and $\cA\rightarrow\overline{\cA}$ are
defined by the expressions 
$\widetilde{\sA}=\sE\sA^{\sT}\sE^{-1}$ and
$\overline{\sA}=\Pi\dot{\sA}\Pi^{-1}$, respectively. Let us find a spinor
representation of the transformation 
$\cA\rightarrow\overline{\widetilde{\cA}}$. The order of the composition of
these transformations is not important,
$\overline{\widetilde{\cA}}=\widetilde{\left(\overline{\cA}\right)}=
\overline{\left(\widetilde{\cA}\right)}$. Indeed,
\[
\overline{\widetilde{\sA}}=\sE\left(\Pi\dot{\sA}\Pi^{-1}\right)^{\sT}\sE^{-1}=
\Pi\left(\sE\sA^{\sT}\sE^{-1}\right)^{\cdot}\Pi^{-1}
\]
or,
\begin{equation}\label{CPT9}
\overline{\widetilde{\sA}}=(\sE\Pi)\left(\dot{\sA}\right)^{\sT}(\sE\Pi)^{-1}=
(\Pi\sE)\left(\sA^{\sT}\right)^{\cdot}(\Pi\sE)^{-1},
\end{equation}
since $\Pi^{-1}=\Pi^{\sT}$ and
$\overline{\widetilde{\sA}}=\Pi\dot{\sE}\left(\sA^{\sT}\right)^{\cdot}
\dot{\sE}^{-1}\Pi^{-1}=\Pi\sE\left(\sA^{\sT}\right)^{\cdot}\sE^{-1}\Pi^{-1}$
in the case when $\dot{\sE}=\sE$ is a real matrix and
$\overline{\widetilde{\sA}}=\Pi\dot{\sE}\left(\sA^{\sT}\right)^\cdot
\dot{\sE}^{-1}\Pi^{-1}=\Pi(-\sE)\left(\sA^{\sT}\right)^\cdot(-\sE^{-1})\Pi^{-1}=
\Pi\sE\left(\sA^{\sT}\right)^\cdot(\Pi\sE)^{-1}$ in the case when
$\dot{\sE}=-\sE$ is a complex matrix. Let $\sS=\Pi\sE$ be a matrix of the
pseudoantiautomorphism $\cA\rightarrow\overline{\widetilde{\cA}}$ in the
spinor representation. Then (\ref{CPT9}) can be rewritten as follows
\begin{equation}\label{CPT10}
\overline{\widetilde{\sA}}=\sS\left(\sA^{\sT}\right)^\cdot\sS^{-1}.
\end{equation}
Since under action of the transformation 
$\cA\rightarrow\overline{\widetilde{\cA}}$ we have $\e_i\rightarrow\e_i$,
in the spinor representation we must demand 
$\cE_i\rightarrow\cE_i$ also, or
\begin{equation}\label{CPT11}
\cE_i\longrightarrow\cE_i=\sS\dot{\cE}^{\sT}_i\sS^{-1}.
\end{equation}
In the case of real subalgebras $\cl_{p,q}$ with the ring $\K\simeq\R$ 
we have $\dot{\cE}_i=\cE_i$ and, therefore, the relation (\ref{CPT11})
takes a form
\begin{equation}\label{CPT12}
\cE_i\longrightarrow\cE_i=\sS\cE^{\sT}_i\sS^{-1}.
\end{equation}
Let $\left\{\cE_{\gamma_i}\right\}$ be a set of symmetric matrices
$\left(\cE^{\sT}_{\gamma_i}=\cE_{\gamma_i}\right)$ and let
$\left\{\cE_{\delta_j}\right\}$ be a set of skewsymmetric matrices
$\left(\cE^{\sT}_{\delta_j}=-\cE_{\delta_j}\right)$ of the spinbasis of
the algebra $\cl_{p,q}$. Then from the relation (\ref{CPT12}) it follows
\[
\cE_{\gamma_i}\longrightarrow\cE_{\gamma_i}=\sS\cE_{\gamma_i}\sS^{-1},\quad
\cE_{\delta_j}\longrightarrow\cE_{\delta_j}=-\sS\cE_{\delta_j}\sS^{-1}.
\]
Whence
\[
\cE_{\gamma_i}\sS=\sS\cE_{\gamma_i},\quad
\cE_{\delta_j}\sS=-\sS\cE_{\delta_j},
\]
that is, the matrix $\sS$ of the pseudoantiautomorphism
$\cA\rightarrow\overline{\widetilde{\cA}}$ in the case of $\K\simeq\R$
commutes with the symmetric part and anticommutes with the skewsymmetric
part of the spinbasis of $\cl_{p,q}$. In virtue of Theorem \ref{tpseudo},
over the ring $\K\simeq\R$
the matrix $\Pi$ of the pseudoautomorphism $\cA\rightarrow\overline{\cA}$
is proportional to the unit matrix, $\Pi\sim\sI$. Therefore, in this case
we have $\sS\sim\sE$ and an explicit form of $\sS$ coincides with $\sE$.

Further, in case of the quaternionic ring $\K\simeq\BH$,
$p-q\equiv 4,6\pmod{8}$, a spinbasis of $\cl_{p,q}$ contains both complex
matrices $\cE_{\alpha_t}$ and real matrices $\cE_{\beta_s}$, among which
there are symmetric and skewsymmetric matrices. It is obvious that the sets
of complex and real matrices do not coincide with the sets of symmetric and
skewsymmetric matrices. Let $\left\{\cE_{\alpha_t}\right\}$ be a complex
part of the spinbasis, then the relation (\ref{CPT11}) takes a form
\begin{equation}\label{CPT13}
\cE_{\alpha_t}\longrightarrow\cE_{\alpha_t}=-\sS\cE^{\sT}_{\alpha_t}\sS^{-1}.
\end{equation}
Correspondingly, let $\left\{\cE_{\alpha_\gamma}\right\}$ and
$\left\{\cE_{\alpha_\delta}\right\}$ be the sets of symmetric and
skewsymmetric matrices of the complex part. Then the relation (\ref{CPT13})
can be written as follows
\[
\cE_{\alpha_\gamma}\longrightarrow\cE_{\alpha_\gamma}=
-\sS\cE_{\alpha_\gamma}\sS^{-1},\quad
\cE_{\alpha_\delta}\longrightarrow\cE_{\alpha_\delta}=
\sS\cE_{\alpha\delta}\sS^{-1}.
\]
Whence
\begin{equation}\label{CPT13'}
\cE_{\alpha_\gamma}\sS=-\sS\cE_{\alpha_\gamma},\quad
\cE_{\alpha_\delta}\sS=\sS\cE_{\alpha_\delta}.
\end{equation}
Therefore, the matrix $\sS$ of the pseudoantiautomorphism
$\cA\rightarrow\overline{\widetilde{\cA}}$ anticommutes with the complex
symmetric matrices and commutes with the complex skewsymmetric matrices of
the spinbasis of $\cl_{p,q}$.

Let us consider now the real part $\left\{\cE_{\beta_s}\right\}$ of the
spinbasis of $\cl_{p,q}$, $p-q\equiv 4,6\pmod{8}$. In this case the
relation (\ref{CPT11}) takes a form
\begin{equation}\label{CPT14}
\cE_{\beta_s}\longrightarrow\cE_{\beta_s}=\sS\cE^{\sT}_{\beta_s}\sS^{-1}.
\end{equation}
Let $\left\{\cE_{\beta_\gamma}\right\}$ and $\left\{\cE_{\beta_\delta}\right\}$
be the sets of real symmetric and real skewsymmetric matrices, respectively.
Then the relation (\ref{CPT14}) can be written as follows
\[
\cE_{\beta_\gamma}\longrightarrow\cE_{\beta_\gamma}=
\sS\cE_{\beta_\gamma}\sS^{-1},\quad
\cE_{\beta_\delta}\longrightarrow\cE_{\beta_\delta}=
-\sS\cE_{\beta_\delta}\sS^{-1}.
\]
Whence
\begin{equation}\label{CPT14'}
\cE_{\beta_\gamma}\sS=\sS\cE_{\beta_\gamma},\quad
\cE_{\beta_\delta}\sS=-\sS\cE_{\beta_\delta}.
\end{equation}
Thus, the matrix $\sS$ of the transformation
$\cA\rightarrow\overline{\widetilde{\cA}}$ commutes with the real
symmetric matrices and anticommutes with the real skewsymmetric matrices
of the spinbasis of $\cl_{p,q}$.

Let us find now an explicit form of the matrix $\sS=\Pi\sE$. In accordance
with Theorem \ref{tpseudo} for the quaternionic types 
$p-q\equiv 4,6\pmod{8}$ the matrix $\Pi$ takes the two different forms:
1) $\Pi=\cE_{\alpha_1}\cE_{\alpha_2}\cdots\cE_{\alpha_a}$ is the product
of complex matrices at $a\equiv 0\pmod{2}$; 2)
$\Pi=\cE_{\beta_1}\cE_{\beta_2}\cdots\cE_{\beta_b}$ is the product of real
matrices at $b\equiv 1\pmod{2}$. In turn, for the matrix $\sE$ of the
antiautomorphism $\cA\rightarrow\widetilde{\cA}$ over the ring $\K\simeq\BH$
(see Theorem \ref{tautr}) we have the following two forms:
1) $\sE=\cE_{j_1}\cE_{j_2}\cdots\cE_{j_k}$ is the product of skewsymmetric
matrices at $k\equiv 0\pmod{2}$; 2) $\sE=\cE_{i_1}\cE_{i_2}\cdots
\cE_{i_{p+q-k}}$ is the product of symmetric matrices at $k\equiv 1\pmod{2}$.
Thus, in accordance with definition $\sS=\Pi\sE$ we have four different
products: 
$\sS=\cE_{\alpha_1\alpha_2\cdots\alpha_a}\cE_{j_1j_2\cdots j_k}$,
$\sS=\cE_{\alpha_1\alpha_2\cdots\alpha_a}\cE_{i_1i_2\cdots i_{p+q-k}}$,
$\sS=\cE_{\beta_1\beta_2\cdots\beta_b}\cE_{j_1j_2\cdots j_k}$,
$\sS=\cE_{\beta_1\beta_2\cdots\beta_b}\cE_{i_1i_2\cdots i_{p+q-k}}$.
It is obvious that in the given products there are identical matrices.

Let us examine the first product $\sS=\cE_{\alpha_1\alpha_2\cdots\alpha_a}
\cE_{j_1j_2\cdots j_k}$. Since in this case $\Pi$ contains all the complex
matrices of the spinbasis, among which there are symmetric and skewsymmetric
matrices, and $\sE$ contains all the skewsymmetric matrices of the spinbasis,
then $\Pi$ and $\sE$ contain a quantity of identical matrices
(complex skewsymmetric matrices). Let $m$ be a number of the complex
skewsymmetric matrices of the spinbasis of the algebra $\cl_{p,q}$,
$p-q\equiv 4,6\pmod{8}$. Then the product $\sS=\Pi\sE$ takes a form
\[
\cE_{\alpha_1}\cE_{\alpha_2}\cdots\cE_{j_1}\cE_{j_2}\cdots\cE_{j_k}=
(-1)^{\cN}\sigma(i_1)\sigma(i_2)\cdots\sigma(i_m)
\cE_{c_1}\cE_{c_2}\cdots\cE_{c_s},
\]
where the indices $c_1,\ldots, c_s$ present itself a totality of the
indices $\alpha_1,\ldots,\alpha_a, j_1,\ldots,j_k$ obtained after
removal of the indices occured twice; $\cN$ is a number of inversions.

First of all, let us remark that $\cE_{c_1}\cE_{c_2}\cdots\cE_{c_s}$ is
an even product, since the original product $\Pi\sE$ is the even product
also. Besides, the product $\cE_{c_1}\cE_{c_2}\cdots\cE_{c_s}$ contains
all the complex symmetric matrices and all the real skewsymmetric matrices
of the spinbasis.

Let us find now permutation conditions between the matrix
$\sS=\cE_{c_1}\cE_{c_2}\cdots\cE_{c_s}$ and the units of the spinbasis of
$\cl_{p,q}$, $p-q\equiv 4,6\pmod{8}$. For the complex symmetric matrices
$\cE_{\alpha_\gamma}$ and complex skewsymmetric matrices $\cE_{\alpha_\delta}$
we have
\begin{gather}
\sS\cE_{\alpha_\gamma}=(-1)^{s-\gamma}\cE_{c_1}\cE_{c_2}\cdots\cE_{c_{s-2}},
\nonumber\\
\cE_{\alpha_\gamma}\sS=(-1)^{\gamma-1}\cE_{c_1}\cE_{c_2}\cdots\cE_{c_{s-2}},
\label{CPT15}\\
\sS\cE_{\alpha_\delta}=(-1)^s\cE_{\alpha_\delta}\sS,
\label{CPT16}
\end{gather}
that is, the matrix $\sS$ always anticommutes with the complex symmetric
matrices and always commutes with the complex skewsymmetric matrices,
since $s\equiv 0\pmod{2}$. Further, for the real symmetric matrices
$\cE_{\beta_\gamma}$ and real skewsymmetric matrices $\cE_{\beta_\delta}$
we have
\begin{gather}
\sS\cE_{\beta_\gamma}=(-1)^s\cE_{\beta_\gamma}\sS,\label{CPT17}\\
\sS\cE_{\beta_\delta}=(-1)^{s-\delta}\cE_{c_1}\cE_{c_2}\cdots\cE_{c_{s-2}},
\nonumber\\
\cE_{\beta_\delta}\sS=(-1)^{\delta-1}\cE_{c_1}\cE_{c_2}\cdots\cE_{c_{s-2}}.
\label{CPT18}
\end{gather}
Therefore, $\sS$ always commutes with the real symmetric matrices and
always anticommutes with the real skewsymmetric matrices.

A comparison of the obtained conditions (\ref{CPT15})--(\ref{CPT18}) with
(\ref{CPT13'}) and (\ref{CPT14'}) shows that 
$\sS=\cE_{c_1}\cE_{c_2}\cdots\cE_{c_s}$ automatically satisfies the
conditions, which define the matrix of the pseudoantiautomorphism
$\cA\rightarrow\overline{\widetilde{\cA}}$.

Let examine the second product 
$\sS=\cE_{\alpha_1\alpha_2\cdots\alpha_a}\cE_{i_1i_2\cdots i_{p+q-k}}$.
In this case $\Pi$ contains all the complex matrices of the spinbasis,
among which there are symmetric and skewsymmetric matrices, and $\sE$
contains all the symmetric matrices of the spinbasis. Thus, $\Pi$ and $\sE$
contain a quantity of identical complex symmetric matrices.
Let $l$ be a number of the complex symmetric matrices, then for the
product $\sS=\Pi\sE$ we obtain
\[
\cE_{\alpha_1}\cE_{\alpha_2}\cdots\cE_{\alpha_a}\cE_{i_1}\cE_{i_2}\cdots
\cE_{i_{p+q-k}}=(-1)^{\cN}\sigma(i_1)\sigma(i_2)\cdots\sigma(i_l)
\cE_{d_1}\cE_{d_2}\cdots\cE_{d_g},
\]
where $d_1,\ldots, d_g$ present itself a totality of the indices
$\alpha_1,\ldots,\alpha_a,i_1,\ldots,i_{p+q-k}$ obtained
after removal of the indices
occured twice. The product $\sS=\cE_{d_1}\cE_{d_2}\cdots\cE_{d_g}$ is odd,
since the original product $\Pi\sE$ is odd also. Besides,
$\cE_{d_1}\cE_{d_2}\cdots\cE_{d_g}$ contains all the complex skewsymmetric
matrices and all the real symmetric matrices of the spinbasis.

Let us find permutation conditions of the matrix 
$\sS=\cE_{d_1}\cE_{d_2}\cdots\cE_{d_g}$ with the units of the spinbasis of
$\cl_{p,q}$. For the complex part of the spinbasis we have
\begin{gather}
\sS\cE_{\alpha_\gamma}=(-1)^g\cE_{\alpha_\gamma}\sS,\label{CPT19}\\
\sS\cE_{\alpha_\delta}=(-1)^{g-\delta}\cE_{d_1}\cE_{d_2}\cdots\cE_{d_{g-2}},
\nonumber\\
\cE_{\alpha_\delta}\sS=(-1)^{\delta-1}\cE_{d_1}\cE_{d_2}\cdots\cE_{d_{g-2}},
\label{CPT20}
\end{gather}
that is, the matrix $\sS$ anticommutes with the complex symmetric
matrices and commutes with the complex skewsymmetric matrices, since
$g\equiv 1\pmod{2}$. For the real part of the spinbasis of $\cl_{p,q}$
we obtain
\begin{gather}
\sS\cE_{\beta_\gamma}=(-1)^{g-\gamma}\cE_{d_1}\cE_{d_2}\cdots\cE_{d_{g-2}},
\nonumber\\
\cE_{\beta_\gamma}\sS=(-1)^{\gamma-1}\cE_{d_1}\cE_{d_2}\cdots\cE_{d_{g-2}},
\label{CPT21}\\
\sS\cE_{\beta_\delta}=(-1)^g\cE_{\beta_\delta}\sS.\label{CPT22}
\end{gather}
Therefore, $\sS$ commutes with the real symmetric matrices and anticommutes
with the real skewsymmetric matrices.

Comparing the obtained conditions (\ref{CPT19})--(\ref{CPT22}) with
the conditions (\ref{CPT13'}) and (\ref{CPT14'}) we see that 
$\sS=\cE_{d_1}\cE_{d_2}\cdots\cE_{d_g}$ automatically satisfies the
conditions which define the matrix of the transformation
$\cA\rightarrow\overline{\widetilde{\cA}}$.

Let us consider now the third product 
$\sS=\cE_{\beta_1\beta_2\cdots\beta_b}\cE_{j_1j_2\cdots k}$.
In this product the matrix $\Pi$ of the pseudoautomorphism
$\cA\rightarrow\overline{\cA}$ contains all the real matrices of the
spinbasis, among which there are symmetric and skewsymmetric matrices,
and the matrix $\sE$ of $\cA\rightarrow\widetilde{\cA}$ contains all the
skewsymmetric matrices of the spinbasis, among which there are both the real
and complex matrices. Therefore, $\sS$ contains a quantity of identical
real skewsymmetric matrices. Let $u$ be a number of the real skewsymmetric
matrices of the spinbasis of the algebra $\cl_{p,q}$, $p-q\equiv 4,6\pmod{8}$,
then for the product $\sS$ we obtain
\[
\cE_{\beta_1}\cE_{\beta_2}\cdots\cE_{\beta_b}\cE_{j_1}\cE_{j_2}\cdots
\cE_{j_k}=(-1)^{\cN}\sigma(i_1)\sigma(i_2)\cdots\sigma(i_u)
\cE_{e_1}\cE_{e_2}\cdots\cE_{e_h},
\]
where the indices $e_1,e_2,\ldots, e_n$ present itself a totality of the
indices $\beta_1,\ldots,\beta_b,j_1,\ldots,j_k$ obtained
after removal of the
indices occurred twice. The product $\sS=\cE_{e_1}\cE_{e_2}\cdots\cE_{e_h}$
is odd, since the original product $\Pi\sE$ is odd also. It is easy to see
that $\cE_{e_1}\cE_{e_2}\cdots\cE_{e_h}$ contains all the real symmetric
matrices and all the complex skewsymmetric matrices of the spinbasis.
Therefore, the matrix $\sS=\cE_{e_1}\cE_{e_2}\cdots\cE_{e_h}$ is similar to
the matrix $\sS=\cE_{d_1}\cE_{d_2}\cdots\cE_{d_g}$, and its permutation
conditions with the units of the spinbasis of $\cl_{p,q}$ are equivalent to
the relations (\ref{CPT19})--(\ref{CPT22}).

Finally, let us examine the fourth product
$\sS=\cE_{\beta_1\beta_2\cdots\beta_b}\cE_{i_1i_2\cdots i_{p+q-k}}$.
In turn, this product contains a quantity of identical real symmetric
matrices. Let $v$ be a number of the real symmetric matrices of the spinbasis
of $\cl_{p,q}$, then
\[
\cE_{\beta_1}\cE_{\beta_2}\cdots\cE_{\beta_b}
\cE_{i_1}\cE_{i_2}\cdots\cE_{i_{p+q-k}}=(-1)^{\cN}\sigma(i_1)\sigma(i_2)
\cdots\sigma(i_v)\cE_{f_1}\cE_{f_2}\cdots\cE_{f_w},
\]
where $f_1,\ldots,f_w$ present itself a totality of the indices
$\beta_1,\ldots,\beta_b,i_1,\ldots,i_{p+q-k}$ obtained
after removal of the indices
occurred twice. The product $\sS=\cE_{f_1}\cE_{f_2}\cdots\cE_{f_w}$ is even,
since the original product $\Pi\sE$ is even also. It is easy to see that
$\cE_{f_1}\cE_{f_2}\cdots\cE_{f_w}$ contains all the real skewsymmetric
matrices and all the complex symmetric matrices of the spinbasis.
Therefore, the matrix $\sS=\cE_{f_1}\cE_{f_2}\cdots\cE_{f_w}$ is similar to
the matrix $\sS=\cE_{c_1}\cE_{c_2}\cdots\cE_{c_s}$, and its permutation
conditions with the units of the spinbasis are equivalent to the relations
(\ref{CPT15})--(\ref{CPT18}).

Thus, from the four products we have only two non-equivalent products.
The squares of the non-equivalent matrices 
$\sS=\cE_{c_1}\cE_{c_2}\cdots\cE_{c_s}$ ($s\equiv 0\pmod{2}$) and
$\sS=\cE_{d_1}\cE_{d_2}\cdots\cE_{d_g}$ ($g\equiv 1\pmod{2}$) are
\begin{gather}
\sS^2=\left(\cE_{c_1}\cE_{c_2}\cdots\cE_{c_s}\right)^2=
\begin{cases}
+\sI, & \text{if $u+l\equiv 0,4\pmod{8}$},\\
-\sI, & \text{if $u+l\equiv 2,6\pmod{8}$};
\end{cases}\label{CPT22'}\\
\sS^2=\left(\cE_{d_1}\cE_{d_2}\cdots\cE_{d_g}\right)^2=
\begin{cases}
+\sI, & \text{if $m+v\equiv 1,5\pmod{8}$},\\
-\sI, & \text{if $m+v\equiv 3,7\pmod{8}$}.
\end{cases}\label{CPT22''}
\end{gather}
\subsection{Pseudoantiautomorphism 
$\cA\rightarrow\overline{\widetilde{\cA^\star}}$}
The pseudoantiautomorphism\index{pseudoantiautomorphism}
$\cA\rightarrow\overline{\widetilde{\cA^\star}}$,
that defines the $CPT$-transformation, is a composition of the
pseudoautomorphism $\cA\rightarrow\overline{\cA}$, antiautomorphism
$\cA\rightarrow\widetilde{\cA}$ and automorphism $\cA\rightarrow\cA^\star$.
Under action of the automorphism $\cA\rightarrow\cA^\star$ the units of
$\cl_{p,q}$ change the sign, $\e_i\rightarrow-\e_i$. In turn, under action
of the transformations $\cA\rightarrow\overline{\cA}$ and
$\cA\rightarrow\widetilde{\cA}$ the units remain unaltered,
$\e_i\rightarrow\e_i$. Therefore, under action of the pseudoantiautomorphism
$\cA\rightarrow\overline{\widetilde{\cA^\star}}$ the units change the sign,
$\e_i\rightarrow-\e_i$.

The spinor representations of the transformations $\cA\rightarrow\cA^\star$,
$\cA\rightarrow\widetilde{\cA}$ and $\cA\rightarrow\overline{\cA}$
have the form: $\sA^\star=\sW\sA\sW^{-1}$, 
$\widetilde{\sA}=\sE\sA^{\sT}\sE^{-1}$ and 
$\overline{\sA}=\Pi\dot{\sA}\Pi^{-1}$. Let us find a matrix of the
transformation $\cA\rightarrow\overline{\widetilde{\cA^\star}}$.
We will consider the pseudoantiautomorphism
$\cA\rightarrow\overline{\widetilde{\cA^\star}}$ as a composition of the
pseudoautomorphism $\cA\rightarrow\overline{\cA}$ with the antiautomorphism
$\cA\rightarrow\widetilde{\cA^\star}$. The spinor representation of
$\cA\rightarrow\widetilde{\cA^\star}$ is 
$\widetilde{\cA^\star}=\sC\sA^{\sT}\sC^{-1}$, where $\sC=\sE\sW$. Since
$\overline{\widetilde{\cA^\star}}=
\widetilde{\left(\overline{\cA}\right)^\star}=
\overline{\left(\widetilde{\cA^\star}\right)}$, then
\[
\overline{\widetilde{\sA^\star}}=
\sC\left(\Pi\dot{\sA}\Pi^{-1}\right)^{\sT}\sC^{-1}=
\Pi\left(\sC\sA^{\sT}\sC^{-1}\right)^{\cdot}\Pi^{-1},
\]
or
\begin{equation}\label{CPT23}
\overline{\widetilde{\sA^\star}}=(\sC\Pi)\dot{\sA}^{\sT}(\sC\Pi)^{-1}=
(\Pi\sC)\dot{\sA}^{\sT}(\Pi\sC)^{-1},
\end{equation}
since $\Pi^{-1}=\Pi^{\sT}$ and $\overline{\widetilde{\sA^\star}}=
\Pi\dot{\sC}\left(\sA^{\sT}\right)^{\cdot}\dot{\sC}^{-1}\Pi^{-1}=
\Pi\sC\dot{\sA}^{\sT}\sC^{-1}\Pi^{-1}$ in the case when $\dot{\sC}=\sC$ is
a real matrix, and also $\overline{\widetilde{\sA^\star}}=\Pi\dot{\sC}
\left(\sA^{\sT}\right)^{\cdot}\dot{\sC}^{-1}\Pi^{-1}=\Pi(-\sC)
\left(\sA^{\sT}\right)^{\cdot}\left(-\sC^{-1}\right)\Pi^{-1}=
\Pi\sC\dot{\sA}^{\sT}\sC^{-1}\Pi^{-1}$ in the case when $\dot{\sC}=-\sC$ is
a complex matrix.

Let $\sF=\Pi\sC$ (or $\sF=\Pi\sE\sW$) be a matrix of the
pseudoantiautomorphism $\cA\rightarrow\overline{\widetilde{\cA^\star}}$.
Then the relation (\ref{CPT23}) can be written as follows
\begin{equation}\label{CPT24}
\overline{\widetilde{\sA^\star}}=\sF\dot{\sA}^{\sT}\sF^{-1}.
\end{equation}
Since under action of the transformation
$\cA\rightarrow\overline{\widetilde{\cA^\star}}$ we have
$\e_i\rightarrow-\e_i$, in the spinor representation we must demand
$\cE_i\rightarrow-\cE_i$ also, or
\begin{equation}\label{CPT25}
\cE_i\longrightarrow-\cE_i=\sF\dot{\cE}^{\sT}_i\sF^{-1}.
\end{equation}
In case of the real subalgebras $\cl_{p,q}$ with the ring $\K\simeq\R$,
$p-q\equiv 0,2\pmod{8}$, we have $\dot{\cE}_i=\cE_i$ for all matrices of
the spinbasis and, therefore, the relation (\ref{CPT25}) takes a form
\begin{equation}\label{CPT26}
\cE_{i}\longrightarrow-\cE_i=\sF\cE^{\sT}_i\sF^{-1}.
\end{equation}
Let $\left\{\cE_{\gamma_i}\right\}\cup\left\{\cE_{\delta_j}\right\}$ be
a spinbasis of the algebra $\cl_{p,q}$ over the ring $\K\simeq\R$,
($\cE^{\sT}_{\gamma_i}=\cE_{\gamma_i},\,\cE^{\sT}_{\delta_j}=-\cE_{\delta_j}$).
Then the relation (\ref{CPT26}) can be written in the form
\[
\cE_{\gamma_i}\longrightarrow-\cE_{\gamma_i}=\sF\cE_{\gamma_i}\sF^{-1},\quad
\cE_{\delta_j}\longrightarrow\cE_{\delta_j}=\sF\cE_{\delta_j}\sF^{-1}.
\]
Whence
\[
\cE_{\gamma_i}\sF=-\sF\cE_{\gamma_i},\quad
\cE_{\delta_j}\sF=\sF\cE_{\delta_j},
\]
that is, the matrix $\sF$ of the pseudoantiautomorphism
$\cA\rightarrow\overline{\widetilde{\cA^\star}}$ in case of the ring
$\K\simeq\R$ anticommutes with the symmetric part of the spinbasis of
$\cl_{p,q}$ and commutes with the skewsymmetric part. In virtue of
Theorem \ref{tpseudo} the matrix $\Pi$ of the pseudoautomorphism
$\cA\rightarrow\overline{\cA}$ over the ring $\K\simeq\R$ is proportional
to the unit matrix, $\Pi\simeq\sI$. Therefore, in this case $\sF\sim\sC$
($\sF\sim\sE\sW$) and an explicit form of the matrix $\sF$ coincides with
$\sC$ (see Theorem \ref{tautr}).

In case of the quaternionic ring $\K\simeq\BH$, $p-q\equiv 4,6\pmod{8}$,
the spinbasis of $\cl_{p,q}$ contains both complex matrices
$\cE_{\alpha_t}$ and real matrices $\cE_{\beta_r}$. For the complex part
the relation (\ref{CPT25}) takes a form
\[
\cE_{\alpha_t}\longrightarrow\cE_{\alpha_t}=\sF\cE^{\sT}_{\alpha_t}\sF^{-1},
\]
Or, taking into account complex symmetric and complex skewsymmetric
components of the spinbasis, we obtain
\[
\cE_{\alpha_\gamma}\longrightarrow\cE_{\alpha_\gamma}=
\sF\cE_{\alpha_\gamma}\sF^{-1},\quad
\cE_{\alpha_\delta}\longrightarrow\cE_{\alpha_\delta}=
-\sF\cE_{\alpha_\delta}\sF^{-1}.
\]
Whence
\begin{equation}\label{CPT27}
\cE_{\alpha_\gamma}\sF=\sF\cE_{\alpha_\gamma},\quad
\cE_{\alpha_\delta}\sF=-\sF\cE_{\alpha_\delta}.
\end{equation}
For the real part of the spinbasis of $\cl_{p,q}$ from (\ref{CPT25})
we obtain
\[
\cE_{\beta_r}\longrightarrow-\cE_{\beta_r}=\sF\cE^{\sT}_{\beta_r}\sF^{-1},
\]
or, taking into account symmetric and skewsymmetric components of the real
part of the spinbasis, we find
\[
\cE_{\beta_\gamma}\longrightarrow-\cE_{\beta_\gamma}=
\sF\cE_{\beta_\gamma}\sF^{-1},\quad
\cE_{\beta_\delta}\longrightarrow\cE_{\beta_\delta}=
\sF\cE_{\beta_\delta}\sF^{-1}.
\]
Whence
\begin{equation}\label{CPT28}
\cE_{\beta_\gamma}\sF=-\sF\cE_{\beta_\gamma},\quad
\cE_{\beta_\delta}\sF=\sF\cE_{\beta_\delta}.
\end{equation}
Therefore, the matrix $\sF$ of the pseudoantiautomorphism
$\cA\rightarrow\overline{\widetilde{\cA^\star}}$ commutes with the complex
symmetric and real skewsymmetric matrices, and also $\sF$ anticommutes
with the complex skewsymmetric and real symmetric matrices of the
spinbasis of $\cl_{p,q}$, $p-q\equiv 4,6\pmod{8}$.

Let us find an explicit form of the matrix $\sF=\Pi\sC$. It is easy to see
that in virtue of $\sF=\Pi\sC=(\Pi\sE)\sW=\sS\sW$ the matrix $\sF$ is a dual
with respect to the matrix $\sS$ of the pseudoantiautomorphism
$\cA\rightarrow\overline{\widetilde{\cA}}$. In accordance with Theorem
\ref{tpseudo} for the quaternionic types $p-q\equiv 4,6\pmod{8}$ the matrix
$\Pi$ has two different forms: $\Pi=\cE_{\alpha_1}\cE_{\alpha_2}\cdots
\cE_{\alpha_a}$ ($\dot{\cE}_{\alpha_t}=-\cE_{\alpha_t}$), $a\equiv 0\pmod{2}$;
$\Pi=\cE_{\beta_1}\cE_{\beta_2}\cdots\cE_{\beta_b}$ 
($\dot{\cE}_{\beta_r}=\cE_{\beta_r}$), $b\equiv 1\pmod{2}$. In turn,
for the quaternionic types the matrix $\sC$ of the antiautomorphism
$\cA\rightarrow\widetilde{\cA^\star}$ has the two forms (see Theorem \ref{tautr}):
1) $\sC=\cE_{i_1}\cE_{i_2}\cdots\cE_{i_{p+q-k}}$ is the product of all
symmetric matrices of the spinbasis of $\cl_{p,q}$ at $p+q-k\equiv 0\pmod{2}$;
2) $\sC=\cE_{j_1}\cE_{j_2}\cdots\cE_{j_k}$ is the product of all
skewsymmetric matrices of the spinbasis at $k\equiv 1\pmod{2}$. Thus,
in accordance with definition $\sF=\Pi\sC$ we have four products:
$\sF=\cE_{\alpha_1\alpha_2\cdots\alpha_a}\cE_{i_1i_2\cdots i_{p+q-k}}$,
$\sF=\cE_{\alpha_1\alpha_2\cdots\alpha_a}\cE_{j_1j_2\cdots j_k}$,
$\sF=\cE_{\beta_1\beta_2\cdots\beta_b}\cE_{i_1i_2\cdots i_{p+q-k}}$,
$\sF=\cE_{\beta_1\beta_2\cdots\beta_b}\cE_{j_1j_2\cdots j_k}$.

Let us examine the first product
$\sF=\cE_{\alpha_1\alpha_2\cdots\alpha_a}\cE_{i_1i_2\cdots i_{p+q-k}}$.
In this case $\Pi$ contains all the complex matrices of the spinbasis, among
which there are symmetric and skewsymmetric matrices. In turn, $\sC$ contains
all the symmetric matrices, among which there are both complex and real
matrices. It is obvious that in this case $\Pi\sC$ contains a quantity
of identical complex symmetric matrices. Therefore, the product $\sF$
consists of all the complex skewsymmetric matrices and all the real
symmetric matrices of the spinbasis. The product $\sF$ is even, since the
original product $\Pi\sC$ is even also. It is easy to see that $\sF$
coincides with the product $\cE_{d_1}\cE_{d_2}\cdots\cE_{d_g}$ at
$g\equiv 0\pmod{2}$.

Let us find permutation conditions of the matrix 
$\sF=\cE_{d_1}\cE_{d_2}\cdots\cE_{d_g}$ with the units of the spinbasis of
$\cl_{p,q}$, $p-q\equiv 4,6\pmod{8}$. For the complex and real parts
we obtain
\begin{gather}
\sF\cE_{\alpha_\gamma}=(-1)^g\cE_{\alpha_\gamma}\sF,\label{CPT29}\\
\sF\cE_{\alpha_\delta}=(-1)^{g-\delta}\cE_{d_1}\cE_{d_2}\cdots\cE_{d_{g-2}},
\nonumber\\
\cE_{\alpha_\delta}\sF=(-1)^{\delta-1}\cE_{d_1}\cE_{d_2}\cdots\cE_{d_{g-2}},
\label{CPT30}\\
\sF\cE_{\beta_\gamma}=(-1)^{g-\gamma}\cE_{d_1}\cE_{d_2}\cdots\cE_{d_{g-2}},
\nonumber\\
\cE_{\beta_\gamma}\sF=(-1)^{\gamma-1}\cE_{d_1}\cE_{d_2}\cdots\cE_{d_{g-2}},
\label{CPT31}\\
\sF\cE_{\beta_\delta}=(-1)^g\cE_{\beta_\delta}\sF.\label{CPT32}
\end{gather}
Therefore, since $g\equiv 0\pmod{2}$ the matrix $\sF$ always commutes
with the complex symmetric and real skewsymmetric matrices and always
anticommutes with the complex skewsymmetric and real symmetric matrices of
the spinbasis. A comparison of the permutation conditions 
(\ref{CPT29})--(\ref{CPT32}) with the conditions (\ref{CPT27})--(\ref{CPT28})
shows that $\sF=\cE_{d_1}\cE_{d_2}\cdots\cE_{d_g}$ at $g\equiv 0\pmod{2}$
automatically satisfies the conditions which define the matrix of the
pseudoantiautomorphism
$\cA\rightarrow\overline{\widetilde{\cA^\star}}$.

Let examine the second product
$\sF=\cE_{\alpha_1\alpha_2\cdots\alpha_a}\cE_{j_1j_2\cdots j_k}$.
This product contains all the complex part of the spinbasis and all the
skewsymmetric matrices. Therefore, in the product $\Pi\sC$ there is a
quantity of identical complex skewsymmetric matrices. The product
$\sF$ is odd and consists of all the complex symmetric and real skewsymmetric
matrices of the spinbasis. It is easy to see that in this case $\sF$
coincides with the product $\cE_{c_1}\cE_{c_2}\cdots\cE_{c_s}$ at
$s\equiv 1\pmod{2}$. Permutation conditions of the matrix
$\sF=\cE_{c_1}\cE_{c_2}\cdots\cE_{c_s}$ with the units of the spinbasis are
\begin{gather}
\sF\cE_{\alpha_\gamma}=(-1)^{s-\gamma}\cE_{c_1}\cE_{c_2}\cdots\cE_{c_{s-2}},
\nonumber\\
\cE_{\alpha_\gamma}\sF=(-1)^{\gamma-1}\cE_{c_1}\cE_{c_2}\cdots\cE_{c_{s-2}},
\label{CPT33}\\
\sF\cE_{\alpha_\delta}=(-1)^s\cE_{\alpha_\delta}\sF,\label{CPT34}\\
\sF\cE_{\beta_\gamma}=(-1)^s\cE_{\beta_\gamma}\sF,\label{CPT35}\\
\sF\cE_{\beta_\delta}=(-1)^{s-\delta}\cE_{c_1}\cE_{c_2}\cdots\cE_{c_{s-2}},
\nonumber\\
\cE_{\beta_\delta}\sF=(-1)^{\delta-1}\cE_{c_1}\cE_{c_2}\cdots\cE_{c_{s-2}}.
\label{CPT36}
\end{gather}
Therefore, since $s\equiv 1\pmod{2}$ the matrix $\sF$ always commutes with
the complex symmetric and real skewsymmetric matrices and always anticommutes
with the complex skewsymmetric and real symmetric matrices of the spinbasis.
Comparing the conditions (\ref{CPT33})--(\ref{CPT36}) with the conditions
(\ref{CPT27}) and (\ref{CPT28}) we see that
$\sF=\cE_{c_1}\cE_{c_2}\cdots\cE_{c_s}$ at $s\equiv 1\pmod{2}$ 
identically satisfies the conditions which define the matrix of the
transformation
$\cA\rightarrow\overline{\widetilde{\cA^\star}}$.

The third product 
$\sF=\Pi\sC=\cE_{\beta_1\beta_2\cdots\beta_b}\cE_{i_1i_2\cdots i_{p+q-k}}$
contains all the real part and all the symmetric matrices of the spinbasis.
Therefore, in the product $\Pi\sC$ there is a quantity of identical real
symmetric matrices. Thus, the product $\sF$ is odd and consists of all
the real skewsymmetric and complex symmetric matrices of the spinbasis.
It is easy to see that we came again to the matrix
$\sF=\cE_{c_1}\cE_{c_2}\cdots\cE_{c_s}$ ($s\equiv 1\pmod{2}$) with the
permutation conditions (\ref{CPT33})--(\ref{CPT36}).

Finally, the fourth product
$\sF=\cE_{\beta_1\beta_2\cdots\beta_b}\cE_{j_1j_2\cdots j_k}$
contains all the real part and all the skewsymmetric matrices of the spinbasis.
Therefore, in the product $\Pi\sC$ there is a quantity of identical real
skewsymmetric matrices. This product is equivalent to the matrix
$\sF=\cE_{d_1}\cE_{d_2}\cdots\cE_{d_g}$ ($g\equiv 0\pmod{2}$) with the
permutation conditions (\ref{CPT29})--(\ref{CPT32}).

As with the pseudoantiautomorphism
$\cA\rightarrow\overline{\widetilde{\cA}}$,
from the four products we have only two non-equivalent products.
Let us find squares of the non-equivalent matrices
$\sF=\cE_{d_1}\cE_{d_2}\cdots\cE_{d_g}$ ($g\equiv 0\pmod{2}$) and
$\sF=\cE_{c_1}\cE_{c_2}\cdots\cE_{c_s}$ ($s\equiv 1\pmod{2}$):
\begin{gather}
\sF^2=\left(\cE_{d_1}\cE_{d_2}\cdots\cE_{d_g}\right)^2=
\begin{cases}
+\sI, & \text{if $m+v\equiv 0,4\pmod{8}$},\\
-\sI, & \text{if $m+v\equiv 2,6\pmod{8}$};
\end{cases}\label{CPT36'}\\
\sF^2=\left(\cE_{c_1}\cE_{c_2}\cdots\cE_{c_s}\right)^2=
\begin{cases}
+\sI, & \text{if $u+l\equiv 3,7\pmod{8}$},\\
-\sI, & \text{if $u+l\equiv 1,5\pmod{8}$}.
\end{cases}\label{CPT36''}
\end{gather}

\section{The structure of $\sExt(\C_n)$}
As noted previously, the group $\sExt(\C_n)$ is 
a finite group\index{group!finite}
of order eight. This group contains as a subgroup the automorphism group
$\sAut_\pm(\C_n)$ (reflection group). Moreover, in the case of $\Pi\sim\sI$
(the subalgebra $\cl_{p,q}$ has the ring $\K\simeq\R$, $p-q\equiv 0,2\pmod{8}$)
the group $\sExt(\C_n)$ is reduced to its subgroup
$\sAut_\pm(\C_n)$. The structure of the groups $\sAut_\pm(\C_n)$,
$\sAut_\pm(\cl_{p,q})$ is studied in detail (see Theorems \ref{taut} and
\ref{tautr}).

There are six finite groups of order eight (see Table 1). One is cyclic
and two are direct group products of cyclic groups, hence these three are
Abelian. The remaining three groups are 
the quaternion group\index{group!quaternionic} $Q_4$ with
elements $\{\pm 1,\pm\bi,\pm\bj,\pm\bk\}$, 
the dihedral group\index{group!dihedral} $D_4$, and
the group $\overset{\ast}{\dZ}_4\otimes\dZ_2$. All these groups are
non--Abelian. As is known, an important property of each finite group is its
order structure.\index{structure!order}
The order of a particular element $\alpha$ in the group
is the smallest integer $p$ for which $\alpha^p=1$. The following table
lists the number of distinct elements in each group which have order 2, 4,
or 8 (the identity 1 is the only element of order 1).
\begin{center}{\renewcommand{\arraystretch}{1.4}
\begin{tabular}{|l|c|ccc|}\hline
   &  & \multicolumn{3}{l}{Order structure}\vline\\ 
   & Type & 2 & 4 & 8 \\ \hline
1. $\dZ_2\otimes\dZ_2\otimes\dZ_2$ & Abelian & 7 & & \\ 
2. $\dZ_4\otimes\dZ_2$ & & 3 & 4 & \\ 
3. $\dZ_8$ & & 1 & 2 & 4\\ \hline
4. $D_4$ & Non--Abelian & 5 & 2 & \\
5. $Q_4$ & & 1 & 6 & \\
6. $\overset{\ast}{\dZ}_4\otimes\dZ_2$ & & 3 & 4 &\\ \hline
\end{tabular}
}
\end{center}
%\medskip
\begin{center}{\small
{\bf Table 1.} Finite groups of order 8.}
\end{center}
Of course, $\dZ_8$ does not occur as a $G(p,q)$ 
(Salingaros group\index{group!Salingaros}), since
every element of $G(p,q)$ has order 1, 2, or 4. The groups
$\dZ_4\otimes\dZ_2$ and $\overset{\ast}{\dZ}_4\otimes\dZ_2$ have the same
order structure, but their signatures $(a,b,c,d,e,f,g)$ are different.
Moreover, the group $\overset{\ast}{\dZ}_4\otimes\dZ_2$ presents a first
example of the finite group of order 8 which has an important physical
meaning.\\[0.2cm]
%\begin{ex}
%\label{Ex:3.1}

{\it Example 1.}
Let us consider a Dirac algebra\index{algebra!Dirac} $\C_4$. In the algebra
$\C_4$ we can evolve four different 
real subalgebras\index{subalgebras!real} $\cl_{1,3}$,
$\cl_{3,1}$, $\cl_{4,0}$, $\cl_{0,4}$. 
Let us evolve the spacetime algebra\index{algebra!spacetime}
$\cl_{1,3}$. The algebra $\cl_{1,3}$ has the quaternionic division ring
$\K\simeq\BH$ ($p-q\equiv 6\pmod{8}$) and, therefore, admits the following
spinor representation (the well known $\gamma$-basis):
\begin{gather}
\gamma_0=\begin{pmatrix}
1 & 0 & 0 & 0\\
0 & 1 & 0 & 0\\
0 & 0 &-1 & 0\\
0 & 0 & 0 &-1
\end{pmatrix},\quad\gamma_1=\begin{pmatrix}
0 & 0 & 0 & 1\\
0 & 0 & 1 & 0\\
0 &-1 & 0 & 0\\
-1& 0 & 0 & 0
\end{pmatrix},\nonumber\\
\gamma_2=\begin{pmatrix}
0 & 0 & 0 &-i\\
0 & 0 & i & 0\\
0 & i & 0 & 0\\
-i& 0 & 0 & 0
\end{pmatrix},\quad\gamma_3=\begin{pmatrix}
0 & 0 & 1 & 0\\
0 & 0 & 0 &-1\\
-1& 0 & 0 & 0\\
0 & 1 & 0 & 0
\end{pmatrix}.\label{GammaB}
\end{gather}
The famous Dirac equation\index{equation!Dirac}
in the $\gamma$--basis looks like
\begin{equation}\label{Diraceq}
\left(i\gamma_0\frac{\partial}{\partial x_0}-
i\boldsymbol{\gamma}\frac{\partial}{\partial\bx}-m\right)\psi(x_0,\bx)=0.
\end{equation}
The invariance of the Dirac equation with respect to $P$--, $T$--, and
$C$--transformations leads to the following representation
(see, for example, \cite{BLP89} and also many other textbooks on quantum
field theory):
\[
P\sim\gamma_0,\quad T\sim\gamma_1\gamma_3,\quad C\sim\gamma_2\gamma_0.
\]
Thus, we can form a finite group of order 8 ($CPT$--group)
\begin{multline}
\{1,\,P,\,T,\,PT,\,C,\,CP,\,CT,\,CPT\}\sim\\
\sim\left\{1,\,\gamma_0,\,\gamma_1\gamma_3,\,\gamma_0\gamma_1\gamma_3,\,
\gamma_2\gamma_0,\,\gamma_2,\,\gamma_2\gamma_0\gamma_1\gamma_3,\,
\gamma_2\gamma_1\gamma_3\right\}.
\label{DirG2}
\end{multline}
It is easy to verify that a Cayley tableau\index{tableau!Cayley}
of this group has a form
\begin{center}{\renewcommand{\arraystretch}{1.4}
\begin{tabular}{|c||c|c|c|c|c|c|c|c|}\hline
  & $1$ & $\gamma_0$ & $\gamma_{13}$ & $\gamma_{013}$ & $\gamma_{20}$ &
$\gamma_2$ & $\gamma_{2013}$ & $\gamma_{213}$\\ \hline\hline
$1$  & $1$ & $\gamma_0$ & $\gamma_{13}$ & $\gamma_{013}$ & $\gamma_2$ &
$\gamma_2$ & $\gamma_{2013}$ & $\gamma_{213}$\\ \hline
$\gamma_0$ & $\gamma_0$ & $1$ & $\gamma_{013}$ & $\gamma_{13}$ & $-\gamma_2$ &
$-\gamma_{20}$ & $-\gamma_{213}$ & $-\gamma_{2013}$\\ \hline
$\gamma_{13}$ & $\gamma_{13}$ & $\gamma_{013}$ & $-1$ & $-\gamma_0$ &
$\gamma_{2013}$ & $\gamma_{213}$ & $-\gamma_{20}$ & $-\gamma_2$\\ \hline
$\gamma_{013}$ & $\gamma_{013}$ & $\gamma_{13}$ & $-\gamma_0$ &
 $-1$ & $-\gamma_{213}$ & $-\gamma_{2013}$ & $\gamma_2$ & 
$\gamma_{20}$\\ \hline
$\gamma_{20}$ & $\gamma_{20}$ & $\gamma_2$ & $\gamma_{2013}$ &
$\gamma_{213}$ & $1$ & $\gamma_0$ & $\gamma_{13}$ &
$\gamma_{013}$\\ \hline
$\gamma_2$ & $\gamma_2$ & $\gamma_{20}$ & $\gamma_{213}$ & $\gamma_{2013}$ &
$-\gamma_0$ & $-1$ & $-\gamma_{013}$ & $-\gamma_{13}$\\ \hline
$\gamma_{2013}$ & $\gamma_{2013}$ & $\gamma_{213}$ & $-\gamma_{20}$ &
$\gamma_2$ & $\gamma_{13}$ & $\gamma_{013}$ & $-1$ & $-\gamma_0$\\ \hline
$\gamma_{213}$ & $\gamma_{213}$ & $\gamma_{2013}$ & $-\gamma_2$ &
$-\gamma_{20}$ & $-\gamma_{013}$ & $-\gamma_{13}$ & $\gamma_0$ & $1$\\ \hline
\end{tabular}\;\;$\sim$
}
\end{center}
\begin{center}{\renewcommand{\arraystretch}{1.4}
\begin{tabular}{|c||c|c|c|c|c|c|c|c|}\hline
     & $1$  & $P$  & $T$  & $PT$ & $C$  & $CP$ & $CT$ & $CPT$ \\ \hline\hline
$1$  & $1$  & $P$  & $T$  & $PT$ & $C$  & $CP$ & $CT$ & $CPT$ \\ \hline
$P$  & $P$  & $1$  & $PT$ & $T$  & $-CP$ & $-C$  & $-CPT$& $-CT$\\ \hline
$T$  & $T$  & $PT$ & $-1$  & $-P$  & $CT$ & $CPT$& $-C$  & $-CP$\\ \hline
$PT$ & $PT$ & $T$  & $-P$  & $-1$  & $-CPT$& $-CT$ & $CP$ & $C$\\ \hline
$C$  & $C$  & $CP$ & $CT$ & $CPT$& $1$  & $P$  & $T$  & $PT$\\ \hline
$CP$ & $CP$ & $C$  & $CPT$& $CT$ & $-P$  & $-1$  & $-PT$ & $-T$\\ \hline
$CT$ & $CT$ & $CPT$& $-C$  & $CP$ & $T$  & $PT$ & $-1$  & $-P$\\ \hline
$CPT$& $CPT$& $CT$ & $-CP$ & $-C$  & $-PT$ & $-T$  & $P$  & $1$\\ \hline
\end{tabular}
}
\end{center}
\begin{sloppypar}
Hence it follows that the $CPT$--group (\ref{DirG2}) is a non--Abelian
finite group of the order structure (3,4). In more details, it is the group
$\overset{\ast}{\dZ}_4\otimes\dZ_2$ with the signature
$(+,-,-,+,-,-,+)$.\end{sloppypar}
%\end{ex}
\begin{theorem}\label{tautext}
Let $\C_n$ be the Clifford algebra over the field $\F=\C$ and let
$\sExt(\C_n)=\{\sI,\,\sW,\,\sE,\,\sC,\,\Pi,\,\sK,\,\sS,\,\sF\}$
be an extended automorphism group\index{group!automorphism!extended}
of the algebra $\C_n$, where
$\sW$, $\sE$, $\sC$, $\Pi$, $\sK$, $\sS$, $\sF$ are spinor representations
of the transformations $\cA\rightarrow\cA^\star$,
$\cA\rightarrow\widetilde{\cA}$, $\cA\rightarrow\widetilde{\cA^\star}$,
$\cA\rightarrow\overline{\cA}$, $\cA\rightarrow\overline{\cA^\star}$,
$\cA\rightarrow\overline{\widetilde{\cA}}$,
$\cA\rightarrow\overline{\widetilde{\cA^\star}}$. Then over the field $\F=\C$
in dependence on a division ring structure of the real subalgebras
$\cl_{p,q}\subset\C_n$ there exist following isomorphisms between
finite groups and groups $\sExt(\C_n)$:\\[0.2cm]
1) $\K\simeq\R$, types $p-q\equiv 0,2\pmod{8}$.\\
In this case the matrix $\Pi$ of the pseudoautomorphism
$\cA\rightarrow\overline{\cA}$ is proportional to the unit matrix
(identical transformation) and the extended automorphism group
$\sExt(\C_n)$ is reduced to the group of fundamental automorphisms
$\sAut_\pm(\C_n)$.\\[0.2cm]
2) $\K\simeq\BH$, types $p-q\equiv 4,6\pmod{8}$.\\
In dependence on a spinbasis structure of the subalgebra $\cl_{p,q}$ there
exist the following group isomorphisms:
$\sExt_-(\C_n)\simeq\dZ_2\otimes\dZ_2\otimes\dZ_2$ with the
signature $(+,+,+,+,+,+,+)$ and $\sExt(\C_n)\simeq\dZ_4\otimes\dZ_2$
with $(+,+,+,-,-,-,-)$ for the type $p-q\equiv 4\pmod{8}$ if
$m,v,l,u\equiv 0\pmod{2}$, where $m$ and $l$ are quantities of
complex skewsymmetric and
symmetric matrices, and $u$ and $v$ are quantities of 
real skewsymmetric and symmetric
matrices of the spinbasis of $\cl_{p,q}$. Correspondingly, at
$m,v,l,u\equiv 0\pmod{2}$ there exist Abelian groups
$\sExt_-(\C_n)\simeq\dZ_4\otimes\dZ_2$ with $(+,-,-,d,e,f,g)$ for the
type $p-q\equiv 4\pmod{8}$ and $\sExt_-(\C_n)\simeq\dZ_4\otimes\dZ_2$
with $(-,+,-,d,e,f,g)$, $(-,-,+,d,e,f,g)$ for the type $p-q\equiv 6\pmod{8}$,
where among the symbols $d,e,f,g$ there are two pluses and two minuses.
\begin{sloppypar}
If $m,v,l,u\equiv 1\pmod{2}$ or if among $m,v,l,u$ there are both even and
odd numbers, then there exists the non--Abelian group
$\sExt_+(\C_n)\simeq Q_4$ with the signatures $(-,-,-,d,e,f,g)$ for
the type $p-q\equiv 6\pmod{8}$, where among $d,e,f,g$ there are one plus and
three minuses. And also there exist $\sExt_+(\C_n)\simeq D_4$ with
$(+,-,+,d,e,f,g)$, $(+,+,-,d,e,f,g)$ for the type $p-q\equiv 4\pmod{8}$ and
$\sExt_+(\C_n)\simeq D_4$ with $(-,+,+,d,e,f,g)$ for the type
$p-q\equiv 6\pmod{8}$, where among the symbols $d,e,f,g$ there are
three pluses and one minus. Besides, there exist the groups
$\sExt_+(\C_n)\simeq Q_4$ with the signatures $(+,-,-,-,-,-,-)$ for
$p-q\equiv 4\pmod{8}$ and $(-,+,-,-,-,-,-)$, $(-,-,+,-,-,-,-)$ for
$p-q\equiv 6\pmod{8}$. And also there exist the groups
$\sExt_+(\C_n)\simeq D_4$ with $(+,+,+,d,e,f,g)$, $(+,-,-,+,+,+,+)$ for
the type $p-q\equiv 4\pmod{8}$ and $(a,b,c,+,+,+,+)$ for the type
$p-q\equiv 6\pmod{8}$ (among the symbols $a,b,c$ there are one plus and two
minuses, and among $d,e,f,g$ there are two pluses and two minuses).
There exists the non--Abelian group $\sExt_+(\C_n)\simeq
\overset{\ast}{\dZ}_4\otimes\dZ_2$ with the signatures $(+,-,-,d,e,f,g)$ for
the type $p-q\equiv 4\pmod{8}$ and $(-,+,-,d,e,f,g)$, $(-,-,+,d,e,f,g)$
for the type $p-q\equiv 6\pmod{8}$, where among $d,e,f,g$ there are two
pluses and two minuses. And also there exist $\sExt_+(\C_n)\simeq
\overset{\ast}{\dZ}_4\otimes\dZ_2$ with $(+,-,+,d,e,f,g)$, $(+,+,-,d,e,f,g)$
for $p-q\equiv 4\pmod{8}$ and $(-,+,+,d,e,f,g)$ for $p-q\equiv 6\pmod{8}$,
where among $d,e,f,g$ there are one plus and three minuses. Finally, for
the type $p-q\equiv 6\pmod{8}$ there exists $\sExt_+(\C_n)\simeq
\overset{\ast}{\dZ}_4\otimes\dZ_2$ with $(-,-,-,d,e,f,g)$, where among the
symbols $d,e,f,g$ there are three pluses and one minus.
The full number of the different signatures $(a,b,c,d,e,f,g)$ is equal to 64.
\end{sloppypar}
\end{theorem}
\begin{proof}
First of all, it is necessary to define permutation relations between
the elements of the group $\sExt$. We start with the matrix of the
pseudoautomorphism $\cA\rightarrow\overline{\cA}$ (permutation relations
between the elements $\sW$, $\sE$ and $\sC$ are found in Theorem \ref{tautr}).
As is known, for the types $p-q\equiv 4,6\pmod{8}$ the matrix $\Pi$
exists in the two different forms: 1) $\Pi=\cE_{\alpha_1}\cE_{\alpha_2}\cdots
\cE_{\alpha_a}$ is the product of all complex matrices of the spinbasis at
$a\equiv 0\pmod{2}$; 2) $\Pi=\cE_{\beta_1}\cE_{\beta_2}\cdots\cE_{\beta_b}$ is
the product of all real matrices of the spinbasis at $b\equiv 1\pmod{2}$.

Let us consider permutation relations of $\Pi$ with the matrix $\sK$ of the
pseudoautomorphism\index{pseudoautomorphism}
$\cA\rightarrow\overline{\cA^\star}$. The matrix $\sK$
also exists in the two different forms: $\sK=\cE_{\alpha_1}\cE_{\alpha_2}\cdots
\cE_{\alpha_a}$ at $a\equiv 1\pmod{2}$ and $\sK=\cE_{\beta_1}\cE_{\beta_2}
\cdots\cE_{\beta_b}$ at $b\equiv 0\pmod{2}$. In virtue of the definition
$\sK=\Pi\sW$, where $\sW=\cE_1\cE_2\cdots\cE_{p+q}$ is the spinor
representation of the automorphism $\cA\rightarrow\cA^\star$, the matrix
$\Pi=\cE_{\alpha_1}\cE_{\alpha_2}\cdots\cE_{\alpha_a}$ corresponds to a
matrix $\sK=\cE_{\beta_1}\cE_{\beta_2}\cdots\cE_{\beta_b}$, since
$n=p+q$ is always even for the types $p-q\equiv 4,6\pmod{8}$. Correspondingly,
for the matrix $\Pi=\cE_{\beta_1}\cE_{\beta_2}\cdots\cE_{\beta_b}$ we obtain
$\sK=\cE_{\alpha_1}\cE_{\alpha_2}\cdots\cE_{\alpha_a}$, where
$a,b\equiv 1\pmod{2}$. It is easy to see that in both cases we have a
relation
\begin{equation}\label{CPT37}
\Pi\sK=(-1)^{ab}\sK\Pi,
\end{equation}
that is, at $a,b\equiv 0\pmod{2}$ the matrices $\Pi$ and $\sK$ always
commute and at $a,b\equiv 1\pmod{2}$ always anticommute.

Let us find now permutation relations of $\Pi$ with the matrix $\sS$ of
the pseudoantiautomorphism 
$\cA\rightarrow\overline{\widetilde{\cA}}$. As is known, the matrix $\sS$
exists in the two non-equivalent forms: 
1) $\sS=\cE_{c_1}\cE_{c_2}\cdots\cE_{c_s}$ is the product of all complex
symmetric and real skewsymmetric matrices at $s\equiv 0\pmod{2}$;
2) $\sS=\cE_{d_1}\cE_{d_2}\cdots\cE_{d_g}$ is the product of all complex
skewsymmetric and real symmetric matrices at $g\equiv 1\pmod{2}$.
From $\sS=\Pi\sE$ it follows that 
$\Pi=\cE_{\alpha_1}\cE_{\alpha_2}\cdots\cE_{\alpha_a}$ corresponds to
$\sS=\cE_{c_1}\cE_{c_2}\cdots\cE_{c_s}$ if 
$\sE=\cE_{j_1}\cE_{j_2}\cdots\cE_{j_k}$ and to
$\sS=\cE_{d_1}\cE_{d_2}\cdots\cE_{d_g}$ if
$\sE=\cE_{i_1}\cE_{i_2}\cdots\cE_{i_{p+q-k}}$. In turn, the matrix
$\Pi=\cE_{\beta_1}\cE_{\beta_2}\cdots\cE_{\beta_b}$ corresponds to
$\sS=\cE_{d_1}\cE_{d_2}\cdots\cE_{d_g}$ if
$\sE=\cE_{j_1}\cE_{j_2}\cdots\cE_{j_k}$ and to
$\sS=\cE_{c_1}\cE_{c_2}\cdots\cE_{c_s}$ if
$\sE=\cE_{i_1}\cE_{i_2}\cdots\cE_{i_{p+q-k}}$. Thus, taking into account
that $\sS=\Pi\sE$, we obtain
\begin{eqnarray}
\Pi\sS&=&(-1)^{\frac{a(a-1)}{2}+\tau}\cE_{j_1}\cE_{j_2}\cdots\cE_{j_k},
\nonumber\\
\sS\Pi&=&(-1)^{\frac{a(a-1)}{2}+\tau-m+ak}\cE_{j_1}\cE_{j_2}\cdots
\cE_{j_k}\label{CPT38}
\end{eqnarray}
for the matrices $\Pi=\cE_{\alpha_1}\cE_{\alpha_2}\cdots\cE_{\alpha_a}$ and
$\sS=\cE_{c_1}\cE_{c_2}\cdots\cE_{c_s}$, where $m$ is the number of
complex skewsymmetric matrices of the spinbasis of $\cl_{p,q}$,
$p-q\equiv 4,6\pmod{8}$. Since a comparison $ak\equiv 0\pmod{2}$ holds
always, then the matrices $\Pi$ and $\sS$ commute at $m\equiv 0\pmod{2}$ and
anticommute at $m\equiv 1\pmod{2}$. Correspondingly,
\begin{eqnarray}
\Pi\sS&=&(-1)^{\frac{a(a-1)}{2}+\tau}\cE_{i_1}\cE_{i_2}\cdots
\cE_{i_{p+q-k}},\nonumber\\
\sS\Pi&=&(-1)^{\frac{a(a-1)}{2}+\tau-l+a(p+q-k)}\cE_{i_1}\cE_{i_2}\cdots
\cE_{i_{p+q-k}}\label{CPT39}
\end{eqnarray}
for the matrices $\Pi=\cE_{\alpha_1}\cE_{\alpha_2}\cdots\cE_{\alpha_a}$ and
$\sS=\cE_{d_1}\cE_{d_2}\cdots\cE_{d_g}$, where $l$ is the number of complex
symmetric matrices of the spinbasis. Since $a(p+q-k)\equiv 0\pmod{2}$
($a\equiv 0\pmod{2}$, $p+q-k\equiv 1\pmod{2}$), then in this case the
matrices $\Pi$ and $\sS$ commute at $l\equiv 0\pmod{2}$ and anticommute at
$l\equiv 1\pmod{2}$. Further, we have
\begin{eqnarray}
\Pi\sS&=&(-1)^{\frac{b(b-1)}{2}+\tau}\cE_{j_1}\cE_{j_2}\cdots
\cE_{j_k},\nonumber\\
\sS\Pi&=&(-1)^{\frac{b(b-1)}{2}+\tau-u+bk}\cE_{j_1}\cE_{j_2}\cdots
\cE_{j_k}\label{CPT40}
\end{eqnarray}
for the matrices $\Pi=\cE_{\beta_1}\cE_{\beta_2}\cdots\cE_{\beta_b}$ and
$\sS=\cE_{d_1}\cE_{d_2}\cdots\cE_{d_g}$, where $u$ is the number of real
skewsymmetric matrices of the spinbasis. Since $bk\equiv 0\pmod{2}$
($b\equiv 1\pmod{2}$, $k\equiv 0\pmod{2}$), then $\Pi$ and $\sS$ commute at
$u\equiv 0\pmod{2}$ and anticommute at $u\equiv 1\pmod{2}$. Finally,
\begin{eqnarray}
\Pi\sS&=&(-1)^{\frac{b(b-1)}{2}+\tau}\cE_{i_1}\cE_{i_2}\cdots
\cE_{i_{p+q-k}},\nonumber\\
\sS\Pi&=&(-1)^{\frac{b(b-1)}{2}+\tau-v+b(p+q-k)}\cE_{i_1}\cE_{i_2}\cdots
\cE_{i_{p+q-k}}\label{CPT41}
\end{eqnarray}
for the matrices $\Pi=\cE_{\beta_1}\cE_{\beta_2}\cdots\cE_{\beta_b}$ and
$\sS=\cE_{c_1}\cE_{c_2}\cdots\cE_{c_s}$, where $v$ is the number of real
symmetric matrices. Therefore, permutation conditions of the matrices
$\Pi$ and $\sS$ in this case have the form $b(p+q-k)\equiv v\pmod{2}$, that is,
$\Pi$ and $\sS$ commute at $v\equiv 1\pmod{2}$ and anticommute at
$v\equiv 0\pmod{2}$, since $b,p+q-k\equiv 1\pmod{2}$.

Now we find permutation conditions of $\Pi$ with the matrix $\sF$ of the
pseudoantiautomorphism
$\cA\rightarrow\overline{\widetilde{\cA^\star}}$. In turn, the matrix
$\sF$ exists also in the two non-equivalent forms: 
$\sF=\cE_{d_1}\cE_{d_2}\cdots\cE_{d_g}$ at $g\equiv 0\pmod{2}$ and
$\sF=\cE_{c_1}\cE_{c_2}\cdots\cE_{c_s}$ at $s\equiv 1\pmod{2}$.
From the definition $\sF=\Pi\sC$ it follows that
$\Pi=\cE_{\alpha_1}\cE_{\alpha_2}\cdots\cE_{\alpha_a}$ corresponds to
$\sF=\cE_{d_1}\cE_{d_2}\cdots\cE_{d_g}$ if
$\sC=\cE_{i_1}\cE_{i_2}\cdots\cE_{i_{p+q-k}}$ and to
$\sF=\cE_{c_1}\cE_{c_2}\cdots\cE_{c_s}$ if
$\sC=\cE_{j_1}\cE_{j_2}\cdots\cE_{j_k}$. The matrix
$\Pi=\cE_{\beta_1}\cE_{\beta_2}\cdots\cE_{\beta_b}$ corresponds to
$\sF=\cE_{c_1}\cE_{c_2}\cdots\cE_{c_s}$ if
$\sC=\cE_{i_1}\cE_{i_2}\cdots\cE_{i_{p+q-k}}$ and to
$\sF=\cE_{d_1}\cE_{d_2}\cdots\cE_{d_g}$ if
$\sC=\cE_{j_1}\cE_{j_2}\cdots\cE_{j_k}$.
Thus, taking into account that $\sF=\Pi\sC$, we obtain
\begin{eqnarray}
\Pi\sF&=&(-1)^{\frac{a(a-1)}{2}+\tau}\cE_{i_1}\cE_{i_2}\cdots
\cE_{i_{p+q-k}},\nonumber\\
\sF\Pi&=&(-1)^{\frac{a(a-1)}{2}+\tau-l+a(p+q-k)}\cE_{i_1}\cE_{i_2}\cdots
\cE_{i_{p+q-k}}\label{CPT42}
\end{eqnarray}
for the matrices $\Pi=\cE_{\alpha_1}\cE_{\alpha_2}\cdots\cE_{\alpha_a}$ and
$\sF=\cE_{d_1}\cE_{d_2}\cdots\cE_{d_g}$. It is easy to see that
$\Pi$ and $\sF$ commute at $l\equiv 0\pmod{2}$ and anticommute at
$l\equiv 1\pmod{2}$, since $a,p+q-k\equiv 0\pmod{2}$. Analogously,
\begin{eqnarray}
\Pi\sF&=&(-1)^{\frac{a(a-1)}{2}+\tau}\cE_{j_1}\cE_{j_2}\cdots
\cE_{j_k},\nonumber\\
\sF\Pi&=&(-1)^{\frac{a(a-1)}{2}+\tau-m+ak}\cE_{j_1}\cE_{j_2}\cdots
\cE_{j_k}\label{CPT43}
\end{eqnarray}
for $\Pi=\cE_{\alpha_1}\cE_{\alpha_2}\cdots\cE_{\alpha_a}$ and
$\sF=\cE_{c_1}\cE_{c_2}\cdots\cE_{c_s}$. Therefore, $\Pi$ and $\sF$
commute at $m\equiv 0\pmod{2}$ and anticommute at $m\equiv 1\pmod{2}$, since
$a\equiv 0\pmod{2}$, $k\equiv 1\pmod{2}$. Further, we have
\begin{eqnarray}
\Pi\sF&=&(-1)^{\frac{b(b-1)}{2}+\tau}\cE_{i_1}\cE_{i_2}\cdots
\cE_{i_{p+q-k}},\nonumber\\
\sF\Pi&=&(-1)^{\frac{b(b-1)}{2}+\tau-v+b(p+q-k)}\cE_{i_1}\cE_{i_2}\cdots
\cE_{i_{p+q-k}}\label{CPT44}
\end{eqnarray}
for the matrices $\Pi=\cE_{\beta_1}\cE_{\beta_1}\cdots\cE_{\beta_b}$ and
$\sF=\cE_{c_1}\cE_{c_2}\cdots\cE_{c_s}$. In this case, $\Pi$ and $\sF$
commute at $v\equiv 0\pmod{2}$ and anticommute at $v\equiv 1\pmod{2}$,
since $b\equiv 1\pmod{2}$, $p+q-k\equiv 0\pmod{2}$. Finally,
\begin{eqnarray}
\Pi\sF&=&(-1)^{\frac{b(b-1)}{2}+\tau}\cE_{j_1}\cE_{j_2}\cdots\cE_{j_k},
\nonumber\\
\sF\Pi&=&(-1)^{\frac{b(b-1)}{2}+\tau-u+bk}\cE_{j_1}\cE_{j_2}\cdots
\cE_{j_k}\label{CPT45}
\end{eqnarray}
for $\Pi=\cE_{\beta_1}\cE_{\beta_2}\cdots\cE_{\beta_b}$ and
$\sF=\cE_{d_1}\cE_{d_2}\cdots\cE_{d_g}$. Therefore, in this case
permutation conditions of the matrices $\Pi$ and $\sF$ have the form
$bk\equiv u\pmod{2}$, that is, $\Pi$ and $\sF$ commute at $u\equiv 1\pmod{2}$
and anticommute at $u\equiv 0\pmod{2}$, since $b,k\equiv 1\pmod{2}$.

Let us define now permutation conditions of $\Pi$ with the matrices
$\sW$, $\sE$ and $\sC$ of the transformations $\cA\rightarrow\cA^\star$,
$\cA\rightarrow\widetilde{\cA}$ and $\cA\rightarrow\widetilde{\cA^\star}$,
respectively. First of all, according to Theorem \ref{tpseudo} in the
case of subalgebras $\cl_{p,q}$ with the real ring $\K\simeq\R$
(types $p-q\equiv 0,2\pmod{8}$) the matrix $\Pi$ is proportional to the
unit matrix and, therefore, $\Pi$ commutes with $\sW$, $\sE$ and $\sC$.
In case of the ring $\K\simeq\BH$ (types $p-q\equiv 4,6\pmod{8}$) the
matrix $\Pi$ exists in the two forms: $\Pi_a$ at $a\equiv 0\pmod{2}$ and
$\Pi_b$ at $b\equiv 1\pmod{2}$. Since $a+b=p+q$, then the matrix $\sW$
can be represented by a product 
$\cE_{\alpha_1}\cE_{\alpha_2}\cdots\cE_{\alpha_a}\cE_{\beta_1}\cE_{\beta_2}
\cdots\cE_{\beta_b}$ and for 
$\Pi=\cE_{\alpha_1}\cE_{\alpha_2}\cdots\cE_{\alpha_a}$ we have
\begin{eqnarray}
\Pi\sW&=&(-1)^{\frac{a(a-1)}{2}}\sigma(\alpha_1)\sigma(\alpha_2)\cdots
\sigma(\alpha_a)\cE_{\beta_1}\cE_{\beta_2}\cdots\cE_{\beta_b},\nonumber\\
\sW\Pi&=&(-1)^{\frac{a(a-1)}{2}+ba}\sigma(\alpha_1)\sigma(\alpha_2)\cdots
\sigma(\alpha_a)\cE_{\beta_1}\cE_{\beta_2}\cdots\cE_{\alpha_b}.\label{CPT46}
\end{eqnarray}
Hence it follows that $\Pi$ and $\sW$ commute at $ab\equiv 0\pmod{2}$
and anticommute at $ab\equiv 1\pmod{2}$, but $a\equiv 0\pmod{2}$ and,
therefore, the matrices $\Pi$ and $\sW$ always commute in this case.
Taking $\Pi=\cE_{\beta_1}\cE_{\beta_2}\cdots\cE_{\beta_b}$, we find the
following conditions
\begin{eqnarray}
\Pi\sW&=&(-1)^{\frac{b(b-1)}{2}+ab}\sigma(\beta_1)\sigma(\beta_2)\cdots
\sigma(\beta_b)\cE_{\alpha_1}\cE_{\alpha_2}\cdots\cE_{\alpha_a},\nonumber\\
\sW\Pi&=&(-1)^{\frac{b(b-1)}{2}}\sigma(\beta_1)\sigma(\beta_2)\cdots
\sigma(\beta_b)\cE_{\alpha_1}\cE_{\alpha_2}\cdots\cE_{\alpha_a}.\label{CPT47}
\end{eqnarray}
Hence it follows that $ab\equiv 1\pmod{2}$, since in this case
$a,b\equiv 1\pmod{2}$ ($p+q=a+b$ is an even number). Therefore, at
$b\equiv 1\pmod{2}$ the matrices $\Pi$ and $\sW$ always anticommute.

Let us find now permutation conditions between $\Pi$ and the matrix
$\sE$ of the antiautomorphism $\cA\rightarrow\widetilde{\cA}$. As is known
(see Theorem \ref{tautr}), the matrix $\sE$ exists in the two non-equivalent
forms. First of all, if $\Pi=\cE_{\alpha_1}\cE_{\alpha_2}\cdots\cE_{\alpha_a}$
and $\sE=\cE_{j_1}\cE_{j_2}\cdots\cE_{j_k}$, then it is obvious that
$\Pi$ and $\sE$ contain $m$ identical complex skewsymmetric matrices.
We can represent the matrices $\Pi$ and $\sE$ in the form of the following
products: $\Pi=\cE_{\alpha_1}\cE_{\alpha_2}\cdots\cE_{\alpha_l}\cE_{i_1}
\cE_{i_2}\cdots\cE_{i_m}$ and $\sE=\cE_{i_1}\cE_{i_2}\cdots\cE_{i_m}\cE_{j_1}
\cE_{j_2}\cdots\cE_{j_u}$, where $l$ and $u$ are the numbers of complex
symmetric and real skewsymmetric matrices, respectively. Therefore,
\begin{eqnarray}
\Pi\sE&=&(-1)^{\frac{m(m-1)}{2}}\sigma(i_1)\sigma(i_2)\cdots\sigma(i_m)
\cE_{\alpha_1}\cE_{\alpha_2}\cdots\cE_{\alpha_l}\cE_{j_1}\cE_{j_2}\cdots
\cE_{j_u},\nonumber\\
\sE\Pi&=&(-1)^{\frac{m(m-1)}{2}+m(u+l)}\sigma(i_1)\sigma(i_2)\cdots
\sigma(i_m)
\cE_{\alpha_1}\cE_{\alpha_2}\cdots\cE_{\alpha_l}\cE_{j_1}\cE_{j_2}\cdots
\cE_{j_u},\label{CPT48}
\end{eqnarray}
that is, $\Pi$ and $\sE$ commute at $m(u+l)\equiv 0\pmod{2}$ and
anticommute at $m(u+l)\equiv 1\pmod{2}$.

Analogously, if $\Pi=\cE_{\alpha_1}\cE_{\alpha_2}\cdots\cE_{\alpha_a}$ and
$\sE=\cE_{i_1}\cE_{i_2}\cdots\cE_{i_{p+q-k}}$, then it is easy to see that
in this case $\Pi$ and $\sE$ contain $l$ identical complex symmetric
matrices. Then $\Pi=\cE_{\alpha_1}\cE_{\alpha_2}\cdots\cE_{\alpha_m}
\cE_{i_1}\cE_{i_2}\cdots\cE_{i_l}$, $\sE=\cE_{i_1}\cE_{i_2}\cdots\cE_{i_l}
\cE_{i_1}\cE_{i_2}\cdots\cE_{i_v}$ ($v$ is the number of all real
symmetric matrices of the spinbasis) and
\begin{eqnarray}
\Pi\sE&=&(-1)^{\frac{l(l-1)}{2}}\sigma(i_1)\sigma(i_2)\cdots\sigma(i_l)
\cE_{\alpha_1}\cE_{\alpha_2}\cdots\cE_{\alpha_m}\cE_{i_1}\cE_{i_2}\cdots
\cE_{i_v},\nonumber\\
\sE\Pi&=&(-1)^{\frac{l(l-1)}{2}+l(m+v)}\sigma(i_1)\sigma(i_2)\cdots\sigma(i_l)
\cE_{\alpha_1}\cE_{\alpha_2}\cdots\cE_{\alpha_m}\cE_{i_1}\cE_{i_2}\cdots
\cE_{i_v}.\label{CPT49}
\end{eqnarray}
Therefore, in this case $\Pi$ and $\sE$ commute at $l(m+v)\equiv 0\pmod{2}$
and anticommute at $l(m+v)\equiv 1\pmod{2}$.
\begin{sloppypar}
In turn, the matrices $\Pi=\cE_{\beta_1}\cE_{\beta_2}\cdots\cE_{\beta_b}$
and $\sE=\cE_{j_1}\cE_{j_2}\cdots\cE_{j_k}$ contain $u$ identical real
skewsymmetric matrices. Therefore, $\Pi=\cE_{\beta_1}\cE_{\beta_2}\cdots
\cE_{\beta_v}\cE_{i_1}\cE_{i_2}\cdots\cE_{i_u}$, $\sE=\cE_{i_1}\cE_{i_2}\cdots
\cE_{i_u}\cE_{j_1}\cE_{j_2}\cdots\cE_{j_m}$ and\end{sloppypar}
\begin{eqnarray}
\Pi\sE&=&(-1)^{\frac{u(u-1)}{2}}\sigma(i_1)\sigma(i_2)\cdots\sigma(i_u)
\cE_{\beta_1}\cE_{\beta_2}\cdots\cE_{\beta_v}\cE_{j_1}\cE_{j_2}\cdots
\cE_{j_m},\nonumber\\
\sE\Pi&=&(-1)^{\frac{u(u-1)}{2}+u(m+v)}\sigma(i_1)\sigma(i_2)\cdots\sigma(i_u)
\cE_{\beta_1}\cE_{\beta_2}\cdots\cE_{\beta_v}\cE_{j_1}\cE_{j_2}\cdots
\cE_{j_m}.\label{CPT50}
\end{eqnarray}
Hence it follows that $\Pi$ and $\sE$ commute at $u(m+v)\equiv 0\pmod{2}$ and
anticommute at $u(m+v)\equiv 1\pmod{2}$.

Finally, the matrices $\Pi=\cE_{\beta_1}\cE_{\beta_2}\cdots\cE_{\beta_b}$ and
$\sE=\cE_{i_1}\cE_{i_2}\cdots\cE_{i_{p+q-k}}$ contain $v$ identical
real symmetric matrices. Therefore, in this case
$\Pi=\cE_{\beta_1}\cE_{\beta_2}\cdots\cE_{\beta_u}\cE_{i_1}\cE_{i_2}\cdots
\cE_{i_v}$, $\sE=\cE_{i_1}\cE_{i_2}\cdots\cE_{i_v}\cE_{i_1}\cE_{i_2}\cdots
\cE_{i_l}$ and
\begin{eqnarray}
\Pi\sE&=&(-1)^{\frac{v(v-1)}{2}}\sigma(i_1)\sigma(i_2)\cdots\sigma(i_v)
\cE_{\beta_1}\cE_{\beta_2}\cdots\cE_{\beta_u}\cE_{i_1}\cE_{i_2}\cdots
\cE_{i_l},\nonumber\\
\sE\Pi&=&(-1)^{\frac{v(v-1)}{2}+v(u+l)}\sigma(i_1)\sigma(i_2)\cdots\sigma(i_v)
\cE_{\beta_1}\cE_{\beta_2}\cdots\cE_{\beta_u}\cE_{i_1}\cE_{i_2}\cdots
\cE_{i_l},\label{CPT51}
\end{eqnarray}
that is, in this case $\Pi$ and $\sE$ commute at $v(u+l)\equiv 0\pmod{2}$ and
anticommute at $v(u+l)\equiv 1\pmod{2}$.

It is easy to see that permutation conditions of $\Pi$ with the matrix
$\sC$ of the antiautomorphism $\cA\rightarrow\widetilde{\cA^\star}$ are
analogous to the conditions (\ref{CPT48})--(\ref{CPT51}). Indeed,
at $\Pi=\cE_{\alpha_1}\cE_{\alpha_2}\cdots\cE_{\alpha_a}$ and
$\sC=\cE_{i_1}\cE_{i_2}\cdots\cE_{i_{p+q-k}}$ these conditions are analogous
to (\ref{CPT49}), that is, $l(m+v)\equiv 0,1\pmod{2}$. For the matrices
$\Pi=\cE_{\alpha_1}\cE_{\alpha_2}\cdots\cE_{\alpha_a}$ and
$\sC=\cE_{j_1}\cE_{j_2}\cdots\cE_{j_k}$ we obtain the condition (\ref{CPT48}),
that is, $m(u+l)\equiv 0,1\pmod{2}$. In turn, for
$\Pi=\cE_{\beta_1}\cE_{\beta_2}\cdots\cE_{\beta_b}$ and 
$\sC=\cE_{i_1}\cE_{i_2}\cdots\cE_{i_{p+q-k}}$ we have the condition
(\ref{CPT51}), that is, $v(u+l)\equiv 0,1\pmod{2}$. Finally, the matrices
$\Pi=\cE_{\beta_1}\cE_{\beta_2}\cdots\cE_{\beta_b}$ and
$\sC=\cE_{j_1}\cE_{j_2}\cdots\cE_{j_k}$ correspond to (\ref{CPT50}) with
$u(m+v)\equiv 0,1\pmod{2}$.

Let us consider now permutation conditions of the matrix $\sK$ of 
$\cA\rightarrow\overline{\cA^\star}$ with other elements
of the group $\sExt(\C_n)$. As it has been shown previously, the structure
of $\sK$ is analogous to the structure of $\Pi$, that is,
$\sK=\cE_{\alpha_1}\cE_{\alpha_2}\cdots\cE_{\alpha_a}$ at $a\equiv 1\pmod{2}$
and $\sK=\cE_{\beta_1}\cE_{\beta_2}\cdots\cE_{\beta_b}$ at $b\equiv 0\pmod{2}$.
Therefore, permutation conditions for $\sK$ are similar to the conditions
for $\Pi$. Permutation conditions between $\sK$ and $\Pi$ are defined by
the relation (\ref{CPT37}). Coming to the matrix $\sS$ of the
pseudoantiautomorphism $\cA\rightarrow\overline{\widetilde{\cA}}$, we see that
permutation conditions between $\sK$ and $\sS$ are analogous to
(\ref{CPT38})--(\ref{CPT41}). Namely, if $\sK=\cE_{\beta_1}\cE_{\beta_2}
\cdots\cE_{\beta_b}$ and $\sS=\cE_{c_1}\cE_{c_2}\cdots\cE_{c_s}$, then
from (\ref{CPT41}) it follows that $\sK$ and $\sS$ commute at 
$v\equiv 0\pmod{2}$ and anticommute at $v\equiv 1\pmod{2}$, since in this
case $b\equiv 0\pmod{2}$ and $p+q-k\equiv 0\pmod{2}$. Analogously, if
$\sK=\cE_{\alpha_1}\cE_{\alpha_2}\cdots\cE_{\alpha_b}$ and
$\sS=\cE_{d_1}\cE_{d_2}\cdots\cE_{d_g}$, then from (\ref{CPT39}) we find
that $\sK$ and $\sS$ commute at $l\equiv 0\pmod{2}$ and anticommute at
$l\equiv 1\pmod{2}$, since $a\equiv 1\pmod{2}$, $p+q-k\equiv 0\pmod{2}$ for
this case. Further, if $\sK=\cE_{\beta_1}\cE_{\beta_2}\cdots\cE_{\beta_b}$
and $\sS=\cE_{d_1}\cE_{d_2}\cdots\cE_{d_g}$, then form (\ref{CPT40}) we
find that in this case $\sK$ and $\sS$ commute at $u\equiv 0\pmod{2}$
and anticommute at $u\equiv 1\pmod{2}$, since $b\equiv 0\pmod{2}$,
$k\equiv 1\pmod{2}$. Finally, if $\sK=\cE_{\alpha_1}\cE_{\alpha_2}\cdots
\cE_{\alpha_a}$ and $\sS=\cE_{c_1}\cE_{c_2}\cdots\cE_{c_s}$, then from
(\ref{CPT38}) it follows that $\sK$ and $\sS$ commute at $m\equiv 1\pmod{2}$
and anticommute at $m\equiv 0\pmod{2}$, since in this case
$a,k\equiv 1\pmod{2}$.

In like manner we can find permutation conditions of $\sK$ with the matrix
$\sF$ of the pseudoantiautomorphism
$\cA\rightarrow\overline{\widetilde{\cA^\star}}$. Indeed, for
$\sK=\cE_{\beta_1}\cE_{\beta_2}\cdots\cE_{\beta_b}$ and
$\sF=\cE_{d_1}\cE_{d_2}\cdots\cE_{d_g}$ from (\ref{CPT45}) it follows
that $\sK$ and $\sF$ commute at $u\equiv 0\pmod{2}$ and anticommute at
$u\equiv 1\pmod{2}$, since in this case $b,k\equiv 0\pmod{2}$. Further, if
$\sK=\cE_{\alpha_1}\cE_{\alpha_2}\cdots\cE_{\alpha_a}$ and
$\sF=\cE_{c_1}\cE_{c_2}\cdots\cE_{c_s}$, then from (\ref{CPT43}) we find
that $\sK$ and $\sF$ commute at $m\equiv 0\pmod{2}$ and anticommute at
$m\equiv 1\pmod{2}$, since $a\equiv 1\pmod{2}$, $k\equiv 0\pmod{2}$.
In turn, for the matrices $\sK=\cE_{\beta_1}\cE_{\beta_2}\cdots\cE_{\beta_b}$
and $\sF=\cE_{c_1}\cE_{c_2}\cdots\cE_{c_s}$ from (\ref{CPT44}) we obtain that
$\sK$ and $\sF$ commute at $v\equiv 0\pmod{2}$ and anticommute at
$v\equiv 1\pmod{2}$, since $b\equiv 0\pmod{2}$, $p+q-k\equiv 1\pmod{2}$.
Finally, if $\sK=\cE_{\alpha_1}\cE_{\alpha_2}\cdots\cE_{\alpha_a}$ and
$\sF=\cE_{d_1}\cE_{d_2}\cdots\cE_{d_g}$, then from (\ref{CPT42}) it follows
that $\sK$ and $\sF$ commute at $l\equiv 1\pmod{2}$ and anticommute at
$l\equiv 0\pmod{2}$, since in this case $a,p+q-k\equiv 1\pmod{2}$.

It is easy to see that in virtue of similarity of the matrices $\sK$ and $\Pi$,
permutation conditions of $\sK$ with the elements of $\sAut_\pm(\cl_{p,q})$
are analogous to the conditions for $\Pi$. Indeed, if
$\sK=\cE_{\beta_1}\cE_{\beta_2}\cdots\cE_{\beta_b}$, then from (\ref{CPT47})
it follows that $\sK$ and $\sW$ always commute, since in this case
$a,b\equiv 0\pmod{2}$. In turn, if $\sK=\cE_{\alpha_1}\cE_{\alpha_2}\cdots
\cE_{\alpha_a}$, then from (\ref{CPT46}) we see that $\sK$ and $\sW$
always anticommute, since $a,b\equiv 1\pmod{2}$. Further, if
$\sK=\cE_{\beta_1}\cE_{\beta_2}\cdots\cE_{\beta_b}$ and
$\sE=\cE_{j_1}\cE_{j_2}\cdots\cE_{j_k}$, then from (\ref{CPT50}) it follows
that $\sK$ and $\sE$ commute at $u(m+v)\equiv 0\pmod{2}$ and anticommute at
$u(m+v)\equiv 1\pmod{2}$. For the matrices
$\sK=\cE_{\alpha_1}\cE_{\alpha_2}\cdots\cE_{\alpha_a}$ and
$\sE=\cE_{j_1}\cE_{j_2}\cdots\cE_{j_k}$ from (\ref{CPT48}) it follows
$m(u+l)\equiv 0,1\pmod{2}$. Correspondingly, for the matrices
$\sK=\cE_{\beta_1}\cE_{\beta_2}\cdots\cE_{\beta_b}$ and
$\sE=\cE_{i_1}\cE_{i_2}\cdots\cE_{i_{p+q-k}}$ from (\ref{CPT51}) we obtain
the condition $v(u+l)\equiv 0,1\pmod{2}$. For $\sK=\cE_{\alpha_1}\cE_{\alpha_2}
\cdots\cE_{\alpha_a}$ and $\sE=\cE_{i_1}\cE_{i_2}\cdots\cE_{i_{p+q-k}}$ 
from (\ref{CPT49}) we have $l(m+v)\equiv 0,1\pmod{2}$. In turn,
for $\sK=\cE_{\beta_1}\cE_{\beta_2}\cdots\cE_{\beta_b}$ and
$\sC=\cE_{i_1}\cE_{i_2}\cdots\cE_{i_{p+q-k}}$ from (\ref{CPT51}) we obtain
$v(u+l)\equiv 0,1\pmod{2}$, for $\sK=\cE_{\alpha_1}\cE_{\alpha_2}\cdots
\cE_{\alpha_a}$ and $\sC=\cE_{i_1}\cE_{i_2}\cdots\cE_{i_{p+q-k}}$ from
(\ref{CPT49}) we obtain $l(m+v)\equiv 0,1\pmod{2}$, for
$\sK=\cE_{\beta_1}\cE_{\beta_2}\cdots\cE_{\beta_b}$ and
$\sC=\cE_{j_1}\cE_{j_2}\cdots\cE_{j_k}$ from (\ref{CPT50}) it follows
$u(m+v)\equiv 0,1\pmod{2}$ and, finally, for
$\sK=\cE_{\alpha_1}\cE_{\alpha_2}\cdots\cE_{\alpha_a}$ and
$\sC=\cE_{j_1}\cE_{j_2}\cdots\cE_{j_k}$ from (\ref{CPT48}) we have
$m(u+l)\equiv 0,1\pmod{2}$. 

Let consider permutation conditions of the matrix $\sS$ of the
transformation $\cA\rightarrow\overline{\widetilde{\cA}}$ with other
elements of the group $\sExt(\C_n)$. Permutation conditions of $\sS$
with the matrices $\Pi$ and $\sK$ have been defined previously
(see (\ref{CPT38})--(\ref{CPT41})). Now we define permutation conditions of
$\sS$ with the matrix $\sF$ of the pseudoantiautomorphism
$\cA\rightarrow\overline{\widetilde{\cA^\star}}$ and elements of the
subgroup $\sAut_\pm(\cl_{p,q})$. So, let $\sS=\cE_{c_1}\cE_{c_2}\cdots
\cE_{c_s}$ and $\sF=\cE_{d_1}\cE_{d_2}\cdots\cE_{d_g}$, where
$s\equiv 0\pmod{2}$ and $g\equiv 0\pmod{2}$. As is known, in this case
the product $\sS$ contains all complex symmetric matrices and all
real skewsymmetric matrices of the spinbasis. In turn, the product $\sF$
contains all complex skewsymmetric and real symmetric matrices. Therefore,
the product $\sS\sF$ does not contain identical matrices. Then
\begin{equation}\label{CPT52}
\sS\sF=(-1)^{sg}\sF\sS,
\end{equation}
that is, in this case $\sS$ and $\sF$ always commute, since
$s,g\equiv 0\pmod{2}$. If $\sS=\cE_{d_1}\cE_{d_2}\cdots\cE_{d_g}$ and
$\sF=\cE_{c_1}\cE_{c_2}\cdots\cE_{c_s}$, where $s,g\equiv 1\pmod{2}$, then
from (\ref{CPT52}) it follows that $\sS$ and $\sF$ always anticommute.

Further, if $\sS=\cE_{c_1}\cE_{c_2}\cdots\cE_{c_s}$, then permutation
conditions of $\sS$ with the matrix $\sW$ of the automorphism
$\cA\rightarrow\cA^\star$ have the form:
\begin{eqnarray}
\sS\sW&=&(-1)^{\frac{s(s-1)}{2}}\sigma(c_1)\sigma(c_2)\cdots\sigma(c_s)
\cE_{d_1}\cE_{d_2}\cdots\cE_{d_g},\nonumber\\
\sW\sS&=&(-1)^{\frac{s(s-1)}{2}+sg}\sigma(c_1)\sigma(c_2)\cdots\sigma(c_s)
\cE_{d_1}\cE_{d_2}\cdots\cE_{d_g},\label{CPT53}
\end{eqnarray}
that is, in this case $\sS$ always commutes with $\sW$, since
$s,g\equiv 0\pmod{2}$. In turn, the matrix $\sS=\cE_{d_1}\cE_{d_2}\cdots
\cE_{d_g}$ always anticommutes with $\sW$, since $s,g\equiv 1\pmod{2}$.

Let us define now permutation conditions between $\sS$ and $\sE$. If
$\sS=\cE_{c_1}\cE_{c_2}\cdots\cE_{c_s}$ and
$\sE=\cE_{j_1}\cE_{j_2}\cdots\cE_{j_k}$, then the product $\sS\sE$ contains
$u$ identical real skewsymmetric matrices.
Hence it follows that $\sS=\cE_{c_1}\cE_{c_2}\cdots\cE_{c_l}\cE_{i_1}
\cE_{i_2}\cdots\cE_{i_u}$ and $\sE=\cE_{i_1}\cE_{i_2}\cdots\cE_{i_u}
\cE_{j_1}\cE_{j_2}\cdots\cE_{j_m}$. Then
\begin{eqnarray}
\sS\sE&=&(-1)^{\frac{u(u-1)}{2}}\sigma(i_1)\sigma(i_2)\cdots\sigma(i_u)
\cE_{c_1}\cE_{c_2}\cdots\cE_{c_l}\cE_{j_1}\cE_{j_2}\cdots\cE_{j_m},\nonumber\\
\sE\sS&=&(-1)^{\frac{u(u-1)}{2}+u(l+m)}\sigma(i_1)\sigma(i_2)\cdots\sigma(i_u)
\cE_{c_1}\cE_{c_2}\cdots\cE_{c_l}\cE_{j_1}\cE_{j_2}\cdots\cE_{j_m},
\label{CPT54}
\end{eqnarray}
that is, $\sS$ and $\sE$ commute at $u(l+m)\equiv 0\pmod{2}$ and
anticommute at $u(l+m)\equiv 1\pmod{2}$.
\begin{sloppypar}
In turn, the products $\sS=\cE_{d_1}\cE_{d_2}\cdots\cE_{d_g}$ and
$\sE=\cE_{j_1}\cE_{j_2}\cdots\cE_{j_k}$ contain $m$ identical complex
skewsymmetric matrices. Therefore,
$\sS=\cE_{d_1}\cE_{d_2}\cdots\cE_{d_v}\cE_{i_1}\cE_{i_2}\cdots\cE_{i_m}$,
$\sE=\cE_{i_1}\cE_{i_2}\cdots\cE_{i_m}\cE_{j_1}\cE_{j_2}\cdots\cE_{j_u}$ and
\end{sloppypar}
\begin{eqnarray}
\sS\sE&=&(-1)^{\frac{m(m-1)}{2}}\sigma(i_1)\sigma(i_2)\cdots\sigma(i_m)
\cE_{d_1}\cE_{d_2}\cdots\cE_{d_v}\cE_{j_1}\cE_{j_2}\cdots\cE_{j_u},\nonumber\\
\sE\sS&=&(-1)^{\frac{m(m-1)}{2}+m(v+u)}\sigma(i_1)\sigma(i_2)\cdots\sigma(i_m)
\cE_{d_1}\cE_{d_2}\cdots\cE_{d_v}\cE_{j_1}\cE_{j_2}\cdots\cE_{j_u},
\label{CPT55}
\end{eqnarray}
that is, permutation conditions between $\sS$ and $\sE$ in this case
have the form $m(v+u)\equiv 0,1\pmod{2}$.
\begin{sloppypar}
Further, the products $\sS=\cE_{d_1}\cE_{d_2}\cdots\cE_{d_g}$ and
$\sE=\cE_{i_1}\cE_{i_2}\cdots\cE_{i_{p+q-k}}$ contain $v$ identical real
symmetric matrices. Then $\sS=\cE_{d_1}\cE_{d_2}\cdots\cE_{d_m}\cE_{i_1}
\cE_{i_2}\cdots\cE_{i_v}$,
$\sE=\cE_{i_1}\cE_{i_2}\cdots\cE_{i_v}\cE_{i_1}\cE_{i_2}\cdots\cE_{i_l}$ and
\end{sloppypar}
\begin{eqnarray}
\sS\sE&=&(-1)^{\frac{v(v-1)}{2}}\sigma(i_1)\sigma(i_2)\cdots\sigma(i_v)
\cE_{d_1}\cE_{d_2}\cdots\cE_{d_m}\cE_{i_1}\cE_{i_2}\cdots\cE_{i_l},\nonumber\\
\sE\sS&=&(-1)^{\frac{v(v-1)}{2}+v(m+l)}\sigma(i_1)\sigma(i_2)\cdots\sigma(i_v)
\cE_{d_1}\cE_{d_2}\cdots\cE_{d_m}\cE_{i_1}\cE_{i_2}\cdots\cE_{i_l},
\label{CPT56}
\end{eqnarray}
that is, permutation conditions between $\sS$ and $\sE$ are
$v(m+l)\equiv 0,1\pmod{2}$.
\begin{sloppypar}
Finally, the products $\sS=\cE_{c_1}\cE_{c_2}\cdots\cE_{c_s}$ and
$\sE=\cE_{i_1}\cE_{i_2}\cdots\cE_{i_{p+q-k}}$ contain $l$ identical complex
symmetric matrices and, therefore,
$\sS=\cE_{c_1}\cE_{c_2}\cdots\cE_{c_u}\cE_{i_1}\cE_{i_2}\cdots\cE_{i_l}$,
$\sE=\cE_{i_1}\cE_{i_2}\cdots\cE_{i_l}\cE_{i_1}\cE_{i_2}\cdots\cE_{i_v}$.
Then\end{sloppypar}
\begin{eqnarray}
\sS\sE&=&(-1)^{\frac{l(l-1)}{2}}\sigma(i_1)\sigma(i_2)\cdots\sigma(i_l)
\cE_{c_1}\cE_{c_2}\cdots\cE_{c_u}\cE_{i_1}\cE_{i_2}\cdots\cE_{i_v},\nonumber\\
\sE\sS&=&(-1)^{\frac{l(l-1)}{2}+l(u+v)}\sigma(i_1)\sigma(i_2)\cdots\sigma(i_l)
\cE_{c_1}\cE_{c_2}\cdots\cE_{c_u}\cE_{i_1}\cE_{i_2}\cdots\cE_{i_v}
\label{CPT57}
\end{eqnarray}
and permutation conditions for $\sS$ and $\sE$ in this case have the form
$l(u+v)\equiv 0,1\pmod{2}$.

It is easy to see that permutation conditions between $\sS$ and $\sC$ are
analogous to (\ref{CPT54})--(\ref{CPT57}). Indeed, if
$\sS=\cE_{c_1}\cE_{c_2}\cdots\cE_{c_s}$ and
$\sC=\cE_{i_1}\cE_{i_2}\cdots\cE_{i_{p+q-k}}$, then from (\ref{CPT57})
it follows the comparison $l(u+v)\equiv 0,1\pmod{2}$. In turn,
if $\sS=\cE_{d_1}\cE_{d_2}\cdots\cE_{d_g}$ and 
$\sC=\cE_{i_1}\cE_{i_2}\cdots\cE_{i_{p+q-k}}$, then from (\ref{CPT56})
we obtain $v(m+l)\equiv 0,1\pmod{2}$. Further, for
$\sS=\cE_{d_1}\cE_{d_2}\cdots\cE_{d_g}$ and
$\sC=\cE_{j_1}\cE_{j_2}\cdots\cE_{j_k}$ from (\ref{CPT55}) it follows that
$m(v+u)\equiv 0,1\pmod{2}$. Analogously, for $\sS=\cE_{c_1}\cE_{c_2}\cdots
\cE_{c_s}$ and $\sC=\cE_{j_1}\cE_{j_2}\cdots\cE_{j_k}$ from (\ref{CPT54})
we have $u(l+m)\equiv 0,1\pmod{2}$.

Finally, let us consider permutation conditions of the matrix $\sF$ of
$\cA\rightarrow\overline{\widetilde{\cA^\star}}$ with other elements
of the group $\sExt(\C_n)$. Permutation conditions of $\sF$ with the
matrices $\Pi$, $\sK$ and $\sS$ have been found previously (see
(\ref{CPT42})--(\ref{CPT45}), (\ref{CPT52})). Now we define permutation
conditions between $\sF$ and the elements of the subgroups
$\sAut_\pm(\cl_{p,q})$. It is easy to see that permutation conditions between
$\sF$ and $\sW$ are equivalent to (\ref{CPT53}), that is, 
$\sF=\cE_{d_1}\cE_{d_2}\cdots\cE_{d_g}$ always commute with $\sW$,
since $s,g\equiv 0\pmod{2}$, and $\sF=\cE_{c_1}\cE_{c_2}\cdots\cE_{c_s}$
always anticommute with $\sW$, since in this case $s,g\equiv 1\pmod{2}$.
In turn, permutation conditions between $\sF$ and $\sE$ are equivalent to
(\ref{CPT54})--(\ref{CPT57}). Indeed, if
$\sF=\cE_{d_1}\cE_{d_2}\cdots\cE_{d_g}$ and 
$\sE=\cE_{j_1}\cE_{j_2}\cdots\cE_{j_k}$, then from (\ref{CPT55}) it follows
the comparison $m(v+u)\equiv 0,1\pmod{2}$. For
$\sF=\cE_{c_1}\cE_{c_2}\cdots\cE_{c_s}$ and
$\sE=\cE_{j_1}\cE_{j_2}\cdots\cE_{j_k}$ from (\ref{CPT54}) we have
$u(l+m)\equiv 0,1\pmod{2}$. Analogously, for
$\sF=\cE_{c_1}\cE_{c_2}\cdots\cE_{c_s}$ and
$\sE=\cE_{i_1}\cE_{i_2}\cdots\cE_{i_{p+q-k}}$ from (\ref{CPT57}) we obtain
$l(u+v)\equiv 0,1\pmod{2}$, and for
$\sF=\cE_{d_1}\cE_{d_2}\cdots\cE_{d_g}$, 
$\sE=\cE_{i_1}\cE_{i_2}\cdots\cE_{i_{p+q-k}}$ from (\ref{CPT56}) it follows
$v(m+l)\equiv 0,1\pmod{2}$. It is easy to see that permutation conditions
between $\sF$ and $\sC$ are equivalent to (\ref{CPT54})--(\ref{CPT57}).
Namely, for $\sF=\cE_{d_1}\cE_{d_2}\cdots\cE_{d_g}$, 
$\sC=\cE_{i_1}\cE_{i_2}\cdots\cE_{i_{p+q-k}}$ from (\ref{CPT56}) we obtain
$v(m+l)\equiv 0,1\pmod{2}$, for $\sF=\cE_{c_1}\cE_{c_2}\cdots\cE_{c_s}$,
$\sC=\cE_{i_1}\cE_{i_2}\cdots\cE_{i_{p+q-k}}$ from (\ref{CPT57}) it follows
$l(u+v)\equiv 0,1\pmod{2}$. Correspondingly, for
$\sF=\cE_{c_1}\cE_{c_2}\cdots\cE_{c_s}$, 
$\sC=\cE_{j_1}\cE_{j_2}\cdots\cE_{j_k}$ from (\ref{CPT54}) we find
$u(l+m)\equiv 0,1\pmod{2}$, and for
$\sF=\cE_{d_1}\cE_{d_2}\cdots\cE_{d_g}$,
$\sC=\cE_{j_1}\cE_{j_2}\cdots\cE_{j_k}$ from (\ref{CPT55}) we see that
$m(v+u)\equiv 0,1\pmod{2}$.

Now, we are in a position to define a detailed classification for extended
automorphism groups $\sExt(\C_n)$. First of all, since for the subalgebras
$\cl_{p,q}$ over the ring $\K\simeq\R$ the group $\sExt(\C_n)$ is reduced
to $\sAut_\pm(\C_n)$, then all essentially different groups $\sExt(\C_n)$
correspond to the subalgebras $\cl_{p,q}$ with the quaternionic ring 
$\K\simeq\BH$, $p-q\equiv 4,6\pmod{8}$. Let us classify the groups
$\sExt(\C_n)$ with respect to their subgroups $\sAut_\pm(\cl_{p,q})$.
Taking into account the structure of $\sAut_\pm(\cl_{p,q})$ at
$p-q\equiv 4,6\pmod{8}$ (see Theorem \ref{tautr}) we obtain for the group
$\sExt(\C_n)=\left\{\sI,\sW,\sE,\sC,\Pi,\sK,\sS,\sF\right\}$ the following
possible realizations:
\begin{gather}
\sExt^1(\C_n)=\left\{\sI,\cE_{12\cdots p+q},\cE_{j_1j_2\cdots j_k},
\cE_{i_1i_2\cdots i_{p+q-k}},\cE_{\alpha_1\alpha_2\cdots\alpha_a},
\cE_{\beta_1\beta_2\cdots\beta_b},\cE_{c_1c_2\cdots c_s},
\cE_{d_1d_2\cdots d_g}\right\},\nonumber\\
\sExt^2(\C_n)=\left\{\sI,\cE_{12\cdots p+q},\cE_{j_1j_2\cdots j_k},
\cE_{i_1i_2\cdots i_{p+q-k}},\cE_{\beta_1\beta_2\cdots\beta_b},
\cE_{\alpha_1\alpha_2\cdots\alpha_a},\cE_{d_1d_2\cdots d_g},
\cE_{c_1c_2\cdots c_s}\right\},\nonumber\\
\sExt^3(\C_n)=\left\{\sI,\cE_{12\cdots p+q},\cE_{i_1i_2\cdots i_{p+q-k}},
\cE_{j_1j_2\cdots j_k}, \cE_{\alpha_1\alpha_2\cdots\alpha_a},
\cE_{\beta_1\beta_2\cdots\beta_b},\cE_{d_1d_2\cdots d_g},
\cE_{c_1c_2\cdots c_s}\right\},\nonumber\\
\sExt^4(\C_n)=\left\{\sI,\cE_{12\cdots p+q},\cE_{i_1i_2\cdots i_{p+q-k}},
\cE_{j_1j_2\cdots j_k}, \cE_{\beta_1\beta_2\cdots\beta_b},
\cE_{\alpha_1\alpha_2\cdots\alpha_a},\cE_{c_1c_2\cdots c_s},
\cE_{d_1d_2\cdots d_g}\right\}.\nonumber
\end{gather}
The groups $\sExt^1(\C_n)$ and $\sExt^2(\C_n)$ have Abelian subgroups
$\sAut_-(\cl_{p,q})$ ($\dZ_2\otimes\dZ_2$ or $\dZ_4$). In turn, the groups
$\sExt^3(\C_n)$ and $\sExt^4(\C_n)$ have only non-Abelian subgroups
$\sAut_+(\cl_{p,q})$ ($Q_4/\dZ_2$ or $D_4/\dZ_2$).

Let us start with the group $\sExt^1(\C_n)$. All the elements of
$\sExt^1(\C_n)$ are even products, that is, $p+q\equiv 0\pmod{2}$,
$k\equiv 0\pmod{2}$, $a\equiv 0\pmod{2}$, $b\equiv 0\pmod{2}$,
$s\equiv 0\pmod{2}$ and $g\equiv 0\pmod{2}$. At this point, the elements
$\sI$, $\sW$, $\sE$, $\sC$ form Abelian subgroups $\dZ_2\otimes\dZ_2$ or
$\dZ_4$ (see Theorem \ref{tautr}). In virtue of (\ref{CPT37}) the element
$\Pi$ commutes with $\sK$, and from (\ref{CPT38}) it follows that $\Pi$
commutes with $\sS$ at $m\equiv 0\pmod{2}$. From (\ref{CPT42}) we see that
$\Pi$ commutes with $\sF$ at $l\equiv 0\pmod{2}$, and from (\ref{CPT46}) it
follows that $\Pi$ commutes with $\sW$. The conditions (\ref{CPT48})
show that $\Pi$ commutes with $\sE$ at $m(u+l)\equiv 0\pmod{2}$ and
commutes with $\sC$ at $l(m+v)\equiv 0\pmod{2}$. Further, the element
$\sK$ commutes with $\sS$ at $v\equiv 0\pmod{2}$ and commutes with $\sF$ at
$u\equiv 0\pmod{2}$. And also $\sK\in\sExt^1$ always commutes with $\sW$
and commutes correspondingly with $\sE$ and $\sC$ at $u(m+v)\equiv 0\pmod{2}$
and $v(u+l)\equiv 0\pmod{2}$. From (\ref{CPT52}) and (\ref{CPT53})
it follows that the element $\sS$ always commutes with $\sF$ and $\sW$.
The conditions (\ref{CPT54}) and (\ref{CPT57}) show that $\sS$ commutes
with $\sE$ and $\sC$ correspondingly at $u(l+m)\equiv 0\pmod{2}$ and
$l(u+v)\equiv 0\pmod{2}$. In virtue of (\ref{CPT53}) the element $\sF$
always commutes with $\sW$. Finally, from (\ref{CPT55}) and (\ref{CPT56})
it follows that $\sF$ commutes with $\sE$ and $\sC$ correspondingly at
$m(v+u)\equiv 0\pmod{2}$ and $v(m+l)\equiv 0\pmod{2}$. Thus,
the group $\sExt^1(\C_n)$ is Abelian at $m,l,u,v\equiv 0\pmod{2}$.
In case of the subgroup $\sAut_-(\cl_{p,q})\simeq\dZ_2\otimes\dZ_2$
($p-q\equiv 4\pmod{8}$) we obtain an Abelian group
$\sExt^1_-(\C_n)\simeq\dZ_2\otimes\dZ_2\otimes\dZ_2$ with the signature
$(+,+,+,+,+,+,+)$ for the elements $\Pi$, $\sK$, $\sS$ and $\sF$ with
positive squares ($m-l\equiv 0,4\pmod{8}$, $v-u\equiv 0,4\pmod{8}$,
$u+l\equiv 0,4\pmod{8}$ and $m+v\equiv 0,4\pmod{8}$).
It is easy to see that for the type $p-q\equiv 4\pmod{8}$ there exists also
$\sExt^1_-(\C_n)\simeq\dZ_4\otimes\dZ_2$ with the signature
$(+,+,+,-,-,-,-)$ and the subgroup $\dZ_2\otimes\dZ_2$, where
$m-l\equiv 2,6\pmod{8}$, $v-u\equiv 2,6\pmod{8}$, $u+l\equiv 2,6\pmod{8}$ and
$m+v\equiv 2,6\pmod{8}$. Further, for the type $p-q\equiv 4\pmod{8}$
there exist Abelian groups $\sExt^1_-(\C_n)\simeq\dZ_4\otimes\dZ_2$ with
the signatures $(+,-,-,d,e,f,g)$ and subgroups $\dZ_4$, where among the
symbols $d$, $e$, $f$, $g$ there are two pluses and two minuses.
Correspondingly, at $m,v,l,u\equiv 0\pmod{2}$ for the type
$p-q\equiv 6\pmod{8}$ there exist Abelian groups 
$\sExt^1_-(\C_n)\simeq\dZ_4\otimes\dZ_2$ with the signatures
$(-,+,-,d,e,f,g)$ and $(-,-,+,d,e,f,g)$ if $m-u\equiv 0,1,4,5\pmod{8}$,
$v-l\equiv 2,3,6,7\pmod{8}$ and $m-u\equiv 2,3,6,7\pmod{8}$,
$v-l\equiv 0,1,4,5\pmod{8}$.

It is easy to see that from the comparison $k,a,b,s,g\equiv 0\pmod{2}$
it follows that the numbers $m$, $v$, $l$, $u$ are simultaneously even or
odd. The case $m,v,l,u\equiv 0\pmod{2}$, considered previously, gives rise
to the Abelian groups $\sExt^1_-$. In contrast to this, the case
$m,v,l,u\equiv 1\pmod{2}$ gives rise to non-Abelian groups $\sExt^1_+$
with Abelian subgroups $\dZ_4$ and $\dZ_2\otimes\dZ_2$ (it follows from
(\ref{CPT37})--(\ref{CPT38})). Namely, in this case we have the group
$\sExt^1_+\simeq\overset{\ast}{\dZ}_4\otimes\dZ_2$ with the subgroup $\dZ_4$
for $p-q\equiv 4,6\pmod{8}$ (it should be noted that signatures of
the groups $\dZ_4\otimes\dZ_2$ and $\overset{\ast}{\dZ}_4\otimes\dZ_2$
do not coincide), the group $\sExt^1_+\simeq Q_4$ with the subgroup
$\dZ_4$ for $p-q\equiv 4,6\pmod{8}$ and $\sExt^1_+\simeq D_4$ with
$\dZ_2\otimes\dZ_2$ for the type $p-q\equiv 4\pmod{8}$.

Let us consider now the group $\sExt^2(\C_n)$. In this case among the
elements of $\sExt^2$ there are both even and odd elements:
$k\equiv 0\pmod{2}$, $b\equiv 1\pmod{2}$, $a\equiv 1\pmod{2}$,
$g\equiv 1\pmod{2}$ and $s\equiv 1\pmod{2}$. At this point, the elements
$\sI$, $\sW$, $\sE$, $\sC$ form Abelian subgroups $\dZ_2\otimes\dZ_2$
and $\dZ_4$. In virtue of (\ref{CPT37}) the element $\Pi$ always
anticommutes with $\sK$, and from (\ref{CPT52}) it follows that the
elements $\sS$ and $\sF$ always anticommute. Therefore, all the groups
$\sExt^2$ are non-Abelian. Among these groups there are the following
isomorphisms: $\sExt^2_+\simeq\overset{\ast}{\dZ}_4\otimes\dZ_2$ with
the signatures $(+,-,-,d,e,f,g)$ for the type $p-q\equiv 4\pmod{8}$ and
$(-,+,-,d,e,f,g)$, $(-,-,+,d,e,f,g)$ for $p-q\equiv 6\pmod{8}$, where among
the symbols $d$, $e$, $f$, $g$ there are two pluses and two minuses;
$\sExt^2_+\simeq Q_4$ with $(+,-,-,-,-,-,-)$ for $p-q\equiv 4\pmod{8}$ and
$(-,+,-,-,-,-,-)$, $(-,-,+,-,-,-,-)$ for $p-q\equiv 6\pmod{8}$;
$\sExt^2_+\simeq D_4$ with $(a,b,c,+,+,+,+)$ and $(+,+,+,d,e,f,g)$, where
among $a$, $b$, $c$ there are two minuses and one plus, and among
$d$, $e$, $f$, $g$ there are two pluses and two minuses. For all the
groups $\sExt^2$ among the numbers $m$, $v$, $u$, $l$ there are both even and
odd numbers.
\begin{sloppypar}
Let consider the group $\sExt^3(\C_n)$. First of all, the groups
$\sExt^3$ contain non-Abelian subgroups $\sAut_+(\cl_{p,q})$ (the elements
$\sE$ and $\sC$ are odd). Therefore, all the groups $\sExt^3$ are
non-Abelian. Among these groups there are the following isomorphisms:
$\sExt^3_+\simeq D_4$ with $(+,-,+,d,e,f,g)$ and $(+,+,-,d,e,f,g)$ for the
type $p-q\equiv 4\pmod{8}$, where among $d$, $e$, $f$, $g$ there are
three pluses and one minus; $\sExt^3_+\simeq Q_4$ with $(-,-,-,d,e,f,g)$,
where among $d$, $e$, $f$, $g$ there are one plus and three minuses;
$\sExt^3_+\simeq D_4$ with $(-,+,+,d,e,f,g)$, where among $d$, $e$, $f$, $g$
there are three pluses and one minus (the type $p-q\equiv 6\pmod{8}$).
Besides, there exist the groups $\sExt^3_+\simeq\overset{\ast}{\dZ}_4
\otimes\dZ_2$ with the signatures $(+,-,+,d,e,f,g)$, $(+,+,-,d,e,f,g)$
for the type $p-q\equiv 4\pmod{8}$ and $(-,-,-,d,e,f,g)$,
$(-,+,+,d,e,f,g)$ for $p-q\equiv 6\pmod{8}$, where among $d$, $e$, $f$, $g$
there are one plus and three minuses.
\end{sloppypar}
Finally, let us consider the group $\sExt^4(\C_n)$. These groups contain
non--Abelian subgroups $\sAut_+(\cl_{p,q})$ and, therefore, all
$\sExt^4$ are non--Abelian. The isomorphism structure of $\sExt^4$ is
similar to $\sExt^3$.

It is easy to see that a full number of all possible signatures
$(a,b,c,d,e,f,g)$ is equal to $2^7=128$. At this point, we have eight 
signature types: (seven `$+$'), (one `$-$', six `$+$'),
(two `$-$', five `$+$'), (three `$-$', four `$+$'),
(four `$-$', three `$+$'), (five `$-$', two `$+$'),
(six `$-$', one `$+$'), (seven `$-$'). However, only four types from
enumerated above correspond to finite groups of order 8:
(seven `$+$') $\rightarrow\dZ_2\otimes\dZ_2\otimes\dZ_2$,
(two `$-$', five `$+$') $\rightarrow D_4$,
(four `$-$', three `$+$') $\rightarrow\dZ_4\otimes\dZ_2$
($\overset{\ast}{\dZ}_4\otimes\dZ_2$) and
(six `$-$', one `$+$') $\rightarrow Q_4$. Therefore, for the group
$\sExt$ there exist 64 different realizations.
\end{proof}
%\begin{ex} 
{\it Example 2.}
Let us study an extended automorphism group of the Dirac
algebra\index{algebra!Dirac} $\C_4$. 
We evolve in $\C_4$ the real subalgebra with the
quaternionic ring. Let it be the spacetime algebra $\cl_{p,q}$ with a
spinbasis defined by the matrices (\ref{GammaB}). We define now elements
of the group $\sExt(\C_4)$. First of all, the matrix of the automorphism
$\cA\rightarrow\cA^\star$ has a form: $\sW=\gamma_0\gamma_1\gamma_2\gamma_3$.
Further, since
\[
\gamma^{\sT}_0=\gamma_0,\quad\gamma^{\sT}_1=-\gamma_1,\quad
\gamma^{\sT}_2=-\gamma_2,\quad\gamma^{\sT}_3=-\gamma_3,
\]
then in accordance with Theorem \ref{tautr} the matrix $\sE$ of the
antiautomorphism $\cA\rightarrow\widetilde{\cA}$ is an even product of
skewsymmetric matrices of the spinbasis (\ref{GammaB}), that is,
$\sE=\gamma_1\gamma_3$. From the definition $\sC=\sE\sW$ we find that
the matrix of the antiautomorphism $\cA\rightarrow\widetilde{\cA^\star}$
has a form $\sC=\gamma_0\gamma_2$. $\gamma$--basis contains three
real matrices $\gamma_0$, $\gamma_1$ and $\gamma_3$, therefore, for the
matrix of the pseudoautomorphism $\cA\rightarrow\overline{\cA}$ we obtain
$\Pi=\gamma_0\gamma_1\gamma_3$ (see Theorem \ref{tpseudo}). Further,
in accordance with $\sK=\Pi\sW$ for the matrix of the pseudoautomorphism
$\cA\rightarrow\overline{\cA^\star}$ we have $\sK=\gamma_2$. Finally,
for the pseudoantiautomorphisms 
$\cA\rightarrow\overline{\widetilde{\cA}}$,
$\cA\rightarrow\overline{\widetilde{\cA^\star}}$ from the definitions
$\sS=\Pi\sE$, $\sF=\Pi\sC$ it follows $\sS=\gamma_0$,
$\sF=\gamma_1\gamma_2\gamma_3$. Thus, we come to the following extended
automorphism group:\index{group!automorphism!extended}
\begin{multline}
\sExt(\C_4)\simeq\{\sI,\,\sW,\,\sE,\,\sC,\,\Pi,\,\sK,\,\sS,\,\sF\}\simeq\\
\{\sI,\,\gamma_0\gamma_1\gamma_2\gamma_3,\,\gamma_1\gamma_3,\,
\gamma_0\gamma_2,\,\gamma_0\gamma_1\gamma_3,\,\gamma_2,\,\gamma_0,\,
\gamma_1\gamma_2\gamma_3\}.\label{Dirac3}
\end{multline}
The Cayley tableau of this group has a form
\begin{center}{\renewcommand{\arraystretch}{1.4}
\begin{tabular}{|c||c|c|c|c|c|c|c|c|}\hline
  & $\sI$ & $\gamma_{0123}$ & $\gamma_{13}$ & $\gamma_{02}$ & $\gamma_{013}$ &
$\gamma_{2}$ & $\gamma_{0}$ & $\gamma_{123}$\\ \hline\hline
$\sI$  & $\sI$ & $\gamma_{0123}$ & $\gamma_{13}$ & $\gamma_{02}$ & $\gamma_{013}$ &
$\gamma_{2}$ & $\gamma_{0}$ & $\gamma_{123}$\\ \hline
$\gamma_{0123}$ & $\gamma_{0123}$ & $-\sI$ & $\gamma_{02}$ & $-\gamma_{012}$ 
& $-\gamma_2$ &
$\gamma_{013}$ & $-\gamma_{123}$ & $\gamma_{0}$\\ \hline
$\gamma_{13}$ & $\gamma_{13}$ & $\gamma_{02}$ & $-\sI$ & $-\gamma_{0123}$ &
$-\gamma_{0}$ & $-\gamma_{123}$ & $\gamma_{013}$ & $\gamma_2$\\ \hline
$\gamma_{02}$ & $\gamma_{02}$ & $-\gamma_{13}$ & $-\gamma_{0123}$ &
$\sI$ & $\gamma_{123}$ & $-\gamma_{0}$ & $-\gamma_2$ & 
$\gamma_{013}$\\ \hline
$\gamma_{013}$ & $\gamma_{013}$ & $\gamma_2$ & $-\gamma_{0}$ &
$-\gamma_{123}$ & $-\sI$ & $-\gamma_{0123}$ & $\gamma_{13}$ &
$\gamma_{02}$\\ \hline
$\gamma_2$ & $\gamma_2$ & $-\gamma_{013}$ & $-\gamma_{123}$ & $\gamma_{0}$ &
$\gamma_{0123}$ & $-\sI$ & $-\gamma_{02}$ & $\gamma_{13}$\\ \hline
$\gamma_{0}$ & $\gamma_{0}$ & $\gamma_{123}$ & $\gamma_{013}$ &
$\gamma_2$ & $\gamma_{13}$ & $\gamma_{02}$ & $\sI$ & $\gamma_{0123}$\\ \hline
$\gamma_{123}$ & $\gamma_{123}$ & $-\gamma_{0}$ & $\gamma_2$ &
$-\gamma_{013}$ & $-\gamma_{02}$ & $\gamma_{13}$ & $-\gamma_{0123}$ & 
$\sI$\\ \hline
\end{tabular}\;\;$\sim$
}
\end{center}
\begin{center}{\renewcommand{\arraystretch}{1.4}
\begin{tabular}{|c||c|c|c|c|c|c|c|c|}\hline
     & $\sI$  & $\sW$  & $\sE$  & $\sC$ & $\Pi$  & $\sK$ & $\sS$ & $\sF$ \\ \hline\hline
$\sI$  & $\sI$  & $\sW$  & $\sE$  & $\sC$ & $\Pi$  & $\sK$ & $\sS$ & $\sF$ \\ \hline
$\sW$  & $\sW$  & $-\sI$  & $\sC$ & $-\Pi$  & $-\sK$ & $\Pi$  & $-\sF$& $\sS$\\ \hline
$\sE$  & $\sE$  & $\sC$ & $-\sI$  & $-\sW$  & $-\sS$ & $-\sF$& $\Pi$  & $\sK$\\ \hline
$\sC$ & $\sC$ & $-\sE$  & $-\sW$  & $\sI$  & $\sF$& $-\sS$ & $-\sK$ & $\Pi$\\ \hline
$\Pi$  & $\Pi$  & $\sK$ & $-\sS$ & $-\sF$& $-\sI$  & $-\sW$  & $\sE$  & $\sC$\\ \hline
$\sK$ & $\sK$ & $-\Pi$  & $-\sF$& $\sS$ & $\sW$  & $-\sI$  & $-\sC$ & $\sE$\\ \hline
$\sS$ & $\sS$ & $\sF$& $\Pi$  & $\sK$ & $\sE$  & $\sC$ & $\sI$  & $\sW$\\ \hline
$\sF$& $\sF$& $-\sS$ & $\sK$ & $-\Pi$  & $-\sC$ & $\sE$  & $-\sW$  & $\sI$\\ \hline
\end{tabular}
}
\end{center}
As follows from this tableau, the group $\sExt(\C_4)$ is non-Abelian.
$\sExt_+(\C_4)$ contains Abelian group of spacetime reflections,
$\sAut_-(\cl_{1,3})\simeq\dZ_4$, as a subgroup. It is easy to see that the group
(\ref{Dirac3}) is a group of the form $\sExt^2_+$ with order structure
$(3,4)$. More precisely, the group (\ref{Dirac3}) is a finite group
$\overset{\ast}{\dZ}_4\otimes\dZ_2$ with the signature
$(-,-,+,-,-,+,+)$.

Coming back to example 1, we see that the groups (\ref{DirG2}) and
(\ref{Dirac3}) are isomorphic,
\[
\{1,\,P,\,T,\,PT,\,C,\,CP,\,CT,\,CPT\}\simeq
\{\sI,\,\sW,\,\sE,\,\sC,\,\Pi,\,\sK,\,\sS,\,\sF\}\simeq
\overset{\ast}{\dZ}_4\otimes\dZ_2.
\]
Moreover, the subgroups of spacetime reflections of these groups are
also isomorphic:
\[
\{1,\,P,\,T,\,PT\}\simeq\{\sI,\,\sW,\sE,\,\sC\}\simeq\dZ_4.
\]
Thus, we come to the following result: the finite group (\ref{DirG2}),
derived from the analysis of invariance properties of the Dirac equation
with respect to discrete transformations $C$, $P$ and $T$, is isomorphic
to an extended automorphism group of the Dirac algebra $\C_4$.
This result allows us to study discrete symmetries and their group
structure for physical fields of any spin (without handling to analysis
of relativistic wave equations).
%\end{ex}

\section{$CPT$--structures}
As it has been shown previously, there exist 64 different signatures
$(a,b,c,d,e,f,g)$ for the extended automorphism group $\sExt(\C_n)$ of the
complex Clifford algebra $\C_n$. At this point, the group of fundamental
automorphisms, $\sAut_\pm(\cl_{p,q})$, which has 8 different signatures
$(a,b,c)$, is defined as a subgroup of $\sExt(\C_n)$. As is known,
the Clifford--Lipschitz group\index{group!Clifford-Lipschitz}
$\pin(n,\C)$ (an universal covering of the
complex orthogonal group $O(n,\C)$) is completely constructed within
the algebra $\C_n$ (see definition (\ref{CL})). If we take into account
spacetime reflections, then according to \cite{Shi58,Shi60,Dab88} there
exist 8 types of universal covering ($PT$--structures) described by the
group $\pin^{a,b,c}(n,\C)$. As it shown in \cite{Var99,Var00}, the
group $\pin^{a,b,c}(n,\C)$ (correspondingly $\pin^{a,b,c}(p,q)$ over the
field $\F=\R$) is completely defined within the algebra $\C_n$
(correspondingly $\cl_{p,q}$) by means of identification of the reflection
subgroup $\{1,P,T,PT\}$ with the automorphism group
$\{\Id,\star,\widetilde{\phantom{cc}},\widetilde{\star}\}$ of $\C_n$
(correspondingly $\cl_{p,q}$). In turn, the pseudoautomorphism
$\cA\rightarrow\overline{\cA}$ of $\C_n$ and the extended automorphism 
group $\{\Id,\star,\widetilde{\phantom{cc}},\widetilde{\star},
\overline{\phantom{cc}},\overline{\star},\overline{\widetilde{\phantom{cc}}},
\overline{\widetilde{\star}}\}\simeq\dZ_2\otimes\dZ_2\otimes\dZ_2$
allows us to give a further generalization of the Clifford--Lipschitz
group (\ref{CL}) (correspondingly (\ref{Pin})). We claim that there
exist 64 types of the universal covering ($CPT$--structures) for the complex
orthogonal group $O(n,\C)$:
\begin{equation}\label{CCL}
\pin^{a,b,c,d,e,f,g}(n,\C)\simeq
\frac{(\spin_+(n,\C)\odot C^{a,b,c,d,e,f,g})}{\dZ_2},
\end{equation}
where $C^{a,b,c,d,e,f,g}$ are five double coverings of the group
$\dZ_2\otimes\dZ_2\otimes\dZ_2$. All the posiible double coverings
$C^{a,b,c,d,e,f,g}$ are given in the following table:
\begin{center}{\renewcommand{\arraystretch}{1.4}
\begin{tabular}{|l|l|l|}\hline
$a\;b\;c\;d\;e\;f\;g$ & $C^{a,b,c,d,e,f,g}$ & Type \\ \hline
$+\;+\;+\;+\;+\;+\;+$ & $\dZ_2\otimes\dZ_2\otimes\dZ_2\otimes\dZ_2$ & Abelian \\ 
three `$+$' and four `$-$' & $\dZ_4\otimes\dZ_2\otimes\dZ_2$ & \\ \hline
one `$+$' and six `$-$' & $Q_4\otimes\dZ_2$ & Non--Abelian \\
five `$+$' and two `$-$' & $D_4\otimes\dZ_2$ &   \\
three `$+$' and four `$-$' & $\overset{\ast}{\dZ}_4\otimes\dZ_2\otimes\dZ_2$
& \\ \hline
\end{tabular}
}
\end{center}
The group (\ref{CCL}) with non--Abelian $C^{a,b,c,d,e,f,g}$ we will call
{\it Cliffordian}\index{group!Cliffordian}
and respectively {\it non--Cliffordian}\index{group!non-Cliffordian} when
$C^{a,b,c,d,e,f,g}$ is Abelian.

Analogously, over the field $\F=\R$ there exist 64 universal coverings
of the real orthogonal group $O(p,q)$:
\[
\rho^{a,b,c,d,e,f,g}:\;\pin^{a,b,c,d,e,f,g}\longrightarrow O(p,q),
\]
where
\begin{equation}\label{RCL}
\pin^{a,b,c,d,e,f,g}(p,q)\simeq
\frac{(\spin_+(p,q)\odot C^{a,b,c,d,e,f,g})}{\dZ_2}.
\end{equation}
It is easy to see that in case of the algebra $\cl_{p,q}$ (or subalgebra
$\cl_{p,q}\subset\C_n$) with the real division ring $\K\simeq\R$,
$p-q\equiv 0,2\pmod{8}$, $CPT$--structures, defined by the groups
(\ref{CCL}) and (\ref{RCL}), are reduced to the eight
Shirokov--D\c{a}browski $PT$--structures.

Further, using the well-known isomorphism (\ref{PinoddC}), we obtain for
the group $O(n,\C)$ with odd dimensionality the following universal
covering:
\[
\pin^{a,b,c,d,e,f,g}(n,\C)\simeq\pin^{a,b,c,d,e,f,g}(n-1,\C)\bigcup
\e_{12\cdots n}\pin^{a,b,c,d,e,f,g}(n-1,\C).
\]
Correspondingly, in virtue of $\cl_{p,q}\simeq\C_{n-1}$ 
($p-q\equiv 3,7\pmod{8}$, $n=p+q$)
and (\ref{Pinodd1})
for the group $O(p,q)$ with odd dimensionality we have
\[
\pin^{a,b,c,d,e,f,g}(p,q)\simeq\pin^{a,b,c,d,e,f,g}(n-1,\C)
\]
for $p-q\equiv 3,7\pmod{8}$ and
\begin{eqnarray}
\pin^{a,b,c,d,e,f,g}(p,q)&\simeq&\pin^{a,b,c,d,e,f,g}(p,q-1)\bigcup
\e_{12\cdots n}\pin^{a,b,c,d,e,f,g}(p,q-1),\nonumber\\
\pin^{a,b,c,d,e,f,g}(p,q)&\simeq&\pin^{a,b,c,d,e,f,g}(q,p-1)\bigcup
\e_{12\cdots n}\pin^{a,b,c,d,e,f,g}(q,p-1)\nonumber
\end{eqnarray}
for the types $p-q\equiv 1,5\pmod{8}$.
\begin{theorem}
Let $\pin^{a,b,c,d,e,f,g}(n,\C)$ be an universal covering of the complex
orthogonal group $O(n,\C)$ of the space $\C^n$ associated with the complex
algebra $\C_n$. Squares of the symbols $a,b,c,d,e,f,g\in\{-,+\}$
correspond to squares of the elements of the finite group
$\sExt=\{\sI,\sW,\sE,\sC,\Pi,\sK,\sS,\sF\}$: $a=\sW^2$, $b=\sE^2$,
$c=\sC^2$, $d=\Pi^2$, $e=\sK^2$, $f=\sS^2$, $g=\sF^2$, where
$\sW$, $\sE$, $\sC$, $\Pi$, $\sK$, $\sS$, $\sF$ are matrices of the
automorphisms $\cA\rightarrow\cA^\star$, $\cA\rightarrow\widetilde{\cA}$,
$\cA\rightarrow\widetilde{\cA^\star}$, $\cA\rightarrow\overline{\cA}$,
$\cA\rightarrow\overline{\cA^\star}$, $\cA\rightarrow
\overline{\widetilde{\cA}}$, 
$\cA\rightarrow\overline{\widetilde{\cA^\star}}$. Then over the field
$\F=\C$ in dependence on a division ring structure $\K=f\cl_{p,q}f$ of
the real subalgebras $\cl_{p,q}\subset\C_n$, there exist the following
universal coverings ($CPT$--structures) of the group $O(n,\C)$:\\[0.2cm]
1) $\K\simeq\R$, $p-q\equiv 0,2\pmod{8}$.\\
In this case $CPT$--structures are reduced to the eight
Shirokov--D\c{a}browski $PT$--structures
\[
\pin^{a,b,c}(n,\C)\simeq\frac{(\spin_+(n,\C)\odot C^{a,b,c})}{\dZ_2},
\]
where $C^{a,b,c}$ are double coverings of the group
$\{1,P,T,PT\}\simeq\{\sI,\sW,\sE,\sC\}\simeq\dZ_2\otimes\dZ_2$.\\[0.2cm]
2) $\K\simeq\BH$, $p-q\equiv 4,6\pmod{8}$.\\
In this case we have 64 universal coverings:\\
a) Non--Cliffordian group\index{group!non-Cliffordian}
\[
\pin^{+,+,+,+,+,+,+}(n,\C)\simeq
\frac{(\spin_+(n,\C)\odot\dZ_2\otimes\dZ_2\otimes\dZ_2\otimes\dZ_2)}{\dZ_2}
\]
exists if the subalgebra $\cl_{p,q}$ admits the type $p-q\equiv 4\pmod{8}$.
Non--Cliffordian groups
\[
\pin^{a,b,c,d,e,f,g}(n,\C)\simeq
\frac{(\spin_+(n,\C)\odot\dZ_4\otimes\dZ_2\otimes\dZ_2)}{\dZ_2}
\]
exist with the signature $(+,+,+,-,-,-,-)$ when $\cl_{p,q}$ has the type
$p-q\equiv 4\pmod{8}$, and also these groups with the
signatures $(a,b,c,d,e,f,g)$ exist when
$p-q\equiv 6\pmod{8}$, where among the symbols $a$, $b$, $c$ there are
two minuses and one plus, and among $d$, $e$, $f$, $g$ -- two pluses and
two minuses.\\
b) Cliffordian groups\index{group!Cliffordian}
\[
\pin^{a,b,c,d,e,f,g}(n,\C)\simeq
\frac{(\spin_+(n,\C)\odot Q_4\otimes\dZ_2)}{\dZ_2}
\]
exist with the signature $(+,-,-,-,-,-,-)$ when $p-q\equiv 4\pmod{8}$ and with
the signatures $(-,+,-,-,-,-,-)$, $(-,-,+,-,-,-,-)$ when
$p-q\equiv 6\pmod{8}$. And also these groups exist with the signature
$(-,-,-,d,e,f,g)$ if $p-q\equiv 6\pmod{8}$, where among the symbols
$d$, $e$, $f$, $g$ there are one plus and three minuses.
Cliffordian groups
\[
\pin^{a,b,c,d,e,f,g}(n,\C)\simeq
\frac{(\spin_+(n,\C)\odot D_4\otimes\dZ_2)}{\dZ_2}
\]
exist with the signatures $(+,-,-,+,+,+,+)$ and $(+,+,+,d,e,f,g)$ when
$\cl_{p,q}$ has the type $p-q\equiv 4\pmod{8}$ and among $d$, $e$, $f$, $g$
there are two minuses and two pluses, and also these groups exist with
$(+,-,+,d,e,f,g)$ and $(+,+,-,d,e,f,g)$, where among $d$, $e$, $f$, $g$
there are three pluses and one minus. Cliffordian groups of this type
exist also with the signatures $(a,b,c,+,+,+,+)$ and $(-,+,+,d,e,f,g)$ when
$p-q\equiv 6\pmod{8}$, where among $a$, $b$, $c$ there are two minuses and
one plus, and among $d$, $e$, $f$, $g$ there are three pluses and one minus.
Cliffordian groups
\[
\pin^{a,b,c,d,e,f,g}(n,\C)\simeq
\frac{(\spin_+(n,\C)\odot\overset{\ast}{\dZ}_4\otimes\dZ_2\otimes\dZ_2)}
{\dZ_2}
\]
exist with the signatures $(+,-,-,d,e,f,g)$ when $p-q\equiv 4\pmod{8}$ and
among the symbols $d$, $e$, $f$, $g$ there are two pluses and two minuses,
and also these groups exist with $(+,-,+,d,e,f,g)$ and $(+,+,-,d,e,f,g)$,
where $p-q\equiv 4\pmod{8}$ and among $d$, $e$, $f$, $g$ there are one plus
and three minuses. Cliffordian groups of this type exist also with
$(-,+,-,d,e,f,g)$ and $(-,-,+,d,e,f,g)$ if $p-q\equiv 6\pmod{8}$, where
among $d$, $e$, $f$, $g$ there are two pluses and two minuses. And also
these groups exist with $(-,-,-,d,e,f,g)$ if $p-q\equiv 6\pmod{8}$, where
among $d$, $e$, $f$, $g$ there are three pluses and one minus.
\end{theorem}
\end{document}